\documentclass[fleqn,useAMS,usenatbib]{mnras}

\usepackage[T1]{fontenc}
\usepackage{ae,aecompl}

\usepackage{graphicx}  

\usepackage[fleqn]{amsmath}
\usepackage{amsfonts}
\usepackage{amssymb}
\usepackage[]{color}

\usepackage{srcltx}

\allowdisplaybreaks[4]
\bibpunct{(}{)}{;}{a}{}{,}

\def\figrelpath{}

\def\bibpath{Hilbert_et_al_Intrinsic_Alignments_in_Illustris}

\newcommand{\satellitename}[1]{\textit{#1}}

\newcommand{\softwarename}[1]{\textsc{#1}}

\newcommand{\vect}[1]{\boldsymbol{#1}}

\newcommand{\N}{\mathbb{N}}

\newcommand{\R}{\mathbb{R}}

\newcommand{\C}{\mathbb{C}}

\newcommand{\ii}{\mathrm{i}}
\newcommand{\ee}{\mathrm{e}}

\newcommand{\parder}[3][]{\frac{\partial^{#1} {#2}}{\partial {#3}^{#1}}}

\newcommand{\totder}[3][]{\frac{\mathrm{d}^{#1} {#2}}{\mathrm{d} {#3}^{#1}}}

\newcommand{\diff}[2][]{\mathrm{d}^{#1}{#2}}
\newcommand{\idiff}[2][]{\!\!\mathrm{d}^{#1}{#2}}

\newcommand{\pdf}{\ensuremath{\operatorname{pdf}}}
\newcommand{\EV}[1]{\left\langle{#1}\right\rangle}
\newcommand{\bEV}[1]{\bigl\langle{#1}\bigr\rangle}

\newcommand{\cconj}[1]{#1^*}
\newcommand{\est}[1]{\hat{#1}}

\newcommand{\ft}[1]{\tilde{#1}}

\newcommand{\DiracDelta}{\delta_\mathrm{D}}
\newcommand{\HeavisideTheta}{\Theta_{\mathrm{H}}}
\newcommand{\const}{\text{const.}}

\newcommand{\kms}{\ensuremath{\mathrm{km\,s}^{-1}}}

\newcommand{\pc}{\ensuremath{\mathrm{pc}}}
\newcommand{\kpc}{\ensuremath{\mathrm{kpc}}}
\newcommand{\Mpc}{\ensuremath{\mathrm{Mpc}}}

\newcommand{\Angs}{\ensuremath{\text{\AA}}}

\newcommand{\Msolar}{\ensuremath{\mathrm{M}_\odot}}

\newcommand{\arcsect}{\ensuremath{\mathrm{arcsec}}}
\newcommand{\arcmint}{\ensuremath{\mathrm{arcmin}}}

\newcommand{\clight}{\ensuremath{\mathrm{c}}}

\newcommand{\HubbleConstant}{H_{0}}

\newcommand{\OmegaMatter}{\Omega_{\mathrm{m}}}
\newcommand{\OmegaBaryon}{\Omega_{\mathrm{b}}}

\newcommand{\RadialFluxWeight}{W}
\newcommand{\ImageSurfaceBrightness}{I}
\newcommand{\ObsImageSurfaceBrightness}{I}
\newcommand{\IntImageSurfaceBrightness}{I^{\text{s}}}

\newcommand{\ObsImageSecondMoment}[1]{Q_{#1}}
\newcommand{\IntImageSecondMoment}[1]{Q^{\text{s}}_{#1}}

\newcommand{\HalfLightRadius}{\theta_{\text{hl}}}

\newcommand{\Mstellar}{M_{*}}
\newcommand{\Msub}{M{_\text{sub}}}
\newcommand{\Msubmin}{M_{\text{sub,min}}}

\newcommand{\fK}{f_K}

\newcommand{\chiD}{\chi_{\mathrm{d}}}
\newcommand{\fD}{f_{\mathrm{d}}}

\newcommand{\zS}{z_{\mathrm{s}}}

\newcommand{\chiS}{\chi_{\mathrm{s}}}
\newcommand{\fS}{f_{\mathrm{s}}}

\newcommand{\zL}{z_{\mathrm{d}}}

\newcommand{\chiL}{\chi_{\mathrm{d}}}
\newcommand{\fL}{f_{\mathrm{d}}}

\newcommand{\fSL}{f_{\mathrm{ds}}}

\newcommand{\vx}{\vect{x}}
\newcommand{\vr}{\vect{r}}
\newcommand{\vk}{\vect{k}}
\newcommand{\vbeta}{\vect{\beta}}
\newcommand{\vtheta}{\vect{\theta}}
\newcommand{\vvartheta}{\vect{\vartheta}}
\newcommand{\vell}{\vect{\ell}}

\newcommand{\geomweight}{q}

\newcommand{\rot}{\text{\textsc{r}}}
\newcommand{\tang}{\text{\textsc{t}}}
\newcommand{\cross}{\!\times\!}

\newcommand{\Npart}[2]{N^{#1}_{\text{p}\,#2}}
\newcommand{\mpart}[2]{m^{#1}_{\text{p}\,#2}}
\newcommand{\xpart}[2]{\vect{x}^{#1}_{\text{p}\,#2}}
\newcommand{\chipart}[2]{\chi^{#1}_{\text{p}\,#2}}

\newcommand{\rhocrit}{\rho_{\text{crit}}}
\newcommand{\rhoMatterMean}{\bar{\rho}_{\mathrm{m}}}
\newcommand{\gravitationalfield}{\phi_{\mathrm{m}}}
\newcommand{\tidalfield}{\tau}

\newcommand{\tidalfieldtoellipticityfactor}{\chi_\tau}

\newcommand{\deltaMatter}{\delta_{\mathrm{m}}}
\newcommand{\ftdeltaMatter}{\ft{\delta}_{\mathrm{m}}}

\newcommand{\deltashearMattersymbol}{\sigma}

\newcommand{\deltashearMatter}{\deltashearMattersymbol}
\newcommand{\ftdeltashearMatter}{\ft{\deltashearMattersymbol}}

\newcommand{\deltashearMattert}{\deltashearMattersymbol_{\tang}}
\newcommand{\deltashearMatterx}{\deltashearMattersymbol_{\cross}}
\newcommand{\deltashearMattertx}{\deltashearMattersymbol_{\tang/\cross}}

\newcommand{\gammat}{\gamma_{\tang}}
\newcommand{\gammax}{\gamma_{\cross}}

\newcommand{\obseta}[2]{\eta^{#1}_{#2}}
\newcommand{\inteta}[2]{\eta^{\text{s}\,#1}_{#2}}

\newcommand{\obse}[2]{e^{#1}_{#2}}
\newcommand{\inte}[2]{e^{\text{s}\,#1}_{#2}}

\newcommand{\obsellraw}[2]{h^{#1}_{#2}}
\newcommand{\intellraw}[2]{h^{\text{s}\,#1}_{#2}}

\newcommand{\obsell}[2]{\varepsilon^{#1}_{#2}}
\newcommand{\intell}[2]{\varepsilon^{\text{s}\,#1}_{#2}}

\newcommand{\obsellt}[2]{\obsell{#1}{\tang\,#2}}
\newcommand{\intellt}[2]{\intell{#1}{\tang\,#2}}

\newcommand{\obsellx}[2]{\obsell{#1}{\cross\,#2}}
\newcommand{\intellx}[2]{\intell{#1}{\cross\,#2}}

\newcommand{\obselltx}[2]{\obsell{#1}{\tang/\cross\,#2}}
\newcommand{\intelltx}[2]{\intell{#1}{\tang/\cross\,#2}}

\newcommand{\ccorrsymbol}{\zeta}
\newcommand{\pcorrsymbol}{w}
\newcommand{\acorrsymbol}{\xi}
 
\newcommand{\ccorr}[2]{\ccorrsymbol_{#1,#2}}
\newcommand{\pcorr}[2]{\pcorrsymbol_{#1,#2}}
\newcommand{\acorr}[2]{\acorrsymbol_{#1,#2}}

\newcommand{\ccorrp}[2]{\ccorrsymbol_{+,#1,#2}}
\newcommand{\pcorrp}[2]{\pcorrsymbol_{+,#1,#2}}

\newcommand{\ccorrpm}[2]{\ccorrsymbol_{\pm,#1,#2}}
\newcommand{\pcorrpm}[2]{\pcorrsymbol_{\pm,#1,#2}}
\newcommand{\acorrpm}[2]{\acorrsymbol_{\pm,#1,#2}}

\newcommand{\estpcorr}[2]{\est{\pcorrsymbol}_{#1,#2}}

\newcommand{\estacorrpm}[2]{\est{\acorrsymbol}_{\pm,#1,#2}}

\newcommand{\seplos}{\Pi}
\newcommand{\sepperp}{r}
\newcommand{\vsepperp}{\vect{r}}

\newcommand{\EstimatorSum}[2]{\est{S}_{#1,#2}}
\newcommand{\EstimatorSumPM}[2]{\est{S}_{\pm, #1,#2}}

\newcommand{\EstimatorSummandEV}[2]{s_{#1,#2}}
\newcommand{\EstimatorSummandPMEV}[2]{s_{\pm,#1,#2}}

\newcommand{\PEstimatorSum}[2]{\est{S}_{#1,#2}}
\newcommand{\PEstimatorSumPM}[2]{\est{S}_{\pm, #1,#2}}

\newcommand{\RandomEstimatorSum}[2]{\est{R}_{#1,#2}}

\newcommand{\RandomEstimatorSummandEV}[2]{r_{#1,#2}}

\newcommand{\FOV}{\mathbb{F}}
\newcommand{\AFOV}{A_{\FOV}}

\newcommand{\ABin}{A_{\Delta}}

\newcommand{\Nsamples}{N_{\text{t}}}

\newcommand{\biasgal}[2]{b^{#1}_{#2}}

\newcommand{\thetagal}[2]{\vtheta^{#1}_{#2}}
\newcommand{\xgal}[2]{\vx^{#1}_{#2}}
\newcommand{\zgal}[2]{z^{#1}_{#2}}
\newcommand{\pzgal}[2]{p^{#1}_{#2}}
\newcommand{\wgal}[2]{u^{#1}_{#2}}

\newcommand{\fgalrep}[2]{\bar{f}^{#1}_{#2}}
\newcommand{\chigalrep}[2]{\bar{\chi}^{#1}_{#2}}
\newcommand{\zgalrep}[2]{\bar{z}^{#1}_{#2}}

\newcommand{\NgalD}[2]{N^{#1}_{\text{d} #2}}
\newcommand{\ngalD}[2]{n^{#1}_{\text{d} #2}}
\newcommand{\meanngalD}[2]{\bar{n}^{#1}_{\text{d} #2}}
\newcommand{\biasgalD}[2]{b^{#1}_{\text{d} #2}}
\newcommand{\deltagalD}[2]{\delta^{#1}_{\text{d} #2}}
\newcommand{\thetagalD}[2]{\thetagal{#1}{\text{d} #2}}
\newcommand{\xgalD}[2]{\xgal{#1}{\text{d} #2}}
\newcommand{\zgalD}[2]{\zgal{#1}{\text{d} #2}}
\newcommand{\pzgalD}[2]{\pzgal{#1}{\text{d} #2}}
\newcommand{\wgalD}[2]{\wgal{#1}{\text{d} #2}}

\newcommand{\NgalS}[2]{N^{#1}_{\text{s} #2}}
\newcommand{\ngalS}[2]{n^{#1}_{\text{s} #2}}
\newcommand{\meanngalS}[2]{\bar{n}^{#1}_{\text{s} #2}}
\newcommand{\biasgalS}[2]{b^{#1}_{\text{s} #2}}
\newcommand{\deltagalS}[2]{\delta^{#1}_{\text{s} #2}}
\newcommand{\thetagalS}[2]{\thetagal{#1}{\text{s} #2}}
\newcommand{\xgalS}[2]{\xgal{#1}{\text{s} #2}}
\newcommand{\zgalS}[2]{\zgal{#1}{\text{s} #2}}
\newcommand{\pzgalS}[2]{\pzgal{#1}{\text{s} #2}}
\newcommand{\wgalS}[2]{\wgal{#1}{\text{s} #2}}
\newcommand{\geomweightS}[2]{\geomweight^{#1}_{\text{s} #2}}

\newcommand{\fgalSrep}[2]{\fgalrep{#1}{\text{s} #2}}
\newcommand{\chigalSrep}[2]{\chigalrep{#1}{\text{s} #2}}
\newcommand{\zgalSrep}[2]{\zgalrep{#1}{\text{s} #2}}

\newcommand{\thetagalDR}[2]{\thetagal{#1}{\text{dr} #2}}
\newcommand{\thetagalSR}[2]{\thetagal{#1}{\text{sr} #2}}
\newcommand{\thetagalSRDR}[2]{\thetagal{#1}{\text{sr/dr} #2}}

\newcommand{\pcorrmm}[1]{\pcorrsymbol_{\delta}^{\mathrm{mm} #1}}
\newcommand{\pcorrmd}[1]{\pcorrsymbol_{\delta}^{\mathrm{md} #1}}

\newcommand{\pcorrtmI}[1]{\pcorrsymbol_{\tang}^{\mathrm{mI} #1}}
\newcommand{\pcorrtdI}[1]{\pcorrsymbol_{\tang}^{\mathrm{dI} #1}}

\newcommand{\pcorrpmGG}[1]{\pcorrsymbol_{\pm}^{\mathrm{GG} #1}}
\newcommand{\pcorrpmGI}[1]{\pcorrsymbol_{\pm}^{\mathrm{GI} #1}}
\newcommand{\pcorrpmII}[1]{\pcorrsymbol_{\pm}^{\mathrm{II} #1}}

\newcommand{\pcorrpGI}[1]{\pcorrsymbol_{+}^{\mathrm{GI} #1}}

\newcommand{\rfsymbol}{\phi}

\newcommand{\rf}[2]{\rfsymbol^{#1}_{#2}}
\newcommand{\rfE}[2]{\rfsymbol^{#1}_{\mathrm{E} #2}}
\newcommand{\rfB}[2]{\rfsymbol^{#1}_{\mathrm{B} #2}}

\newcommand{\rfX}[2]{\rfsymbol^{#1}_{\mathrm{X} #2}}
\newcommand{\rfY}[2]{\rfsymbol^{#1}_{\mathrm{Y} #2}}

\newcommand{\rfr}[2]{\rfsymbol^{#1}_{\rot #2}}

\newcommand{\rft}[2]{\rfsymbol^{#1}_{\tang #2}}

\newcommand{\rfx}[2]{\rfsymbol^{#1}_{\cross #2}}

\newcommand{\ftrf}[2]{\ft{\rfsymbol}^{#1}_{#2}}

\newcommand{\ftrfX}[2]{\ft{\rfsymbol}^{#1}_{\mathrm{X} #2}}
\newcommand{\ftrfY}[2]{\ft{\rfsymbol}^{#1}_{\mathrm{Y} #2}}

\newcommand{\PSEE}[2]{P^{(#1|#2)}_{\mathrm{EE}}}

\newcommand{\PSBB}[2]{P^{(#1|#2)}_{\mathrm{BB}}}
\newcommand{\PSXY}[2]{P^{(#1|#2)}_{\mathrm{XY}}}

\begin{document}

\title[Intrinsic Alignments in Illustris]{Intrinsic Alignments of Galaxies in the Illustris Simulation}
\author[S. Hilbert, et al.]{
Stefan Hilbert$^{1,2}$\thanks{\href{mailto:stefan.hilbert@tum.de}{\texttt{stefan.hilbert@tum.de}}},
Dandan Xu$^{3}$,
Peter Schneider$^{4}$,
Volker Springel$^{3,5}$, \newauthor
Mark Vogelsberger$^{6}$,
and Lars Hernquist$^{7}$
\\$^{1}$ Exzellenzcluster Universe, Boltzmannstr. 2, 85748 Garching, Germany,
\\$^{2}$ Ludwig-Maximilians-Universit{\"a}t, Universit{\"a}ts-Sternwarte, Scheinerstr. 1, 81679 M{\"u}nchen, Germany,
\\$^{3}$ Heidelberg Institute for Theoretical Studies, Schloss-Wolfsbrunnenweg 35, 69118 Heidelberg, Germany,
\\$^{4}$ Argelander-Institut f{\"u}r Astronomie, Auf dem H{\"u}gel 71, 53121 Bonn, Germany,
\\$^{5}$ Zentrum f{\"u}r Astronomie der Universit{\"a}t Heidelberg, Astronomisches Recheninstitut, M{\"o}nchhofstr. 12-14, 69120 Heidelberg, Germany,
\\$^{6}$ Department of Physics, Massachusetts Institute of Technology, 77 Massachusetts Avenue, Cambridge, MA, USA
\\$^{7}$ Harvard-Smithsonian Center for Astrophysics, 60 Garden Street, Cambridge, MA 02138, USA
}

\date{\today}

\pubyear{20??}

\label{firstpage}
\pagerange{\pageref{firstpage}--\pageref{lastpage}}
\maketitle

\begin{abstract}
We study intrinsic alignments (IA) of galaxy image shapes within the Illustris cosmic structure formation simulations.
We investigate how IA correlations depend on observable galaxy properties such as stellar mass, apparent magnitude, redshift, and photometric type, and on the employed shape measurement method. The correlations considered include the matter density-intrinsic ellipticity (mI), galaxy density-intrinsic ellipticity (dI), gravitational shear-intrinsic ellipticity (GI), and intrinsic ellipticity-intrinsic ellipticity (II) correlations.
We find stronger correlations for more massive and more luminous galaxies, as well as for earlier photometric types, in agreement with observations. Moreover, the correlations significantly depend on the choice of shape estimator, even if calibrated to serve as unbiased shear estimators. In particular, shape estimators that down-weight the outer parts of galaxy images produce much weaker IA signals on intermediate and large scales than methods employing flat radial weights. The expected contribution of intrinsic alignments to the observed ellipticity correlation in tomographic cosmic shear surveys may be below one percent or several percent of the full signal depending on the details of the shape measurement method.
A comparison of our results to a tidal alignment model indicates that such a model is able to reproduce the IA correlations well on intermediate and large scales, provided the effect of varying galaxy density is correctly taken into account.
We also find that the GI contributions to the observed ellipticity correlations could be inferred directly from measurements of galaxy density-intrinsic ellipticity correlations, except on small scales, where systematic differences between mI and dI correlations are large.
\end{abstract}


\section{Introduction}
\label{sec:introduction}

Weak gravitational lensing by the large-scale structure, also referred to as \lq{}cosmic shear\rq{}, creates correlations in the observed ellipticities of distant galaxies.
Measuring these correlations provides valuable information about our Universe's matter distribution and geometry. 
Ongoing and planned weak lensing surveys, such as the Dark Energy Survey\footnote{\href{http://www.darkenergysurvey.org}{\texttt{http://www.darkenergysurvey.org}}} (DES), the Kilo Degree Survey\footnote{\href{http://www.astro-wise.org/projects/KIDS}{\texttt{http://www.astro-wise.org/projects/KIDS}}} (KiDS), the Euclid\footnote{\href{http://www.euclid-ec.org}{\texttt{http://www.euclid-ec.org}}} mission, and Large Synoptic Survey Telescope\footnote{\href{http://www.lsst.org}{\texttt{http://www.lsst.org}}} survey, are designed to constrain cosmological parameters through accurate ellipticity correlation measurements with percent-level statistical accuracy. 

To fully exploit these ongoing and future surveys, one must understand the various sources of correlations in the observed galaxy ellipticities. Intrinsic alignments (IA), i.e. alignments of the intrinsic ellipticities of galaxies, are suspected to be a significant source besides gravitational lensing.
Just considering these two sources, the observed ellipticity correlation has contributions from gravitational shear-gravitational shear (GG) correlations, intrinsic ellipticity-intrinsic ellipticity (II) correlations, and gravitational shear-intrinsic ellipticity (GI) correlations. II correlations may arise from physically close galaxies whose shapes are aligned due to common environmental effects such as large-scale tidal fields. GI correlations may arise when the matter structures associated with lower-redshift \lq{}foreground\rq{} galaxies induce gravitational shear in the images of higher-redshift \lq{}background\rq{} galaxies with a preferred direction relative to the foreground galaxy shapes.

Analytical and numerical studies suggest that intrinsic alignments contribute $\sim10\%$ to the cosmic shear signal in medium and deep surveys \citep[e.g.][]{CroftMetzler2000,HeavensRefregierHeymans2000,HirataSeljak2004,HeymansEtal2006,SemboloniEtal2008, JoachimiEtal2013_IA_II}. In observational studies, intrinsic alignments have been clearly detected and likely contribute several percent to the observed ellipticity correlations \citep[e.g.][]{HeymansEtal2004,MandelbaumEtal2006_IA,HirataEtal2007,MandelbaumEtal2011,JoachimiEtal2011,SinghMandelbaumMore2015}. 

Various techniques have been developed to remove the intrinsic alignment effects from weak lensing measurements, including down-weighting schemes \citep[][]{KingSchneider2003,HeymansHeavens2003,TakadaWhite2004}, or nulling and boosting techniques \citep[][]{JoachimiSchneider2008,JoachimiSchneider2010}. All these methods require excellent redshift information of the source galaxies, and weaken cosmological constraints due to the loss of information. Therefore it is preferable to have accurate models of intrinsic alignments, which can be used as priors in the weak lensing analysis to reduce this loss of cosmological information in lensing surveys.

Gravitational tidal fields likely cause alignments of the intrinsic shapes of galaxies and their surrounding matter on larger scales. Some intrinsic alignment models thus employ parametrized relations between the large-scale galaxy shape alignments and the statistical properties of the gravitational tidal field \citep[e.g.][]{CatelanKamionkowskiBlandford2001,CrittendenEtal2001,HirataSeljak2004,BlazekVlahSeljak2015}. The parameters of these relations usually depend strongly on the galaxies' properties and have to be determined by external data \citep[e.g. observations,][]{JoachimiEtal2011}, which limits the predictive power of these tidal field models.

The shapes and spins of dark matter halos and their alignments in cosmological gravity-only $N$-body simulations can be robustly measured \citep[e.g.][]{CroftMetzler2000,HopkinsBahcallBode2005,BailinSteinmetz2005,BettEtal2007}. 
Predictions for the alignments of the luminous shapes of galaxies can then be obtained by augmenting this information with models describing the galaxy content of the halos and how the shapes of the galaxy light relate to the host dark matter halo properties \citep[][]{JoachimiEtal2013_IA_I,JoachimiEtal2013_IA_II}. This approach requires one to also specify the expected alignment of galaxies within their host halo, which is still quite uncertain \citep[][]{MandelbaumEtal2006_ggl_halo_shapes,Bett2012,SchrabbackEtal2015}.

Numerical simulations of cosmic structure formation that include star formation and resolve the luminous shapes of galaxies offer a more direct way to predictions for galaxy shape alignments \citep[][]{ChisariEtal2015,TennetiEtal2015,VelliscigEtal2015b,TennetiMandelbaumDiMatteo2016,ChisariEtal2016}.
Here we report on our study of intrinsic alignments within the Illustris simulations \citep[][]{VogelsbergerEtal2014_Illustris, VogelsbergerEtal2014_Illustris_Nature}. In particular, we investigate the strength of intrinsic alignments of galaxy image ellipticities, when defined such as to provide \emph{unbiased shear estimates}, as a function of \emph{observable} galaxy properties such as apparent magnitude, redshift, and photometric type, and the employed shape measurement method.
We consider the matter density-intrinsic ellipticity (mI) correlations, galaxy density-intrinsic ellipticity (dI) correlations, gravitational shear-intrinsic ellipticity (GI) correlations, and intrinsic ellipticity-intrinsic ellipticity (II) correlations. Based on these findings, we discuss the expected contribution of intrinsic alignments to the observed ellipticity correlation in tomographic cosmic shear surveys, and also how these contributions could be estimated from measurements of galaxy density-intrinsic ellipticity correlations in a model-independent way.
Furthermore, we compare the intrinsic alignments measured in the simulations to predictions from tidal alignment models.

The paper is organized as follows: We provide a brief introduction to our notation and the theory of gravitational lensing and galaxy image shapes in Section~\ref{sec:theory}. The Illustris simulation and the methods we employ to extract galaxy properties and correlations from it are presented in Section~\ref{sec:methods}. Section~\ref{sec:tests} deals with tests of the resulting galaxy properties such as magnitudes, colors, and shapes. In Section~\ref{sec:results} we present our results. The main part of the paper concludes with a summary and discussion in Section~\ref{sec:summary}. 
A more detailed discussion of correlations involving ellipticities can be found in the Appendix.

\section{Theory}
\label{sec:theory}

The observed shape of a galaxy image is influenced by several factors. Every galaxy has a particular intrinsic shape. Gravitational lensing may leave imprints on the observed image shapes. Various processes in the imaging apparatus (e.g. telescope aberrations or atmospheric seeing) contribute to the observed image shape. In the following, we assume that such instrumental effects have been somehow taken care of, and concentrate on gravitational lensing effects and intrinsic shapes.

\subsection{Gravitational lensing}
\label{sec:theory:lensing}

Photons emitted from sources at cosmological distances are deflected on their way toward us by the gravity of intervening matter structures. As a result of this gravitational lensing effect \citep[see, e.g.][for an introduction]{SchneiderKochanekWambsganss_book}, the observed image position $\vtheta=(\theta_1, \theta_2)$ of a source at redshift $\zS$ may thus differ from the source's (usually unobservable) `true' angular position $\vbeta=\bigl(\beta_1(\vtheta,\zS),\beta_2(\vtheta,\zS)\bigr)$. Spatial variations of the light deflection induce image distortions, which can be quantified to leading order by the distortion matrix
\begin{equation}
\label{eq:lens_distortion}
 \left(\parder{\beta_i(\vtheta,\zS)}{\theta_j}\right)_{i,j=1,2} =
\begin{pmatrix}
 1 - \kappa - \gamma_1 & - \gamma_2 - \omega \\
 - \gamma_2 + \omega &  1 - \kappa + \gamma_1
\end{pmatrix},
\end{equation}
which is conventionally decomposed into the convergence $\kappa(\vtheta,\zS)$, the asymmetry $\omega(\vtheta,\zS)$, and the complex shear  $\gamma(\vtheta,\zS) = \gamma_1(\vtheta,\zS) +\ii \gamma_2(\vtheta,\zS)$. A rotated version of the shear $\gamma$ (which transforms like a spin-2 quantity) may be used to define its tangential component $\gammat$ and cross component $\gammax$ relative to a given direction $\vvartheta$:
\begin{equation}
\label{eq:df_shear_tangential_and_cross_component}
\gammat(\vtheta, \zS;\vvartheta) + \ii \gammax(\vtheta, \zS;\vvartheta) = -\ee^{-2\ii\varphi(\vvartheta)} \gamma(\vtheta, \zS),
\end{equation}
where $\varphi(\vvartheta)$ denotes denotes the polar angle of the vector $\vvartheta$, and $\gammat, \gammax \in \R$.

In cosmological models with General Relativity as the theory of gravity, the convergence can be expressed to first order (and to a good approximation in those parts of the sky where lensing effects are weak) by a weighted projection of the matter density contrast along the line of sight:
\begin{equation}
\label{eq:convergence}
  \kappa(\vtheta, \zS) =
  \int_{0}^{\chiS} \idiff[]{\chiL}
  \,\geomweight(\chiL, \zS)\,
 \deltaMatter \big(\fL \vtheta, \chiL, \zL \big)
   ,
\end{equation}
with the source redshift-dependent geometric weight
\begin{equation}
\label{eq:geom_factor}
  \geomweight(\chiL, \zS) =
  \frac{3 \HubbleConstant^2\OmegaMatter}{2\clight^2} 
  \bigl(1 + \zL \bigr)
  \frac{\fSL \fL}{\fS}
.
\end{equation}
Here, $\HubbleConstant$ denotes the Hubble constant, $\OmegaMatter$ denotes the cosmic mean matter density in units of the critical density, and $\clight$ denotes the speed of light. In these equations, we also introduce several convenient abbreviations: $\zL=z(\chiL)$, $\chiL=\chi(\zL)$, $\chiS=\chi(\zS)$, $\fL=\fK(\chiL)$, $\fS=\fK(\chiS)$, and $\fSL=\fK(\chiS-\chiL)$, where $z(\chi)$ denotes the redshift and $\fK(\chi)$ the comoving angular diameter distance for sources at comoving line-of-sight distance $\chi$. Furthermore, $\deltaMatter(\vx, \chi, z)$ denotes the relative matter overdensity at comoving transverse position\footnote{
Note that we use angular coordinates for some quantities such as the convergence $\kappa$, but comoving transverse coordinates for others such as the matter density contrast $\deltaMatter$.
} $\vx$, comoving line-of-sight distance $\chi$, and cosmic epoch expressed by the redshift $z$ (which is tied to the line-of-sight distance $\chi$ for quantities at spacetime points on the observers's backward lightcone).

The asymmetry $\omega$ vanishes in the first-order lensing approximation, and we will ignore it in the subsequent analysis. Furthermore, the shear field and the convergence field obey a one-to-one relation, which becomes a simple phase factor in flat harmonic space (i.e. 2D Fourier space) in the flat-sky approximation:
\begin{equation}
	\ft{\gamma}(\vell, \zS) = \frac{(\ell_1 + \ii \ell_2)^2}{\lvert \ell_1 + \ii \ell_2\rvert^2} \ft{\kappa}(\vell, \zS). 
\end{equation}
The shear can then also be expressed by a weighted projection:
\begin{equation}
\label{eq:shear_as_los_integral}
\gamma(\vtheta, \zS) =
  \int_{0}^{\chiS} \idiff[]{\chiL} 
  \,\geomweight(\chiL, \zS)\,
 \deltashearMatter \big(\fL \vtheta, \chiL, \zL \big)
,
\end{equation}
where the matter shear contrast $\deltashearMatter(\vx, \chi, z)$ is related to the overdensity $\deltaMatter(\vx, \chi, z)$ in two-dimensional transverse harmonic space via 
\begin{equation}
\ftdeltashearMatter(\vk, \chi, z) = \frac{(k_1 + \ii k_2)^2}{\lvert k_1 + \ii k_2 \rvert^2} \ftdeltaMatter(\vk, \chi, z).
\end{equation}
The complex matter shear contrast $\deltashearMatter(\vx, \chi, z)$ can also be decomposed into a tangential and a cross component for a given direction $\vsepperp$ on the sky:
\begin{equation}
  \deltashearMattert(\vx, \chi, z; \vsepperp) + \ii \deltashearMatterx(\vx, \chi, z; \vsepperp) = -\ee^{-2\ii\varphi(\vsepperp)} \deltashearMatter(\vx, \chi, z).
\end{equation}

In the following, we assume the validity of several further approximations related to weak gravitational lensing. We assume that the lens mapping $\vtheta \mapsto \vbeta(\vtheta, \zS)$ for fixed source redshift $\zS$ is one-to-one and smooth. We also assume that on the scales of single galaxy images, the lens mapping is affine and the distortion matrix \eqref{eq:lens_distortion} is constant. Furthermore, we assume that the convergence and shear are small enough such that the reduced shear $g = \gamma / (1 - \kappa)$ can be approximated by the shear, i.e. $g \approx \gamma$, and that $|g| < 1$.

\subsection{Galaxy image ellipticities}
\label{sec:theory:ellipticities}

The ellipticity of an observed galaxy image may be quantified with the help of moments of the observed image light distribution. For a given observed image brightness distribution $\ObsImageSurfaceBrightness(\vtheta)$, image center $\vtheta_0$, and radial and flux weight function\footnote{
For simplicity, we restrict the discussion to simple radial weight functions. Weights used in practice often also depend on the local surface brightness (e.g. via a brightness threshold).
}
 $\RadialFluxWeight(\theta)$, one may define the following second moments:
\begin{equation}
\label{eq:observed_second_moments}
 \ObsImageSecondMoment{ij} = \int\idiff[2]{\vtheta}\,  \theta_i \theta_j \RadialFluxWeight\bigl(|\vtheta|\bigr) \ObsImageSurfaceBrightness(\vtheta + \vtheta_0) 
 , \quad i,j = 1,2.
\end{equation} 
These can then be combined into the observed image ellipticity \citep{SeitzSchneider1997}
\begin{equation}
\label{eq:observed_ellipticity}
 \obseta{}{} = \obseta{}{1} + \ii \obseta{}{2} =
 \frac{\ObsImageSecondMoment{11} - \ObsImageSecondMoment{22} + 2 \ii \ObsImageSecondMoment{12} }
      {\ObsImageSecondMoment{11} + \ObsImageSecondMoment{22} + 2 \sqrt{\ObsImageSecondMoment{11} \ObsImageSecondMoment{22} - \ObsImageSecondMoment{12}^2}}
      .
\end{equation}
For example,  one obtains $|\obseta{}{}| = (1 - r) / (1 + r)$ for a solid ellipse with axis ratio $r \leq 1$ as image, when $\vtheta_0$ is chosen as the center of the ellipse and $W(\theta) = \const$

Another common way to combine the second moments into an observed image ellipticity reads
\begin{equation}
\label{eq:observed_alt_ellipticity}
 \obse{}{} = \obse{}{1} + \ii \obse{}{2} =
 \frac{\ObsImageSecondMoment{11} - \ObsImageSecondMoment{22} + 2 \ii \ObsImageSecondMoment{12} }
      {\ObsImageSecondMoment{11} + \ObsImageSecondMoment{22}}
.
\end{equation}
This yields $|\obse{}{}| = (1 - r^2) / (1 + r^2)$ for an ellipse with axis ratio $r \leq 1$ and $W(\theta) = \const$ Both ellipticity definitions contain the same information, and are related through \citep[e.g.][]{BartelmannSchneider2001_WL_review}:
\begin{equation}
\label{eq:e_from_eps}
	 \obse{}{} = \frac{2\obseta{}{}}{1+|\obseta{}{}|^2}.
\end{equation}
 
One may also define intrinsic image moments $\IntImageSecondMoment{ij}$ and intrinsic ellipticities $\inteta{}{}$ and $\inte{}{}$ derived from the intrinsic light distribution $\IntImageSurfaceBrightness(\vbeta) = \ObsImageSurfaceBrightness\bigl(\vtheta(\vbeta)\bigr)$ one would observe in the absence of gravitational lensing:
 \begin{equation}
\label{eq:intrinsic_second_moments}
  \IntImageSecondMoment{ij} = \int\idiff[2]{\vbeta}\,  \beta_i \beta_j \RadialFluxWeight\bigl(|\vbeta|\bigr) \IntImageSurfaceBrightness(\vbeta + \vbeta_0) 
,
 \end{equation} 
where $\vbeta_0 = \vbeta(\vtheta_0)$ denotes the intrinsic center, and $\RadialFluxWeight(\beta)$ is the same weight function as used for the observed moments~\eqref{eq:observed_second_moments}. The intrinsic moments can then be combined into the intrinsic image ellipticities
 \begin{align}
 \label{eq:intrinsic_ellipticity}
 \inteta{}{} = \inteta{}{1} + \ii \inteta{}{2} &= 
  \frac{\IntImageSecondMoment{11} - \IntImageSecondMoment{22} + 2 \ii \IntImageSecondMoment{12} }
       {\IntImageSecondMoment{11} + \IntImageSecondMoment{22} + 2 \sqrt{\IntImageSecondMoment{11} \IntImageSecondMoment{22} - (\IntImageSecondMoment{12})^2}},
       \\
 \label{eq:intrinsic_alt_ellipticity}
 \inte{}{} = \inte{}{1} + \ii \inte{}{2} &= 
  \frac{\IntImageSecondMoment{11} - \IntImageSecondMoment{22} + 2 \ii \IntImageSecondMoment{12} }
       {\IntImageSecondMoment{11} + \IntImageSecondMoment{22}}
        .
 \end{align}

\subsection{Shear estimators and image ellipticities}
\label{sec:theory:shear_estimators}

The relation between observed and intrinsic image ellipticity of a galaxy depends on the gravitational lensing distortion matrix \eqref{eq:lens_distortion} across the image, on the employed weight function $\RadialFluxWeight$, and also the employed ellipticity definition. In the case of a flat weight,
$\RadialFluxWeight(\theta) = \const$,
and the center of light as image center, Eqs.~\eqref{eq:observed_second_moments} and \eqref{eq:intrinsic_second_moments} yield \lq{}unweighted\rq{} moments. The observed ellipticity~\eqref{eq:observed_ellipticity} and the intrinsic ellipticity~\eqref{eq:intrinsic_ellipticity} are then related by:
\begin{equation}
  \label{eq:obseps_to_inteps_flat_weight}
  \obseta{}{} = \frac{\inteta{}{} + g}{1 + \inteta{}{} \cconj{g}}
  \qquad \Leftrightarrow \qquad
  \inteta{}{} = \frac{\obseta{}{} - g}{1 - {\inteta{}{}} \cconj{g}}
.
\end{equation}
An advantage of $\obseta{}{}$ based on unweighted moments is that for $|g| \leq 1$, one may use the observed ellipticity as an unbiased estimator,
\begin{equation}
 \label{eq:shear_estimator_from_obseps}
 \est{g} = \obseta{}{}
\end{equation} 
for the reduced shear $g$, i.e. $\EV{\est{g}} = \EV{\obseta{}{}} = g$, where angular brackets $\EV{\cdot}$ denote averages over the galaxy intrinsic ellipticity distribution (which is assumed rotation invariant), regardless of the details of the intrinsic ellipticity distribution \citep{SeitzSchneider1997}.

Unweighted moments and ellipticities derived from them are susceptible to image noise in the outer image regions [due to the factor $\theta_i \theta_j$ in the integral in Eq.~\eqref{eq:observed_second_moments}].
Moreover, unweighted moments and ellipticities are strongly affected by errors in the separation of the galaxy light from the light of neighboring galaxies.
Ellipticity estimation from real galaxy images thus often employs down-weighting of the outer regions of the image. The drawback of such radially weighted moments is that they complicate the relation between observed and intrinsic ellipticities. For a general choice of ellipticity definition $\obsellraw{}{}\in\{\obseta{}{}, \obse{}{}\}$ and weight function $\RadialFluxWeight(\theta)$, and sufficiently small reduced shear $g$, the observed ellipticity $\obsellraw{}{}(g)$ as a function of reduced shear $g$ can be approximated by a linear response in $g$:
\begin{equation}
 \obsellraw{}{}(g) \approx \intellraw{}{} + P g,
\end{equation} 
where the intrinsic ellipticity $\intellraw{}{}=\obsellraw{}{}(0)$, and the shear polarizability
\begin{equation}
	P = \left.\parder{\obsellraw{}{}(g)}{g}\right|_{g=0}.
\end{equation}
If $P$ can be estimated with sufficient accuracy (e.g. from the observed galaxy image itself) and is invertible, one may obtain an unbiased shear estimator by applying $P^{-1}$ as a shear polarizability correction factor:
\begin{equation}
\label{eq:shear_estimator_based_on_response}
 \est{g} = 
 P^{-1} \obsellraw{}{} \approx 
 P^{-1} \intellraw{}{} + g
 .
\end{equation} 

If the average shear polarizability $\EV{P}$ for the observed galaxy population is known, another unbiased shear estimator can be constructed that employs $\EV{P}^{-1}$ as polarisability correction:
\begin{equation}
\label{eq:shear_estimator_based_on_mean_response}
\begin{split}
 \est{g} &= 
 \EV{P}^{-1} \obsellraw{}{}
 \\&\approx 
 \EV{P}^{-1} \intellraw{}{} + 
 \EV{P}^{-1} P g.
\end{split}
\end{equation} 

The above image ellipticity-based shear estimators can be treated in a uniform manner,\footnote{
For simplicity, we assume $g\approx \gamma$ and $\EV{\partial\obsellraw{}{}/ \partial g}^{-1} (\partial \obsellraw{}{}/ \partial g) g \approx \gamma$.
}
\begin{equation}
 \label{eq:obsell_to_intell_simple_approx}
\est{g} = \obsell{}{} = \intell{}{} + \gamma,
\end{equation} 
with appropriate definitions for the observed ellipticity $\obsell{}{}$ and intrinsic ellipticity $\intell{}{}$. For example, Eq.~\eqref{eq:shear_estimator_from_obseps} suggests
\begin{equation}
  \obsell{}{} = \obseta{}{}
  \quad\text{and}\quad
  \intell{}{} = \inteta{}{},
\end{equation} 
with $\obseta{}{}$ and $\inteta{}{}$ from unweighted moments.
For $e$ computed from unweighted moments, one obtains \citep[e.g.][]{BernsteinJarvis2002}:\footnote{
This equation usually does \emph{not} hold  when $\obseta{}{}$ is computed from moments for a \emph{non}-uniform radial weight function.
}
\begin{equation}
  \obsell{}{} =  \frac{\obse{}{}}{2 - \EV{|\obse{}{}|^2}} 
  \quad\text{and}\quad
  \intell{}{} =  \frac{\inte{}{}}{2 - \EV{|\inte{}{}|^2}} .
\end{equation} 
A simplified version of the estimator by \citet[][KSB]{KaiserSquiresBroadhurst1995} yields:
\begin{equation}
  \obsell{}{} = P^{-1} \obse{}{}
  \quad\text{and}\quad
  \intell{}{} = P^{-1} \inte{}{} ,
\end{equation} 
with $\obse{}{}$ and $\inte{}{}$ from moments with a Gaussian radial weight function, and $P=(\partial \obse{}{}/\partial g)$ estimated from the individual observed galaxy images.

As for the shear, the ellipticities $\obsell{}{}$ and $\intell{}{}$ can be decomposed into tangential and cross components with respect to a given angular direction $\vvartheta$ or comoving transverse direction $\vsepperp$:
\begin{align}
\obsellt{}{} + \ii \obsellx{}{} &= -\ee^{-2\ii\varphi(\vvartheta)} \obsell{}{} \quad\text{and}\quad
\intellt{}{} + \ii \intellx{}{}  = -\ee^{-2\ii\varphi(\vsepperp )} \intell{}{}
\end{align}
with $\obsellt{}{}$, $\obsellx{}{}$, $\intellt{}{}$, and $\intellx{}{} \in \R$.

\subsection{Density and intrinsic shape correlations}
\label{sec:theory:intrinsic_correlations}

The intrinsic shapes of galaxies may have a preferred orientation towards nearby other galaxies and matter overdensities. The preferred orientation of galaxies towards matter overdensities may be quantified by the correlation\footnote{
We employ the symbol $\ccorrsymbol$ to denote two-point correlation functions of spacetime fields to avoid potential confusion with angular correlations of fields on the sky denoted with the symbol $\acorrsymbol$.}
\begin{multline}
\label{eq:df_pure_corr_matter_density_shape}
  \ccorr{\deltaMatter}{\intellt{}{}} (|\vsepperp|, \seplos, z, z') 
  =
\\
  \EV{\deltaMatter(\vx + \vsepperp, \chi + \seplos, z) \, \intellt{}{}(\vx, \chi, z'; \vsepperp)}
.
\end{multline} 
Here, $\EV{\cdot}$ denotes the expectation for a statistical ensemble of realizations of the galaxy population and matter density fields for a given cosmological model,
and $\intellt{}{}(\vx, \chi, z; \vsepperp)$ denotes the tangential component of the intrinsic ellipticity $\intell{}{}(\vx, \chi, z)$ of a galaxy at spacetime position $(\vx, \chi, z)$ relative to the transverse direction $\vsepperp$.
The first argument $|\vsepperp|$ of $\ccorrsymbol$ indicates that due to statistical isotropy assumed here, the correlation function depends on the transverse separation $\vsepperp$ only through its magnitude $|\vsepperp|$ but not its direction.
To be general, we also consider the possibility of different redshifts $z$ and $z'$ for the two fields entering the correlations.

One can observe galaxy ellipticities $\intell{}{}(\vx, \chi, z)$ only at positions $(\vx, \chi, z)$ where there is a galaxy to measure a shape from. Thus, correlation functions of practical relevance feature $\intell{}{}(\vx, \chi, z)$ only in conjunction with a factor $\bigl[1  + \deltagalS{}{}(\vx, \chi, z) \bigr]$, where $\deltagalS{}{}(\vx, \chi, z)$ denotes the relative galaxy number overdensity of the galaxy population with shape information, e.g.
\begin{multline}
\label{eq:df_corr_matter_density_shape}
  \ccorr{\deltaMatter}{(1 + \deltagalS{}{})\intellt{}{}} (|\vsepperp|, \seplos, z, z') 
  =\\
  \EV{ \deltaMatter(\vx + \vsepperp, \chi + \seplos, z) \, \bigl[1  + \deltagalS{}{}(\vx, \chi, z') \bigr] \intellt{}{}(\vx, \chi, z'; \vsepperp) }
.
\end{multline}

A related correlation substitutes the matter density contrast in Eq.~\eqref{eq:df_corr_matter_density_shape} by the number density contrast $\deltagalD{}{}$ of a suitable density tracer population such as a galaxy population with sufficiently known galaxy bias:
\begin{multline}
\label{eq:df_corr_galaxy_density_shape}
 \ccorr{\deltagalD{}{}}{(1 + \deltagalS{}{}) \intellt{}{}} (|\vsepperp|, \seplos, z, z') 
 = \\
  \EV{ \deltagalD{}{}(\vx + \vsepperp, \chi + \seplos, z) \, \bigl[1  + \deltagalS{}{}(\vx, \chi, z') \bigr] \intellt{}{}(\vx, \chi, z'; \vsepperp) }
.
\end{multline} 
In case the density tracers follow a simple linear deterministic local bias model, i.e. $\deltagalD{}{}(\vx, \chi, z) = \biasgalD{}{} \,\deltaMatter(\vx, \chi, z)$, the above correlations obey 
\begin{equation}
	\ccorr{\deltagalD{}{}}{(1 + \deltagalS{}{})\intellt{}{}} =  \biasgalD{}{}\,\ccorr{\deltaMatter}{(1 + \deltagalS{}{})\intellt{}{}}
.
\end{equation}

Correlations between the shapes of galaxies may be quantified by
\begin{subequations}
\begin{multline}
  \ccorr{\intellt{}{}}{\intellt{}{}} (|\vsepperp|, \seplos, z, z') 
  =
\\
 \EV{ \intellt{}{}(\vx + \vsepperp, \chi + \seplos, z; \vsepperp) \, \intellt{}{}(\vx, \chi, z'; \vsepperp) }
,
\end{multline}
\begin{multline}
  \ccorr{\intellx{}{}}{\intellx{}{}} (|\vsepperp|, \seplos, z, z') 
  =
\\
  \EV{ \intellx{}{}(\vx + \vsepperp, \chi + \seplos, z;\vsepperp) \, \intellx{}{}(\vx, \chi, z'; \vsepperp) }
,
\end{multline} 
and linear combinations of these:
\begin{multline}
  \ccorrpm{\intell{}{}}{\intell{}{}} (\sepperp, \seplos, z, z') 
=
 \\
  \ccorr{\intellt{}{}\!}{\intellt{}{}\!} (\sepperp, \seplos, z, z') 
  \pm
  \ccorr{\intellx{}{}\!}{\intellx{}{}\!} (\sepperp, \seplos, z, z') 
  ,
\end{multline}
\end{subequations}
where $\intellx{}{}(\vx, \chi, z; \vsepperp)$ denotes the cross component of $\intell{}{}(\vx, \chi, z)$ relative to the transverse direction $\vsepperp$.

Analogous (cross)correlations may be defined for the intrinsic ellipticities of two different galaxy populations and between intrinsic galaxy ellipticities and the shear density contrast.
Of particular relevance for intrinsic alignments in cosmic shear (see Section~\ref{sec:theory:ellipticity_correlations}) are the correlations $\ccorrpm{\deltashearMatter}{(1 + \deltagalS{}{}) \intellt{}{}}$ and $\ccorrpm{(1 + \deltagalS{(1)}{}) \intellt{(1)}{}}{(1 + \deltagalS{(2)}{}) \intellt{(2)}{}}$.

When denoting correlations of quantities defined as functions of redshift and angular position (instead of redshift and comoving position), we follow the above pattern. For example, correlations of observed ellipticities read
\begin{subequations}
\begin{align}
\begin{split}
  \acorr{\obsellt{}{}}{\obsellt{}{}} (|\vvartheta|, z, z') 
  &= \EV{ \obsellt{}{} (\vtheta + \vvartheta, z; \vvartheta) \, \obsellt{}{}(\vtheta, z'; \vvartheta) }
,
\end{split}
\\
\begin{split}
  \acorr{\obsellx{}{}}{\obsellx{}{}} (|\vvartheta|, z, z') 
  &= \EV{ \obsellx{}{}(\vtheta + \vvartheta, z; \vvartheta) \, \obsellx{}{}(\vtheta, z'; \vvartheta) }
, \; \text{and}
\end{split}
\\
\begin{split}
  \acorrpm{\obsell{}{}}{\obsell{}{}} (\vartheta, z, z') 
  &= 
  \acorr{\obsellt{}{}}{\obsellt{}{}} (\vartheta, z, z') 
  \pm
  \acorr{\obsellx{}{}}{\obsellx{}{}} (\vartheta, z, z') 
,
\end{split}
\end{align} 
\end{subequations}
where  $\obselltx{}{}(\vtheta, z; \vvartheta)$ denotes the tangential/cross component of the observed ellipticity $\obsell{}{}(\vtheta, z)$ of a galaxy at sky position $\vtheta$ and redshift $z$ relative to the angular direction $\vvartheta$ on the sky.

Expectation values for the ellipticity correlation estimators discussed in this work involve correlations of projected density and ellipticity fields. Such correlations can often be expressed in Limber-type approximations \citep[][]{Limber1953} that feature projected correlation functions (see Appendix~\ref{app:projected_correlations}). 
For the above correlations, one may define corresponding projected correlations by:\footnote{In certain cases (e.g. when the correlations do not decrease sufficiently fast with increasing line-of-sight separation), one may wish to include a non-uniform l.o.s. weighting function in the projection integral.}
\begin{equation}
\label{eq:df_projected_correlation}
  \pcorrsymbol_{\ldots} (\sepperp, z) = \int \idiff{\seplos}\,  \ccorrsymbol_{\ldots} \bigl(\sepperp, \seplos, z, z \bigr) 
.
\end{equation}

\subsection{Observed density-ellipticity correlations}
\label{sec:theory:density_ellipticity_correlations}

Important information on the intrinsic alignment of galaxies may be obtained by estimating the correlation between the number density of a galaxy \lq{}density sample\rq{} (used to trace the overall matter density) and the observed ellipticities of a galaxy \lq{}shape sample\rq{}. Consider $\NgalD{}{}$ galaxies $i=1,\ldots,\NgalD{}{}$ with observed sky positions $\thetagalD{(i)}{}$ in a survey $\FOV$ with solid angle $\AFOV$,  and redshifts $\zgalD{(i)}{}$ drawn from an underlying redshift distribution $\pzgalD{}{}(\zgalD{(i)}{})$ serving as density sample, and $\NgalS{}{}$ galaxies $j=1,\ldots,\NgalS{}{}$ with observed ellipticities $\obsell{(j)}{}$, observed positions $\thetagalS{(j)}{}$ and redshift distribution $\pzgalS{}{}(\zgalS{(j)}{})$ serving as shape sample. The projected galaxy density-ellipticity correlation as a function of comoving transverse separation $\sepperp$ may be then estimated by \citep[e.g.][]{MandelbaumEtal2006_IA}:
\begin{align}
  \label{eq:density_ellipticity_correlation_estimator}
 \estpcorr{\deltagalD{}{} }{\obsellt{}{}} (\sepperp) &= \frac{\EstimatorSum{\deltagalD{}{} }{\obsellt{}{}}(\sepperp)}{\RandomEstimatorSum{1}{1}(\sepperp)}
\text{ with}\\
  \label{eq:density_ellipticity_correlation_estimator_signal_sum}
\EstimatorSum{\deltagalD{}{}}{\obsellt{}{}}(\sepperp) &=
  \sum_{i,j=1}^{\NgalD{}{},\NgalS{}{}} \wgalD{(i)}{} \wgalS{(j)}{}  \,\Delta\bigl(\sepperp, \fD^{(i)} \lvert  \thetagalS{(j)}{} -   \thetagalD{(i)}{} \rvert \bigr) \obsellt{(j|i)}{}
,\\
  \label{eq:density_ellipticity_correlation_estimator_norm_sum}
\RandomEstimatorSum{1}{1}(\sepperp) &=
  \sum_{i,j=1}^{\NgalD{}{},\NgalS{}{}} \wgalD{(i)}{} \wgalS{(j)}{} \,\Delta\bigl(\sepperp, \fD^{(i)} \lvert \thetagalSR{(j)}{} - \thetagalDR{(i)}{} \rvert \bigr)
.
\end{align}
Here, $\wgalD{(i)}{}$ and $\wgalS{(i)}{}$ denote weights,
$\obsellt{(j|i)}{}$ denotes the tangential component of the observed ellipticity of galaxy $j$ in the shape sample relative to the direction towards galaxy $i$ of the density sample,
and the $\thetagalSRDR{(j)}{}$ and denote positions obtained by randomly distributing the ellipticity/density sample galaxy positions within the survey area. 
The bin window function
\begin{equation}
\label{eq:bin_function}
  \Delta\bigl(r, r' \bigr)
  = \begin{cases}
  1 & \text{for } \lvert r' - r \rvert \leq \varDelta(r)/2 \text{ and}\\
  0 & \text{otherwise,}
  \end{cases}
\end{equation}
where $\varDelta(r)$ denotes the bin width (which we assume small compared to scales on which correlations change noticeably).

Assuming Eq.~\eqref{eq:obsell_to_intell_simple_approx} holds, the expectation of the estimator~\eqref{eq:density_ellipticity_correlation_estimator} can be expressed as a sum of a density-gravitational shear contribution (dG) and a density-intrinsic ellipticity contribution (dI): 
\begin{equation}
\label{eq:ev_est_w_delta_obsellt_split_into_dG_and_dI}
\bEV{\estpcorr{\deltagalD{}{} }{\obsellt{}{}} (\sepperp)} =
\bEV{\estpcorr{\deltagalD{}{} }{\obsellt{}{}} (\sepperp)}^{\text{dG}} +
\bEV{\estpcorr{\deltagalD{}{} }{\obsellt{}{}} (\sepperp)}^{\text{dI}} 
.
\end{equation}
For these contributions (see Appendix~\ref{app:density_ellipticity_correlations} for a derivation),
\begin{align}
\label{eq:ev_est_w_delta_obsellt_dG_contribution}
  \bEV{\estpcorr{\deltagalD{}{} }{\obsellt{}{}} (\sepperp)}^{\text{dG}}  &\approx \frac{\EstimatorSummandEV{\deltagalD{}{}}{\obsellt{}{}}^{\text{dG}}(\sepperp)}{\RandomEstimatorSummandEV{1}{1}(\sepperp)}
,\quad\text{and}\\
\label{eq:ev_est_w_delta_obsellt_dI_contribution}
  \bEV{\estpcorr{\deltagalD{}{} }{\obsellt{}{}} (\sepperp)}^{\text{dI}}  &\approx \frac{\EstimatorSummandEV{\deltagalD{}{}}{\obsellt{}{}}^{\text{dI}}(\sepperp)}{\RandomEstimatorSummandEV{1}{1}(\sepperp)} 
,
\end{align}
where
\begin{subequations}
\begin{align}
\label{eq:raw_pair_contribution_to_density_ellipticity_normalization}
  \RandomEstimatorSummandEV{1}{1}(\sepperp)
  &=
    \int\idiff[]{\zgalD{}{}}\, \pzgalD{}{}(\zgalD{}{}) 
    \frac{\ABin(\sepperp, \zgalD{}{})}{\AFOV}
,\\
\label{eq:raw_pair_contribution_to_density_ellipticity_dG_signal}
  \EstimatorSummandEV{\deltagalD{}{}}{\obsellt{}{}}^{\text{dG}}(\sepperp) 
  &\approx
    \int\idiff[]{\zgalD{}{}} \, \pzgalD{}{}(\zgalD{}{}) 
    \frac{\ABin(\sepperp, \zgalD{}{})}{\AFOV}
    \geomweightS{}{}(\chiD^{})\,
    \pcorr{\deltagalD{}{}}{\deltashearMattert} (r,  \zgalD{}{})
,\!\!\\
\label{eq:raw_pair_contribution_to_density_ellipticity_dI_signal}
\begin{split}
  \EstimatorSummandEV{\deltagalD{}{}}{\obsellt{}{}}^{\text{dI}}(\sepperp) 
  & \approx
    \int\idiff[]{\zgalD{}{}}\, \pzgalD{}{}(\zgalD{}{}) 
    \frac{\ABin(r, \zgalD{}{})}{\AFOV}
    \int\idiff[]{\zgalS{}{}}\, \pzgalS{}{}(\zgalS{}{}) 
    \\&\quad\times
    \ccorr{\deltagalD{}{}}{(1 + \deltagalS{}{})\intellt{}{}}\bigl(r, \chiS - \chiD, \zgalD{}{}, \zgalS{}{}\bigr)
.
\end{split}
\end{align}
\end{subequations}
Here,
\begin{equation}
  \ABin(\sepperp, \zgalD{}{}) = 
  \frac{1}{\AFOV}\int_{\FOV} \idiff[2]{\thetagalD{}{}} \int_{\FOV}\idiff[2]{\thetagalS{}{}} \Delta\bigl(\sepperp, \fD \lvert \thetagalS{}{} - \thetagalD{}{} \rvert \bigr)
\end{equation} 
denotes the effective area for a bin at radius $r$ and density sample galaxies at redshift $\zgalD{}{}$, and
\begin{equation}
\label{eq:geometric_weight_for_source_galaxy_sample}
  \geomweightS{}{}(\chiL) 
=
 \frac{3 \HubbleConstant^2\OmegaMatter}{2\clight^2}(1 + \zL) \fL
\int_{\zL}^{\infty}\!\!\diff{\zS}\, \pzgalS{}{}(\zS)  \frac{\fSL}{\fS}
\end{equation}
denotes the source redshift weighted geometric weight.

If the density or shape sample's redshift distribution varies little over the range where the correlation between tracer galaxy density and intrinsic ellipticity is markedly different from zero to permit a Limber-type approximation for the dI term, one obtains:
\begin{equation}
\begin{split}
\label{eq:raw_pair_contribution_to_density_ellipticity_dI_signal_Limber}
 \EstimatorSummandEV{\deltagalD{}{}}{\obsellt{}{}}^{\text{dI}}(\sepperp) 
 & \approx
 \int\idiff[]{\zgalD{}{}}\, \pzgalD{}{}(\zgalD{}{}) 
\frac{\ABin(r, \zgalD{}{})}{\AFOV}
\\&\quad\times
 \pzgalS{}{}(\zgalD{}{}) 
 \biggl(\totder{\chiD}{\zgalD{}{}} \biggr)^{-1}
\pcorr{\deltagalD{}{}}{(1 + \deltagalS{}{})\intellt{}{}}\bigl(r, \zgalD{}{} \bigr)
.
\end{split}
\end{equation}

The estimator~\eqref{eq:density_ellipticity_correlation_estimator} is very similar to estimators commonly used in galaxy-galaxy lensing. A notable difference is the normalization~\eqref{eq:density_ellipticity_correlation_estimator_norm_sum}, which sums over random positions instead of the actual galaxy positions. This feature makes the normalization insensitive to correlations between the density and shape sample's galaxy densities.

\subsection{Observed ellipticity correlations}
\label{sec:theory:ellipticity_correlations}

Measured correlations of observed galaxy image ellipticities are used in gravitational lensing studies to obtain constraints on the spatial correlations of the gravitational shear. However, the observed ellipticity correlation may also contain contributions from intrinsic shape correlations. As part of most weak lensing survey analyses, the ellipticity correlation between two (possibly identical) sets of galaxies is estimated as a function of image separation $\vartheta$. Assume each set $\alpha \in \{ 1, 2 \}$ contains $\NgalS{(\alpha)}{}$ galaxies $i=1,\ldots,\NgalS{(\alpha)}{}$ with observed ellipticities $\obsell{(\alpha,i)}{}$, observed angular positions $\thetagalS{(\alpha,i)}{}$ inside a survey field $\FOV$ with area $\AFOV$, and known probability distribution $\pzgalS{(\alpha)}{}(\zgalS{(\alpha,i)}{})$ for their redshifts $\zgalS{(\alpha,i)}{}$.
A common estimator reads
\begin{align}
  \label{eq:obsell_correlation_estimator}
 \estacorrpm{\obsell{}{}}{\obsell{}{}}^{(1 | 2)} (\vartheta) &= \frac{\EstimatorSumPM{\obsell{}{}}{\obsell{}{}}^{(1 | 2)} (\vartheta)}{\EstimatorSum{1}{1}^{(1 | 2)} (\vartheta)}
\text{ with}\\
  \label{eq:obsell_correlation_estimator_signal_sum}
\begin{split}
 \EstimatorSumPM{\obsell{}{}}{\obsell{}{}}^{(1 | 2)} (\vartheta) &=
  \sum_{i,j=1}^{\NgalS{(1)}{}\!,\NgalS{(2)}{}\!} \wgalS{(1,i)}{} \wgalS{(2,j)}{} \,\Delta\bigl(\vartheta, \lvert \thetagalS{(2,j)}{} - \thetagalS{(1,i)}{} \rvert \bigr)
 \\&\quad\times
 \left( \obsellt{(1,i|2,j)}{} \obsellt{(2,j|1,i)}{} \pm \obsellx{(1,i|2,j)}{} \obsellx{(2,j|1,i)}{} \right)
,
\end{split}
\\
  \label{eq:obsell_correlation_estimator_norm_sum}
\EstimatorSum{1}{1}^{(1 | 2)} (\vartheta) &=
  \sum_{i,j=1}^{\NgalS{(1)}{}\!,\NgalS{(2)}{}\!} \wgalS{(1,i)}{} \wgalS{(2,j)}{} \,\Delta\bigl(\vartheta, \lvert \thetagalS{(2,j)}{} - \thetagalS{(1,i)}{} \rvert \bigr)
.
\end{align}
Here, the $\wgalS{(\alpha,i)}{}$ denote statistical weights. 
Furthermore, $\obselltx{(\alpha,i|\nu,j)}{}$ denotes the tangential/cross component of the observed ellipticity of galaxy $(\alpha,i)$ relative to the direction towards galaxy $(\nu,j)$.

Assuming Eq.~\eqref{eq:obsell_to_intell_simple_approx} holds, the expectation $\bEV{\estacorrpm{\obsell{}{}}{\obsell{}{}}^{(1 | 2)} (\vartheta)}$ can then be separated into four terms:
\begin{equation}
\begin{split}
   \bEV{\estacorrpm{\obsell{}{}}{\obsell{}{}}^{(1|2)} (\vartheta)}
&= 
   \bEV{\estacorrpm{\obsell{}{}}{\obsell{}{}}^{(1|2)} (\vartheta)}^{\text{GG}}
 + \bEV{\estacorrpm{\obsell{}{}}{\obsell{}{}}^{(1|2)} (\vartheta)}^{\text{GI}}
 \\&\quad
 + \bEV{\estacorrpm{\obsell{}{}}{\obsell{}{}}^{(1|2)} (\vartheta)}^{\text{IG}}
 + \bEV{\estacorrpm{\obsell{}{}}{\obsell{}{}}^{(1|2)} (\vartheta)}^{\text{II}}
.
\end{split}
\end{equation}
The gravitational shear-shear (GG), gravitational shear-intrinsic ellipticity (GI), intrinsic ellipticity-gravitational shear (IG), and intrinsic ellipticity-intrinsic ellipticity (II) contributions are given by (see Appendix~\ref{app:observed_ellipticity_correlations} for a derivation):
\begin{equation}
\label{eq:contributions_to_observed_ellipticity_correlations_from_raw_pair_contributions}
   \bEV{\estacorrpm{\obsell{}{}}{\obsell{}{}}^{(1|2)} (\vartheta)}^{\text{XY}}
\approx
\frac
{ \EstimatorSummandPMEV{\obsell{}{}}{\obsell{}{}}^{(1|2)\,\text{XY}}(\vartheta)}
{ \EstimatorSummandEV{1}{1}^{(1|2)}(\vartheta)}
,\quad \mathrm{X},\mathrm{Y} \in \{ \mathrm{G}, \mathrm{I} \}, 
\end{equation}
with
\begin{subequations}
\label{eq:raw_pair_contributions_to_observed_ellipticity_correlations}
\begin{align}
\label{eq:raw_pair_contributions_to_observed_ellipticity_correlations:normalization}
\begin{split}
&
\EstimatorSummandEV{1}{1}^{(1|2)}(\vartheta)
  \approx
  \int\idiff[]{\zgalS{(1)}{}} \pzgalS{(1)}{}(\zgalS{(1)}{}) 
  \int\idiff[]{\zgalS{(2   )}{}} \pzgalS{(2   )}{}(\zgalS{(2   )}{})
\\&\quad\times 
  \left[1 +
  \ccorr{\deltagalS{(1)}{}}{\deltagalS{(2)}{}} \bigl(\fS^{(1)} \vartheta, \chiS^{(2)} - \chiS^{(1)}, \zgalS{(1)}{},  \zgalS{(2)}{} \bigr) \right]
  ,
\end{split}
\\
\label{eq:raw_pair_contributions_to_observed_ellipticity_correlations:GG}
 \begin{split}
  &
  \EstimatorSummandPMEV{\obsell{}{}}{\obsell{}{}}^{(1|2)\,\text{GG}}(\vartheta)
  \approx
   \int\idiff[]{\zgalS{(1)}{}} \pzgalS{(1)}{}(\zgalS{(1)}{}) 
   \int\idiff[]{\zgalS{(2   )}{}} \pzgalS{(2   )}{}(\zgalS{(2   )}{}) 
\\&\quad\times
   \left[1 + \ccorr{\deltagalS{(1)}{}}{\deltagalS{(2)}{}} \bigl(\fS^{(1)} \vartheta, \chiS^{(2)} - \chiS^{(1)}, \zgalS{(1)}{},  \zgalS{(2)}{} \bigr) \right]
  \\&\quad\times
   \int\idiff[]{\chiD} \, \geomweight(\chiD, \zgalS{(1)}{}) \, \geomweight(\chiD, \zgalS{(2)}{}) \,
   \pcorrpm{\deltashearMatter}{\deltashearMatter} \bigl(\fD \vartheta, \zgalD{}{} \bigr) 
   ,
 \end{split}
 \\
\label{eq:raw_pair_contributions_to_observed_ellipticity_correlations:GI}
 \begin{split}
 &
 \EstimatorSummandPMEV{\obsell{}{}}{\obsell{}{}}^{(1|2)\,\text{GI}}(\vartheta)
   \approx
    \int\idiff[]{\zgalS{}{}} \, \pzgalS{(2   )}{}(\zgalS{}{})  \,
   \geomweightS{(1)}{}( \chiS) \,
  \\&\quad\times
 \pcorrpm{\deltashearMatter}{(1 + \deltagalS{(2   )}{}) \intell{(2)}{}} \bigl(\fS \vartheta, \zgalS{}{} \bigr)
   ,
  \end{split}
\\
\label{eq:raw_pair_contributions_to_observed_ellipticity_correlations:IG}
 \begin{split}
 &
 \EstimatorSummandPMEV{\obsell{}{}}{\obsell{}{}}^{(1|2)\,\text{IG}}(\vartheta)
   \approx
   \int\idiff[]{\zgalS{}{}} \, \pzgalS{(1   )}{}(\zgalS{}{}) \,
   \geomweightS{(2)}{}( \chiS) \,
  \\&\quad\times
   \pcorrpm{\deltashearMatter}{(1 + \deltagalS{(1   )}{}) \intell{(1)}{}} \bigl(\fS \vartheta, \zgalS{}{} \bigr)
   ,\quad\text{and}
 \end{split}
 \\
\label{eq:raw_pair_contributions_to_observed_ellipticity_correlations:II}
\begin{split}
 &
 \EstimatorSummandPMEV{\obsell{}{}}{\obsell{}{}}^{(1|2)\,\text{II}}(\vartheta)
   \approx
   \int\idiff[]{\zgalS{(1)}{}} \pzgalS{(1)}{}(\zgalS{(1)}{}) 
   \int\idiff[]{\zgalS{(2)}{}} \pzgalS{(2   )}{}(\zgalS{(2)}{}) \,
  \\&\quad\times
   \ccorrpm{(1 + \deltagalS{(1)}{}) \intell{(1)}{}\!}{(1 + \deltagalS{(2   )}{}) \intell{(2)}{}\!} 
   \bigl( \fS^{(1)}\! \vartheta, \chiS^{(2)} \! -  \chiS^{(1)}\!, \zgalS{(1)}{}\!, \zgalS{(2)}{} \bigr)
  .
 \end{split}
\end{align}
\end{subequations}
Here, $\deltagalS{(\alpha)}{}$ denotes the overdensity of the galaxy density field underlying the distribution of the galaxies $(\alpha,i)$. The general geometric factor $q(\chi, z)$ is given by Eq.~\eqref{eq:geom_factor}. The effective geometric factor $\geomweightS{(\alpha)}{}(\chi)$ for the source galaxy sample $\alpha$ is given by Eq.~\eqref{eq:geometric_weight_for_source_galaxy_sample} with $\pzgalS{(\alpha)}{}$ replacing $\pzgalS{}{}$.

For redshift distributions sufficiently broad to permit a Limber-type approximation, the normalization and the II term can also be expressed in terms of projected correlations:
\begin{subequations}
\label{eq:raw_pair_contributions_to_observed_ellipticity_correlations_broad_pz}
\begin{align}
\label{eq:raw_pair_contributions_to_observed_ellipticity_correlations_broad_pz:normalization}
\begin{split}
 &
\EstimatorSummandEV{1}{1}^{(1|2)}(\vartheta)
  \approx
	1 + 
  \int\idiff[]{\zgalS{}{}}\, \pzgalS{(1)}{}(\zgalS{}{})\, \pzgalS{(2)}{}(\zgalS{}{})
  	\left(\totder{\chiS{}{}}{\zgalS{}{}}\right)^{-1}
\\&\quad\times 
  \pcorr{\deltagalS{(1)}{}}{\deltagalS{(2)}{}} \bigl(\fS \vartheta, \zgalS{}{} \bigr)
	,
  \end{split}
\\
\label{eq:raw_pair_contributions_to_observed_ellipticity_correlations_broad_pz:II}
\begin{split}
 &
 \EstimatorSummandPMEV{\obsell{}{}}{\obsell{}{}}^{(1|2)\,\text{II}}(\vartheta)
   \approx
   \int\idiff[]{\zgalS{}{}}\, \pzgalS{(1)}{}(\zgalS{}{}) \, \pzgalS{(2)}{}(\zgalS{}{}) \,
	\left(\totder{\chiS{}{}}{\zgalS{}{}}\right)^{-1}
  \\&\quad\times
   \pcorrpm{(1 + \deltagalS{(1)}{}) \intell{(1)}{}}{(1 + \deltagalS{(2   )}{}) \intell{(2)}{}} 
   \bigl( \fS \vartheta, \zgalS{}{} \bigr)
  .
 \end{split}
\end{align}
\end{subequations}

If density cross-correlations can be neglected, e.g. when the two sets of galaxies are well separated in redshift, the normalization and GG term reduce to:
\begin{subequations}
\label{eq:raw_pair_contributions_to_observed_ellipticity_correlations_no_density_cross_correlations}
\begin{align}
\label{eq:raw_pair_contributions_to_observed_ellipticity_correlations_no_density_cross_correlations:normalization}
\begin{split}
&
\EstimatorSummandEV{1}{1}^{(1|2)}(\vartheta)
  \approx
  1
  ,
\end{split}
\\
\label{eq:raw_pair_contributions_to_observed_ellipticity_correlations_no_density_cross_correlations:GG}
 \begin{split}
  &
  \EstimatorSummandPMEV{\obsell{}{}}{\obsell{}{}}^{(1|2)\,\text{GG}}(\vartheta)
  \approx \!
   \int\idiff[]{\chiD} \, \geomweightS{(1)}{}(\chiD) \, \geomweightS{(2)}{}(\chiD) \,
   \pcorrpm{\deltashearMatter}{\deltashearMatter} \bigl(\fD \vartheta, \zgalD{}{} \bigr) 
   .\!\!
 \end{split}
\end{align}
\end{subequations}

The above expressions simplify considerably further if all correlations with galaxy overdensities $\deltagalS{(\alpha)}{}$ as factors are neglected (see Appendix~\ref{app:observed_ellipticity_correlations}). 
Although it has been pointed out \citep[e.g. by][]{HirataSeljak2004,Valageas2014,BlazekVlahSeljak2015} that these galaxy density correlations are important for the understanding of observed ellipticity correlations, such correlations are often not explicitly taken into account in works on intrinsic alignment \citep[][]{JoachimiEtal2015_IA_review_part_I,KirkEtal2015_IA_review_part_II,KiesslingEtal2015_IA_review_part_III}.\footnote{
In various studies, however, $\pcorr{\deltagalD{}{}}{(1+\deltagalS{}{})\obsellt{}{}}$,  $\pcorrpm{(1+\deltagalS{(1)}{})\intell{(1)}{}}{(1+\deltagalS{(2)}{})\intell{(2)}{}}$, etc. are measured even though their notation may suggest $\pcorr{\deltagalD{}{}}{\obsellt{}{}}$, $\pcorrpm{\intell{(1)}{}}{\intell{(2)}{}}$ etc.}

\subsection{More on notation}
\label{sec:theory:notation}

Our naming scheme for correlations differs from commonly used notations in the literature on intrinsic alignments. A systematic notation such as ours facilitates the discussion of intrinsic alignments and observed ellipticity correlations. Translation between the notations is straightforward in many cases. For example, the correlation called $\xi_{\delta+}$ in several works \citep[e.g. in][]{JoachimiEtal2015_IA_review_part_I} becomes $\ccorr{\deltaMatter}{\intellt{}{}}$ in our notation. 

Our full naming scheme is somewhat lengthy in some cases (in particular, when used as labels in plots). For correlations involving the matter density field (m), the matter shear field (G), a galaxy density sample (d), or a galaxy shape sample (I), we introduce shorter names: 
\begin{subequations}
\begin{align}
 \pcorrsymbol_{\delta}^{\mathrm{mm}} &= \pcorr{\deltaMatter}{\deltaMatter} 
 ,\\
 \pcorrsymbol_{\delta}^{\mathrm{md}} &= \pcorr{\deltaMatter}{\deltagalD{}{}} 
 ,\\
 \pcorrsymbol_{\delta}^{\mathrm{dd}} &= \pcorr{\deltagalD{}{}}{\deltagalD{}{}} 
,\\
 \pcorrsymbol_{\tang/\cross}^{\mathrm{mG}} &= \pcorr{\deltaMatter}{\deltashearMattertx} 
,\\
 \pcorrsymbol_{\tang/\cross}^{\mathrm{mI}} &= \pcorr{\deltaMatter}{(1+\deltagalS{}{})\intelltx{}{}} 
 ,\\
 \pcorrsymbol_{\tang/\cross}^{\mathrm{dG}} &= \pcorr{\deltagalD{}{}}{\deltashearMattertx} 
 ,\\
 \pcorrsymbol_{\tang/\cross}^{\mathrm{dI}} &= \pcorr{\deltagalD{}{}}{(1+\deltagalS{}{})\intelltx{}{}} 
 ,\\
 \pcorrsymbol_{\pm}^{\mathrm{GG}} &= \pcorrpm{\deltashearMatter}{\deltashearMatter} 
 ,\\
 \pcorrsymbol_{\pm}^{\mathrm{GI}} &= \pcorrpm{\deltashearMatter}{(1+\deltagalS{}{})\intell{}{}} 
 ,\quad \text{and}\\
 \pcorrsymbol_{\pm}^{\mathrm{II}} &= \pcorrpm{(1+\deltagalS{}{})\intell{}{}}{(1+\deltagalS{}{})\intell{}{}} 
.
\end{align}
\end{subequations}

\subsection{Relations for correlation functions}
\label{sec:theory:integral_relations}

The close connection between the convergence field $\kappa$ and the shear field $\gamma$ in weak lensing gives rise to simple integral relations between their correlation functions \citep[e.g.][]{CrittendenEtal2002,SchneiderVanWaerbekeMellier2002}. Since the relation between the matter overdensity $\deltaMatter$ and the matter shear contrast $\deltashearMatter$ parallels the relation between $\kappa$ and $\gamma$, similar relations hold for the projected correlations of $\deltaMatter$ and $\deltashearMatter$:
\begin{subequations}
\begin{align}
\pcorrsymbol_{+}^{\mathrm{GG}} (r,z) &=
\int_{0}^{\infty}\! r' \diff{r'}\, \mathcal{G}_{+,\delta}(r,r')
\pcorrsymbol_{\delta}^{\mathrm{mm}} (r',z)
,\\
\pcorrsymbol_{-}^{\mathrm{GG}} (r,z) &= 
\int_{0}^{\infty}\! r' \diff{r'}\, \mathcal{G}_{-,\delta}(r,r')
\pcorrsymbol_{\delta}^{\mathrm{mm}} (r',z)
,\quad\text{and}\\
\pcorrsymbol_{\tang}^{\mathrm{mG}} (r,z) &= 
\int_{0}^{\infty}\! r' \diff{r'}\, \mathcal{G}_{\tang,\delta}(r,r')
\pcorrsymbol_{\delta}^{\mathrm{mm}} (r',z)
.
\end{align}
\end{subequations}
The kernels read:
\begin{subequations}
\begin{align}
\mathcal{G}_{+,\delta}(r,r') &=
\frac{1}{r} \DiracDelta(r - r')
,\\
\mathcal{G}_{-,\delta}(r,r') &= 
\frac{1}{r} \DiracDelta(r - r') + 
\left(\frac{4}{r^2} - \frac{\! 12 {r'}^2 \! }{r^4}  \right) \! \HeavisideTheta(r - r')
,\\
 \mathcal{G}_{\tang,\delta}(r,r') &=
-\frac{1}{r} \DiracDelta(r - r') + 
\frac{2}{r^2} \HeavisideTheta(r - r')
 .
\end{align}
\end{subequations}
where $\DiracDelta$ denotes the Dirac delta \lq{}function\rq{}, and $\HeavisideTheta$ denotes the Heaviside step function.

The kernels of the inverse transformations are obtained by swapping the arguments, e.g.:
\begin{equation}
\pcorrsymbol_{\delta}^{\mathrm{mm}} (r,z) = 
\int_{0}^{\infty}\!\! r' \diff{r'}\, \mathcal{G}_{\delta,\tang}(r,r')
\pcorrsymbol_{\tang}^{\mathrm{mG}} (r',z)
,
\end{equation} 
where $\mathcal{G}_{\delta, \tang}(r,r') = \mathcal{G}_{\tang,\delta}(r',r)$.

Certain correlations involving intrinsic ellipticities are connected by analogous integral relations. For example (see Appendix~\ref{app:integral_relations} for a derivation):
\begin{equation}
\label{eq:relation_GI_mI}
\pcorrsymbol_{+}^{\mathrm{GI}} (r,z) = 
\int_{0}^{\infty}\!\! r' \diff{r'}\, \mathcal{G}_{+,\tang}(r,r')
\pcorrsymbol_{\tang}^{\mathrm{mI}} (r',z)
,
\end{equation}
where $\mathcal{G}_{+,\tang}(r,r') =  \mathcal{G}_{\delta,\tang}(r,r') =  \mathcal{G}_{\tang,\delta}(r',r)$.
This relation may be used in suitable situations to estimate the GI contribution to observed ellipticity correlations from observed density-ellipticity correlations.

Here, we discuss the application of relation~\eqref{eq:relation_GI_mI} in a very simple case: Consider a higher-redshift galaxy shape sample 1 and a lower-redshift galaxy shape sample 2. Their cross-correlation~\eqref{eq:obsell_correlation_estimator} of observed ellipticities may contain GI contributions stemming from correlations of the matter shear contrast and the galaxy shapes of sample 2.
Assume for simplicity that the shape sample 2 has a narrow redshift distribution $\pzgalS{(2)}{}$ centered around a representative redshift $\zgalSrep{(2)}{}$ with negligible overlap with the redshift distribution $\pzgalS{(1)}{}$ of shape sample 1. Assume that there is also a sample of density tracer galaxies available with galaxy bias $\biasgalD{}{}$, i.e. $\bEV{\deltagalD{}{}\ldots} = \biasgalD{}{} \bEV{\deltaMatter\ldots}$, whose redshift distribution covers the redshift range of shape sample 2, but does not extend to significantly lower redshifts such that dG contributions~\eqref{eq:ev_est_w_delta_obsellt_dG_contribution} to the observed ellipticity-density correlation~\eqref{eq:density_ellipticity_correlation_estimator} between the density sample and shape sample 2 can be neglected. Assume, moreover, that the projected density-intrinsic ellipticity correlation does not vary significantly with redshift in the range where $\pzgalS{(2)}{}>0$. Then, according to Eqs.~\eqref{eq:ev_est_w_delta_obsellt_split_into_dG_and_dI}, \eqref{eq:ev_est_w_delta_obsellt_dI_contribution}, \eqref{eq:raw_pair_contribution_to_density_ellipticity_normalization}, and \eqref{eq:raw_pair_contribution_to_density_ellipticity_dI_signal_Limber}, the observed galaxy density-ellipticity estimator~\eqref{eq:density_ellipticity_correlation_estimator} yields:
\begin{equation}
	 \estpcorr{\deltagalD{}{} }{\obsellt{(2)}{}} (\sepperp) \approx F \,
	 \pcorr{\deltaMatter}{(1+\deltagalS{(2)}{})\intellt{(2)}{}} (\sepperp, \zgalSrep{(2)}{}),
\end{equation}
where
\begin{equation}
\begin{split}
	F &=
	\left[ 
	 \int\idiff[]{\zgalD{}{}}\, \pzgalD{}{}(\zgalD{}{}) 
    \frac{\ABin(\sepperp, \zgalD{}{})}{\AFOV}
		\right]^{-1}
\\&\quad\times	
\biasgalD{}{}
	 \int\idiff[]{\zgalD{}{}}\, \pzgalD{}{}(\zgalD{}{}) 
\frac{\ABin(r, \zgalD{}{})}{\AFOV}
 \pzgalS{(2)}{}(\zgalD{}{}) 
 \biggl(\totder{\chiD}{\zgalD{}{}} \biggr)^{-1}
	.
\end{split}
\end{equation}
Exploiting relation~\eqref{eq:relation_GI_mI}, one obtains:
\begin{multline}
	\pcorrp{\deltashearMatter}{(1+\deltagalS{(2)}{})\intell{(2)}{}} 
	(\sepperp, \zgalSrep{(2)}{}) 
	\approx
\\
\int_{0}^{\infty}\!\! r' \diff{r'}\, \mathcal{G}_{+,\tang}(r,r')
F^{-1} \estpcorr{\deltagalD{}{}} {\obsellt{(2)}{}}  (\sepperp')
.
\end{multline}
Combining this result with Eqs.~\eqref{eq:contributions_to_observed_ellipticity_correlations_from_raw_pair_contributions}, \eqref{eq:raw_pair_contributions_to_observed_ellipticity_correlations:GI}, and \eqref{eq:raw_pair_contributions_to_observed_ellipticity_correlations_no_density_cross_correlations:normalization}, and recalling that $\pzgalS{(2)}{}$ vanishes except for redshifts $\approx \zgalSrep{(2)}{}$,
the shear-intrinsic (GI) contribution to the observed ellipticity correlation between the galaxy shape samples 1 and 2 can be roughly estimated by:
\begin{multline}
 \bEV{\estacorrpm{\obsell{}{}}{\obsell{}{}}^{(1|2)} (\vartheta)}^{\text{GI}}
\approx
\\
F^{-1} \geomweightS{(1)}{} \bigl( \chigalSrep{(2)}{}\bigr) 
	\int_{0}^{\infty}\!\! \sepperp' \diff{\sepperp'}\, \mathcal{G}_{+,\tang}\bigl(\fgalSrep{(2)}{} \vartheta, \sepperp'\bigr)
 \estpcorr{\deltagalD{}{}} {\obsellt{(2)}{}}  (\sepperp')
,
\end{multline}
where $\chigalSrep{(2)}{} = \chi\bigl( \zgalSrep{(2)}{} \bigr)$, and $\fgalSrep{(2)}{} = \fK\bigl( \chigalSrep{(2)}{} \bigr)$.

\section{Methods}
\label{sec:methods}

\subsection{Structure formation simulation}
\label{sec:methods:simulations}

Our studies on intrinsic alignments are based on the Illustris-1 simulation. The Illustris project \citep[][]{VogelsbergerEtal2014_Illustris_Nature, VogelsbergerEtal2014_Illustris, GenelEtal2014_Illustris, SijackiEtal2015_Illustris, NelsonEtal2015} comprises a suite of cosmic structure formation simulations carried out with the moving-mesh hydrodynamics and gravity code \softwarename{AREPO} \citep[][]{Springel2010_Arepo, VogelsbergerEtal2012_Arepo}.
The simulations include various astrophysical processes such as primordial and metal-line cooling of gas with self-shielding corrections, stochastic star formation in dense gas, stellar evolution and feedback from stellar winds and supernovae producing galactic outflows, gas recycling, chemical enrichment, super-massive black hole growth, and feedback from active galactic nuclei \citep[][]{VogelsbergerEtal2013, VogelsbergerEtal2014_Erratum}.

The Illustris simulations assume a spatially flat cold dark matter (CDM) cosmological model with a cosmological constant. The cosmological parameters for the simulations are \citep[][]{VogelsbergerEtal2014_Illustris}:
a Hubble constant $\HubbleConstant = h 100\,\kms \Mpc^{-1}$ with $h=0.704$, a mean matter density parameter $\OmegaMatter=0.2726$, a baryon density parameter $\OmegaBaryon=0.0456$, a matter density power spectrum normalization $\sigma_8 = 0.809$, and a spectral index $n_{\mathrm{s}} = 0.936$.

The highest resolution simulation, Illustris-1, covers a comoving volume of $(106.5\,\Mpc)^3$ with $2\times 1820^3$ resolution elements, resulting in a mass resolution of $6.26\times10^6\,\Msolar$ for the dark matter, and an initial mass resolution of $1.26\times10^6\,\Msolar$ for baryonic matter. At redshift $z=0$, gravitational forces are resolved down to a physical scale of $700\,\pc$, and the smallest hydrodynamical gas cells have an extent of $50\,\pc$. There are about 40 000 well-resolved galaxies at $z=0$. These statistically reproduce many fundamental properties of observed galaxies, such as galaxy luminosity functions and Tully-Fisher relations, as well as the observed mix of galaxy morphologies and colors including early-type, late-type, and irregular galaxies.

\subsection{Galaxy luminosities and apparent magnitudes}
\label{sec:methods:galaxy_luminosities}

In the simulation outputs, galaxies are identified as self-bound structures of gas cells, dark-matter and stellar particles using an updated version of the \softwarename{Subfind} algorithm \citep{SpringelDiMatteoHernquist2005_Subfind}.
To create simulated images for these galaxies (from which we then measure galaxy luminosities and shapes), we first compute raw luminosities $L_{\rm raw}$ of all stellar particles using the \citet{BruzualCharlot2003} stellar population synthesis model \softwarename{galaxev}. Each stellar particle is treated as a single stellar population created in an instantaneous starburst with a \citet{Chabrier2003} initial mass function. Given the star formation time and the stellar metallicity of a stellar particle, $L_{\rm raw}$ measured for any given filter bandpass can be uniquely determined.

We use a semi-analytic approach to take effects of dust attenuation into account. 
We assume a redshift-dependent optical depth $\tau_\lambda^{\rm a}$ due to dust absorption that is modeled as a function of the solar-neighborhood extinction curve $(\frac{A_\lambda}{A_{\rm v}})_{Z_{\odot}}$ (where $Z_{\odot}=0.02$ is the measured solar metallicity), the gas metallicity $Z_{\rm g}$, and the average hydrogen column density $\langle N_{\rm H}\rangle$ \citep[][]{GuiderdoniRoccaVolmerange1987,DevriendtGuiderdoniSadat1999,DevriendtGuiderdoni2000,KitzbichlerWhite2007,GuoWhite2009}:
\begin{equation}
\label{eq:KW07}
\tau_\lambda^{\rm a} = 
\bigg(\frac{A_\lambda}{A_{\rm v}}\bigg)_{Z_{\odot}} (1+z)^{\beta}\bigg(\frac{Z_{\rm  g}}{Z_{\odot}}\bigg)^s\frac{\langle N_{\rm H} \rangle}{2.1\times10^{21}{\rm cm}^{-2}}.
\end{equation}
We adopt the extinction curve $(\frac{A_\lambda}{A_{\rm v}})_{Z_{\odot}}$ from \citet{CardelliClaytonMathis1989}. We use a metallicity-dependence power-law index $s=1.35$ for $\lambda<2000\,\Angs$, and $s=1.6$ for $\lambda>2000\,\Angs$ as obtained by \citet{GuiderdoniRoccaVolmerange1987}. The redshift-dependence parameter $\beta=-0.5$ was originally adopted in \citet{KitzbichlerWhite2007} to reproduce measurements of Lyman-break galaxies at $z \sim 3$.

To also account for scattering by dust, we assume the effective optical depth $\tau_\lambda$ is given by \citep{CalzettiKinneyStorchiBergmann1994}:
\begin{equation}
\label{eq:scatslabtau}
\tau_\lambda = h_{\lambda}\sqrt{1-\omega_{\lambda}}\tau_\lambda^{\rm a}+(1-h_{\lambda})(1-\omega_{\lambda})\tau_\lambda^{\rm a},
\end{equation}
where $\omega_{\lambda}$ is the albedo, defined as the ratio between the scattering and the extinction coefficients. Eq.~\eqref{eq:scatslabtau} interpolates between two extreme cases: isotropic scattering and the forward-only scattering, weighted by $h_\lambda$ and ($1-h_\lambda$), respectively. The weighting parameter $h_{\lambda}$ describes the wavelength-dependent anisotropy scattering.

For each galaxy, the three principal directions of the simulation box are used as viewing directions. Each view of a galaxy is covered by a regular mesh of $100\times100$ cells and a total side length of $6$ times the half-stellar mass radius.  
This mesh is used to map the wavelength-dependent optical depth $\tau_{\lambda}$, which measures the total amount of extinction due to both absorption and scattering along the entire path of a radiation beam.
For each cell, we compute a hydrogen column density $\langle N_{\rm H} \rangle$ by projecting the cold hydrogen mass inside the halo of the galaxy onto the mesh. The optical depth for each mesh cell is then computed from this hydrogen column density via Eqs.~\eqref{eq:KW07} and \eqref{eq:scatslabtau}. The optical depth $\tau_\lambda$ at the position of any given stellar particle inside the mesh coverage is then computed by interpolation. For regions outside the mesh coverage, $\tau_\lambda=0$ has been assumed. 
The observed (i.e. dust-attenuated) luminosity $L_{\rm obs}$ of a stellar particle is then computed assuming a simple slab geometry with constant star and dust density within the slab:
\begin{equation}
L_{\rm obs} = 
\frac{1-\exp(-\tau_\lambda)}{\tau_\lambda} L_{\rm raw}
,
\end{equation}
where $L_{\rm raw}$ denotes the particle's raw (dust-free) luminosity.

The resulting observed luminosities of the stellar particles are then used to compute the observed magnitudes for each simulated galaxy in its viewing directions (galaxy luminosity functions and color distributions are presented in Section~\ref{sec:tests:magnitudes_and_colors}). Furthermore, the apparent magnitudes in the CFHT $u,g,r,i,z$ filters are used to compute galaxy spectral types with the Bayesian photometric redshift code \softwarename{BPZ} \citep[][]{Benitez2000,Benitez2011_BPZ_code}.

\subsection{Galaxy sizes and shapes}
\label{sec:methods:galaxy_shapes}

For each simulated galaxy and viewing direction, we generate an `unlensed' galaxy image from the projected positions and dust-attenuated luminosities of the galaxy's stellar particles. The galaxy image centers are assumed to coincide with the projected galaxy positions as provided by \softwarename{subfind} (i.e. the local minima of the galaxies' gravitational potentials).

The galaxy images are then used as input to compute galaxies' half-light radii, second central moments \eqref{eq:intrinsic_second_moments}, and ellipticities for various radial weight functions. First, we consider a uniform weight
\begin{equation}
\label{eq:uniform_weight}
  \RadialFluxWeight(\theta) = 1,
\end{equation}
from which we compute \lq{}unweighted\rq{} ellipticities.

Furthermore, we consider a weight function with a sharp radial cutoff. Central galaxies in the simulation may be embedded in an extended halo of gravitationally bound stellar particles that are assigned by the simulation analysis software to the central galaxy, but would likely be considered background light in observations. As a simple way to (partially) remove such extended stellar halos, we also compute moments only taking into account stellar particles within a radius $\theta_{\text{cut}}$ corresponding to a comoving projected radius of $r_{\text{cut}}=50\,\kpc$ from the galaxy center:
\begin{equation}
\label{eq:cut_weight}
  \RadialFluxWeight(\theta) = \HeavisideTheta(\theta_{\text{cut}} - \theta),
\end{equation}

Furthermore, we consider a weight function that combines the sharp radial cutoff with a Gaussian whose width is given by the image's half-light radius\footnote{computed from the light distribution, after the sharp cutoff has been applied} $\HalfLightRadius$ \citep[similar to the weights employed in the shear estimation method of][KSB]{KaiserSquiresBroadhurst1995}:
\begin{equation}
\label{eq:cut_and_Gauss_weight}
\RadialFluxWeight(\theta) = 
	 \exp\left( - \frac{1}{2} \frac{\theta^2}{\HalfLightRadius^2} \right)
	\HeavisideTheta(\theta_{\text{cut}} - \theta) 
.
\end{equation}

We estimate the galaxy shear polarizabilities for the various definitions of image ellipticity as follows: We apply small shear values $g_i = \pm 0.01$ to the \lq{}unlensed\rq{} galaxy image (i.e. we compute image positions for the stellar particles assuming a small shear), and then compute the moments and ellipticities from the resulting lensed image. The components of the shear polarizability tensor are then computed from these lensed ellipticity components by finite difference. We check the correctness of the numerical polarizability estimation by comparing its results to known analytic expressions in the case of unweighted moments.

\subsection{Matter density and tidal fields}
\label{sec:methods:densities_and_tidal_fields}

We wish to compare the results for ellipticity correlations computed from the galaxy light distribution in the simulations to the results predicted by adopting a simple tidal field model of galaxy shape alignment. We therefore compute the three-dimensional matter distribution in the simulation, and from that, the tree-dimensional matter overdensity field $\deltaMatter(\vx,\chi,z)$, the peculiar gravitational field $\gravitationalfield(\vx,\chi,z)$, defined here by the Poisson equation
\begin{equation}
	\deltaMatter(\vx,\chi,z) = \left( \parder{^2}{x_1^2} + \parder{^2}{x_2^2} + \parder{^2}{\chi^2}\right)\gravitationalfield(\vx,\chi,z),
\end{equation}
and the tidal field components 
\begin{equation}
\label{eq:tidal_field}
\tidalfield_{ij} (\vx,\chi,z) = \parder{^2 \gravitationalfield(\vx,\chi,z)}{x_i \partial x_j}.
\end{equation}
The tidal field components are used to define a tidal-field model intrinsic ellipticity field:
\begin{multline}
\label{eq:tidal_field_ellipticity}
	\intell{}{}(\vx,\chi,z) = 
	\\
	\tidalfieldtoellipticityfactor \left[ \bigl( \tidalfield_{11}(\vx,\chi,z) - \tidalfield_{22}(\vx,\chi,z) \bigr) + 2 \ii \tidalfield_{12}(\vx,\chi,z) \right],
\end{multline}
where the susceptibility $\tidalfieldtoellipticityfactor$ of the galaxy ellipticity to tidal fields is expressed as
\begin{equation}
\label{eq:tidalfieldtoellipticityfactor}
 \tidalfieldtoellipticityfactor = \frac{A C_1 \rhocrit \OmegaMatter}{D(z)},
\end{equation} 
with an adjustable amplitude parameter $A$, the critical density $\rhocrit$, the linear growth amplitude  $D(z)$ normalized to unity at redhift $z=0$, and a constant $C_1$ chosen such that $C_1 \rhocrit = 0.0134$ \citep[][]{JoachimiEtal2011}.  

To compute the intrinsic ellipticity field of the tidal field model, we first project all matter of a simulation snapshot onto a regular mesh of $1024^3$ pixels covering the simulation box (yielding a spatial resolution of $~100h^{-1}\,\kpc$). The resulting matter density on the mesh is converted to the relative matter overdensity $\deltaMatter$. The values of the gravitational field $\gravitationalfield$ and the tidal field components $\tidalfield_{ij}$ at the mesh points are then computed with Fast Fourier Transform (FFT) methods. The tidal-field components are then used to compute the ellipticity field at the mesh points using Eq.~\eqref{eq:tidal_field_ellipticity}.

We use the tidal field model ellipticity values on the three-dimensional mesh and trilinear interpolation to compute the ellipticity field values at the positions of the galaxies in the simulation. We thus obtain a galaxy catalog with galaxy ellipticities given by the simple tidal field model. From this catolog, we compute tidal field model predictions for the IA correlations as described in the following Section.

We also consider tidal field model predictions for IA correlations under the assumption that the galaxy density is given by a linear deterministic galaxy bias model with number density contrast $\deltagalS{}{} =  \biasgalS{}{}\deltaMatter$, where $\biasgalS{}{}$ denotes an adjustable galaxy bias parameter. This variant is essentially equivalent to the \emph{non-linear linear-alignment} (NLA) model of IA with (if assuming $\biasgalS{}{}>0$) or without (if $\biasgalS{}{}=0$) galaxy density weighting \citep[][]{HirataSeljak2004,BridleKing2007,JoachimiEtal2011,BlazekVlahSeljak2015}. For this variant, the field $(1 + \deltagalS{}{}) \intell{}{} = (1 + \biasgalS{}{} \deltaMatter{}{}) \intell{}{}$,  with $\intell{}{}$ given by Eq.~\eqref{eq:tidal_field_ellipticity}, is computed on the three-dimensional mesh from the values of the matter overdensity $\deltaMatter{}{}$ and tidal field components $\tidalfield_{ij}$ at the mesh points. Projections of the resulting field (used to compute projected correlations as described in the following Section) are then computed by summing the respective field values at the mesh points of the three-dimensional mesh along a principal direction of the simulation box and recording the result in a two-dimensional mesh.

\subsection{Correlations}
\label{sec:methods:correlations}

In this work, we focus on projected correlations involving intrinsic ellipticities as well as on the impact of IA on observed ellipticity correlations in cosmic shear.
We compute projected correlations primarily by correlating projected fields. We analyze the data from a set of simulation snapshots with redshifts between $z=0$ and $z=1.5$.
For each of these snapshots, we compute projections of the matter overdensity $\deltaMatter$, matter shear contrast $\deltashearMatter$, galaxy overdensities $\deltagalD{}{}$ and $\deltagalS{(\alpha)}{}$ (where $\alpha$, $\nu$, etc. $=1,2,\ldots$ may denote different galaxy samples), and galaxy intrinsic ellipticity fields $(1+\deltagalS{(\alpha)}{}) \intell{(\alpha)}{}$ for various galaxy samples\footnote{
Note that computing $[1+\deltagalS{(\alpha)}{}(\vx, \chi, z)] \intell{(\alpha)}{}(\vx, \chi, z)$ requires a value for $\intell{(\alpha)}{}(\vx, \chi, z)$ only where there is a galaxy at $(\vx, \chi, z)$, i.e. where $\deltagalS{(\alpha)}{}(\vx, \chi, z) \neq -1$.
}
along each simulation box axis on regular meshes of $8192^2$ pixels, which implies a resolution $\sim 10h^{-1}\,\kpc$ comoving (see Appendix~\ref{app:projected_correlations_from_simulations} for more information on computing the projected fields and their correlations). We then employ FFTs to compute the two-dimensional correlations of the projected fields. We compute averages of the resulting two-dimensional correlations within 50 logarithmically spaced bins of transverse separation, which together cover a range from 5 to $50\,000h^{-1}\,\kpc$ comoving.\footnote{
This setup for the binning ensures that the radial resolution and maximum range for the projected correlations is not limited by the binning, but by the mesh resolution and simulation box size.} These averages then serve as estimates of the projected correlations $\pcorr{\deltagalS{(\alpha)}{}}{\deltagalS{(\nu)}{}}(\sepperp, z)$, $\pcorr{\deltaMatter}{(1+\deltagalS{(\nu)}{}) \intellt{(\nu)}{}}(\sepperp, z)$, etc. at a discrete set of separations $\sepperp$ (chosen as the mid-values of the radial bins) and redshifts $z$ (given by the snapshot redshifts).

We use the projected correlations estimated from the simulation to compute predictions for the IA contributions to the observed ellipticity correlations. The expected II contributions for broad redshift distributions are computed using Eqs.~\eqref{eq:contributions_to_observed_ellipticity_correlations_from_raw_pair_contributions}, \eqref{eq:raw_pair_contributions_to_observed_ellipticity_correlations_broad_pz:normalization}, and \eqref{eq:raw_pair_contributions_to_observed_ellipticity_correlations_broad_pz:II} together with $\pcorr{\deltagalS{(\alpha)}{}}{\deltagalS{(\nu)}{}}(\sepperp, z)$ and $\pcorrpm{(1+\deltagalS{(\alpha)}{}) \intell{(\alpha)}{}}{(1+\deltagalS{(\nu)}{}) \intell{(\nu)}{}}(\sepperp, z)$ estimated from the simulation. The expected GI contributions for non-overlapping redshift distributions are computed using Eqs.~\eqref{eq:contributions_to_observed_ellipticity_correlations_from_raw_pair_contributions}, \eqref{eq:raw_pair_contributions_to_observed_ellipticity_correlations:GI}, and \eqref{eq:raw_pair_contributions_to_observed_ellipticity_correlations_no_density_cross_correlations:normalization} together with $\pcorrpm{\deltashearMatter}{(1+\deltagalS{(\nu)}{})\intell{(\nu)}{}}(\sepperp, z)$ estimated from the simulation. The integrals in these equation are computed by numerical integration with the projected correlations linearly interpolated between the values measured from the simulation.

To obtain an indication of the statistical error on our estimates for the correlations, we compare measurements using the three different box axes as viewing direction. In plots of the correlation functions below, the error bars indicate the spread between the three directions, and the lines indicate the mean.

Note that due to the limited box size, the projected correlations are likely underestimated for separations of a few $\Mpc$ and beyond. 
How strongly the estimates for correlations involving galaxy shapes are suppressed on larger scales has to be addressed in future work, e.g., employing simulations in larger boxes.
Currently, we can estimate from studies of gravity-only simulations \citep[see][]{AnguloHilbert2015} that $\pcorrpmGG{}$, which enters the prediction~\eqref{eq:raw_pair_contributions_to_observed_ellipticity_correlations:GG} for the GG contribution, is underestimated by $10\%$ or more on separations $\gtrsim 2 h^{-1}\,\Mpc$. We thus employ \softwarename{nicaea} \citep[][]{Nicaea} to compute predictions for the GG contribution to the observed ellipticity correlations instead.\footnote{
Our choice of \softwarename{nicaea} is for aesthetic reasons. We use GG predictions only to assess the relative impact of GI and II contributions on the observed ellipticity correlations, for which $\sim10\%$ accuracy in GG would suffice given the large uncertainty in  GI and II.
}

\section{Tests}
\label{sec:tests}

\subsection{Stellar particle numbers}
\label{sec:results:stellar_particle_distribution}

\begin{figure}
\centerline{\includegraphics[width=\linewidth]{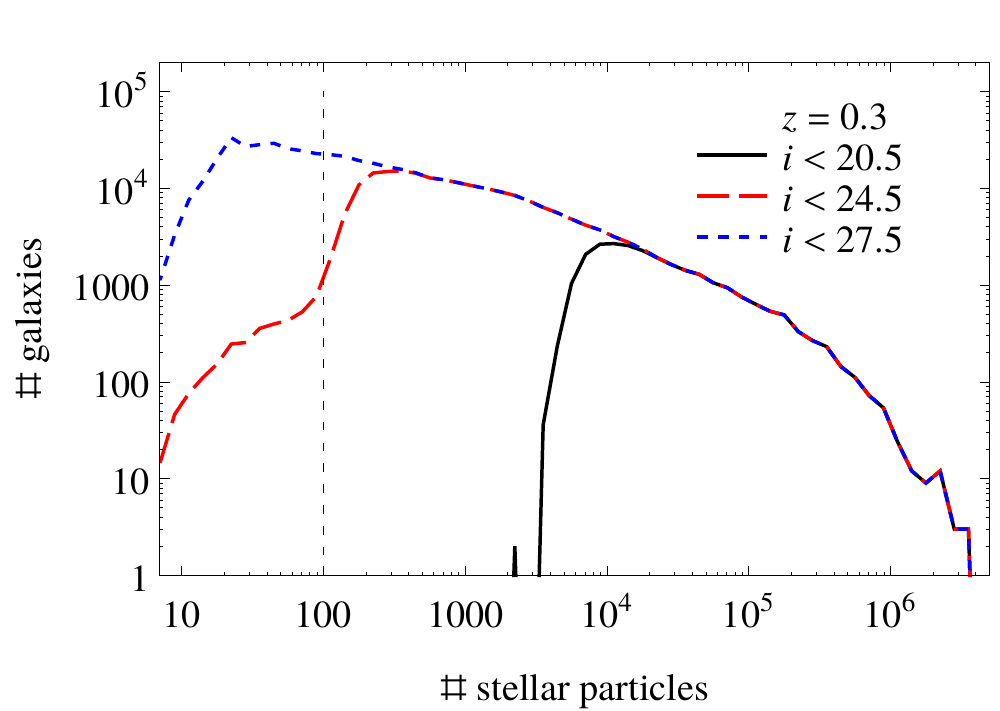}}
\caption{
\label{fig:stellar_particle_distribution}
Distribution of stellar particle numbers in galaxy samples at redshift $z=0.3$ with different apparent $i$-band magnitude limits.
The vertical dashed line indicates the lower limit of 100 stellar particles imposed for the galaxy shape analysis.
}
\end{figure}

We always require that any galaxy we use for shape measurements has at least 100 stellar particles \citep[see, e.g.,][for shape distributions and shape measurement errors as a function of particle number]{ChisariEtal2015, VelliscigEtal2015a}. This lower stellar particle limit corresponds roughly to a minimum stellar mass of $6\times 10^7 h^{-1} \,\Msolar$. The stellar particle limit does not affect samples that are selected according to stellar mass with a minimum mass $\gtrsim 10^8 h^{-1}\,\Msolar$. The minimum stellar particle number also roughly corresponds to a redshift-dependent apparent magnitude. For example, at $z=0.3$ that corresponding apparent magnitude is $i\sim 25$. 

Fig.~\ref{fig:stellar_particle_distribution} illustrates the distribution of stellar particle numbers in samples of galaxies at $z=0.3$ with various magnitude limits. The bright sample with $i<20.5$ does not contain galaxies with less than $3000$ stellar particles, and thus is not affected by the hard cut at 100 stellar particles. The intermediate-limit sample with $i<24.5$ is marginally affected. The faint-limit sample with $i<27.5$ is essentially limited by the stellar particle cut instead of the magnitude limit. Tests at higher redshift show that the $i<27.5$ becomes effectively magnitude-limited only for $z \geq 1$.

This test shows that (according to our 100-particle criterion) the Illustris simulation's mass resolution is high enough to study galaxy shapes for bright samples with $i<20.5$, as well as for deeper samples (at least for $z\geq 0.3$) with an apparent magnitude limit $i< 24.5$ similar to that of, e.g., the Canada-France Hawaii Telescope Lensing Survey\footnote{\href{http://www.cfhtlens.org}{\texttt{http://www.cfhtlens.org}}} \citep[CFHTLenS,][]{HeymansEtal2012} or \satellitename{Euclid}\footnote{\href{http://www.euclid-ec.org/}{\texttt{http://www.euclid-ec.org}}} \citep[][]{LaureijsEtal2011arXiv}. The mass resolution is however not sufficient to study galaxy image shapes for very deep samples (except for $z\geq 1$) with magnitudes $i<27$ such as, e.g. for the The Large Synoptic Survey Telescope\footnote{\href{https://www.lsst.org/}{\texttt{https://www.lsst.org/}}} \citep[LSST,][]{LSST2009arXiv} or Wide Field Infrared Survey Telescope\footnote{\href{http://wfirst.gsfc.nasa.gov}{\texttt{http://wfirst.gsfc.nasa.gov}}} \citep[WFIRST,][]{SpergelEtal2015arXiv}.

\subsection{Galaxy magnitudes and colors}
\label{sec:tests:magnitudes_and_colors}

\begin{figure*}
\centerline{\includegraphics[width=0.85\linewidth]{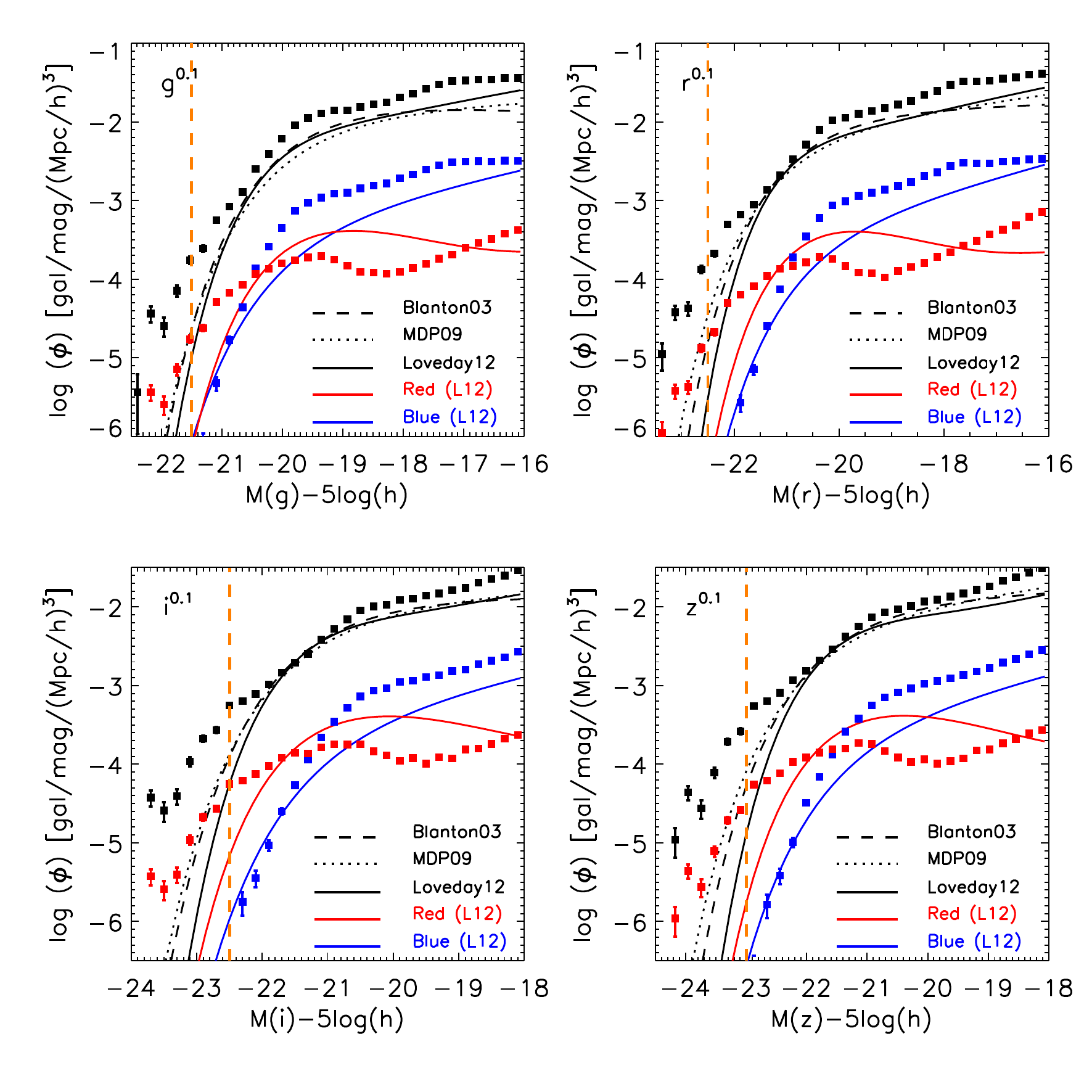}}
\caption{
\label{fig:LF_sdssz0p1}
The luminosity functions of low-redshift Illustris galaxies measured in the blue-shifted SDSS $g^{0.1}$, $r^{0.1}$, $i^{0.1}$, $z^{0.1}$ filter bandpasses. Square symbols (with their Poisson error bars) are for the simulation galaxy data. The blue and red colors represent the `blue' and `red' galaxies according to the color cut given in eq. ~(3) of \citet{LovedayEtal2012}. The solid lines are the double power-law Schechter function fits for the low-redshift GAMA galaxies from \citet{LovedayEtal2012}. The dashed and the dotted lines are single power-law Schechter function fits for the SDSS galaxies from \citet{BlantonEtal2003b} and \citet{MonteroDortaPrada2009}, respectively. The dashed (orange) vertical line indicates the bright-end magnitude limit of the observed galaxy sample.
}
\end{figure*}

Fig.~\ref{fig:LF_sdssz0p1} shows the luminosity functions of low-redshift Illustris galaxies measured in the blue-shifted SDSS $g^{0.1}$, $r^{0.1}$, $i^{0.1}$, $z^{0.1}$ filter bandpasses. Also plotted are the double power-law Schechter function fits by \citet{LovedayEtal2012} for galaxies in the Galaxy and Mass Assembly (GAMA). Using the same galaxy color cut (their eq.~3) to separate `blue' and `red' galaxies, our simulated galaxy sample roughly reproduces the observed luminosity functions for the total, as well as for the blue and red galaxy samples. In particular, the total number and the fractions of blue and red galaxies are well reproduced at the `knees' of the luminosity functions. This is also seen for higher-redshift galaxy samples in Illustris. The bright end of the observed luminosity functions has larger measurement uncertainties, as the derived magnitudes can differ strongly depending on the assumed light profiles \citep[see][for details]{BernardiEtal2013}.

Including the effects of dust in the computation of the galaxy magnitudes is crucial for reaching the level of agreement between observed and simulated galaxy luminosity functions seen in Fig.~\ref{fig:LF_sdssz0p1}. Without dust attenuation, blue galaxies would dominate the simulated galaxy luminosity functions in the shown filter bands to much brighter magnitudes $M\sim-22$ in conflict with the observed galaxy luminosity functions.

\begin{figure}
\includegraphics[width=0.9\linewidth]{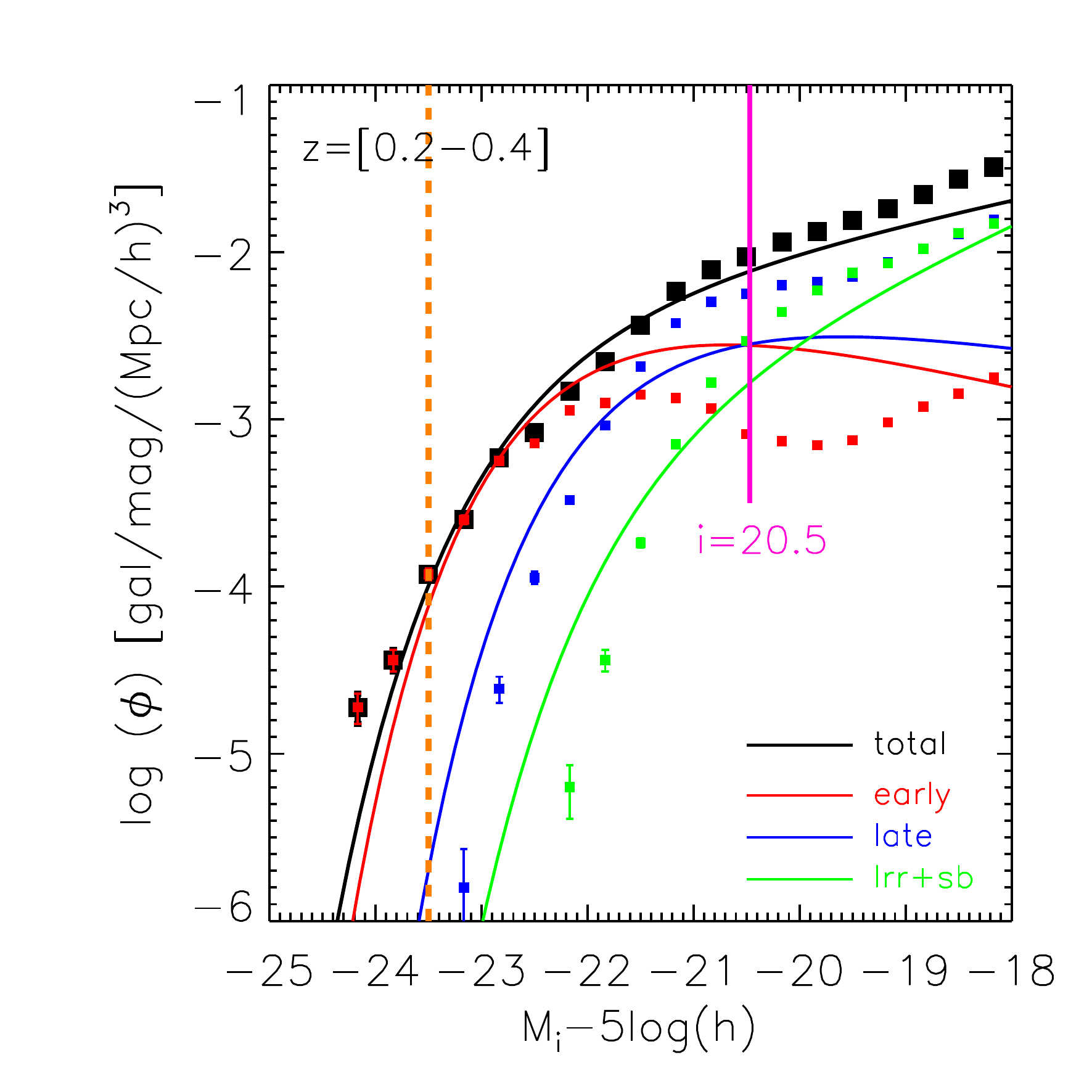}
\caption{
\label{fig:LF_CFHT_z0p3}
CFHT-$i$ band luminosity function of Illustris galaxies at $z=0.2-0.4$. Square symbols (with their Poisson error bars) are for the simulated galaxies. Black, green, blue and red represent
the `total', `irregular + star burst', `late-type' and `early-type' galaxy samples according to their \softwarename{BPZ} spectral types. For comparison, fits by \citet{RamosEtal2011} to observed luminosity function are shown as the solid lines. The dashed vertical line to the left indicates the bright-end magnitude limit of the observed galaxy sample. The solid vertical line on the right indicates an apparent magnitude $i=20.5$.}
\end{figure}
               
\begin{figure}
\includegraphics[width=0.9\linewidth]{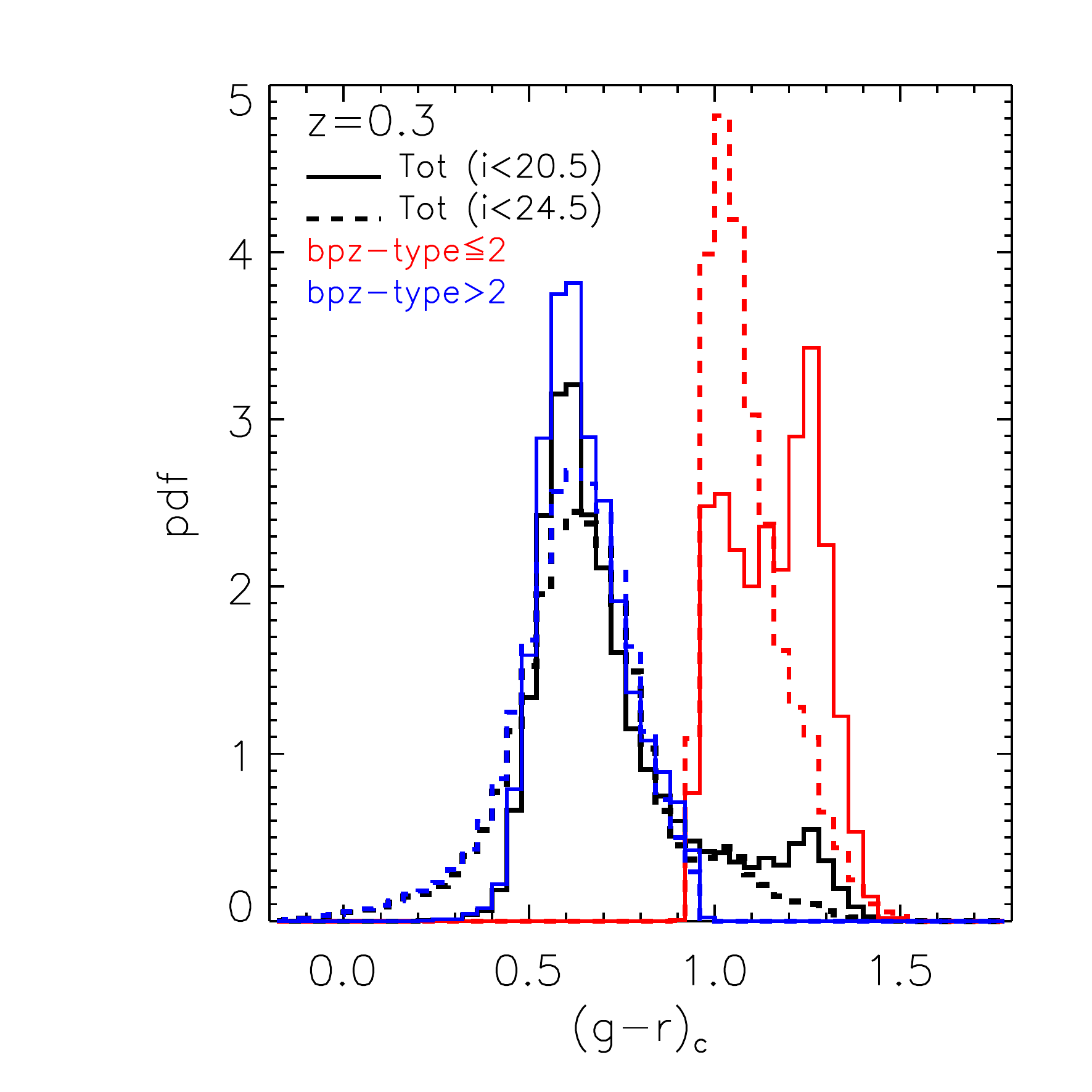}
\caption{
\label{fig:Color_CFHT_z0p3}
Distribution of galaxy color $(g-r)$ for Illustris galaxies at $z=0.3$. Black, blue and red curves represent the total galaxy sample, galaxies with \softwarename{BPZ}-types $>2$ (late-type) and with \softwarename{BPZ}-types $\leqslant2$ (early-type), respectively. The solid and the dashed lines represent the distributions for samples with magnitude limits of $i<20.5$ and $i<24.5$, respectively.}
\end{figure}

The dust treatment also significantly affects the photometric type classification of the simulated galaxies.
Fig.~\ref{fig:LF_CFHT_z0p3} shows the CFHT-$i$ band luminosity functions of Illustris galaxy subsamples selected by their \softwarename{BPZ} type. The Illustris galaxy luminosity functions are compared to the power-law Schechter function fits to observed galaxy samples by \citet{RamosEtal2011}. Once again the fractions of early-type (\softwarename{BPZ}-types $\leq 2$) and late-type (\softwarename{BPZ}-types $>2$) galaxies are roughly reproduced at the `knees' magnitudes. At the fainter end, an excess of late-type galaxies together with a deficit in early-type galaxies from the simulation indicate an overshoot of star-formation activities. This is possibly due to shortcomings of the implemented feedback models \citep[see][]{NelsonEtal2015}.

The $(g-r)$ color distributions presented in Fig.~\ref{fig:Color_CFHT_z0p3} clearly show bimodality with the late-type galaxies bluer than their early-type counterparts. The fraction of galaxies in the simulation classified as early type is somewhat low, e.g. about $9\%$ at redshift $z = 0.3$ and about $5\%$ at $z = 0.6$ for galaxies with apparent magnitude $i<24.5$.

\subsection{Apparent galaxy sizes}
\label{sec:tests:galaxy_sizes}

\begin{figure}
\centerline{\includegraphics[width=\linewidth]{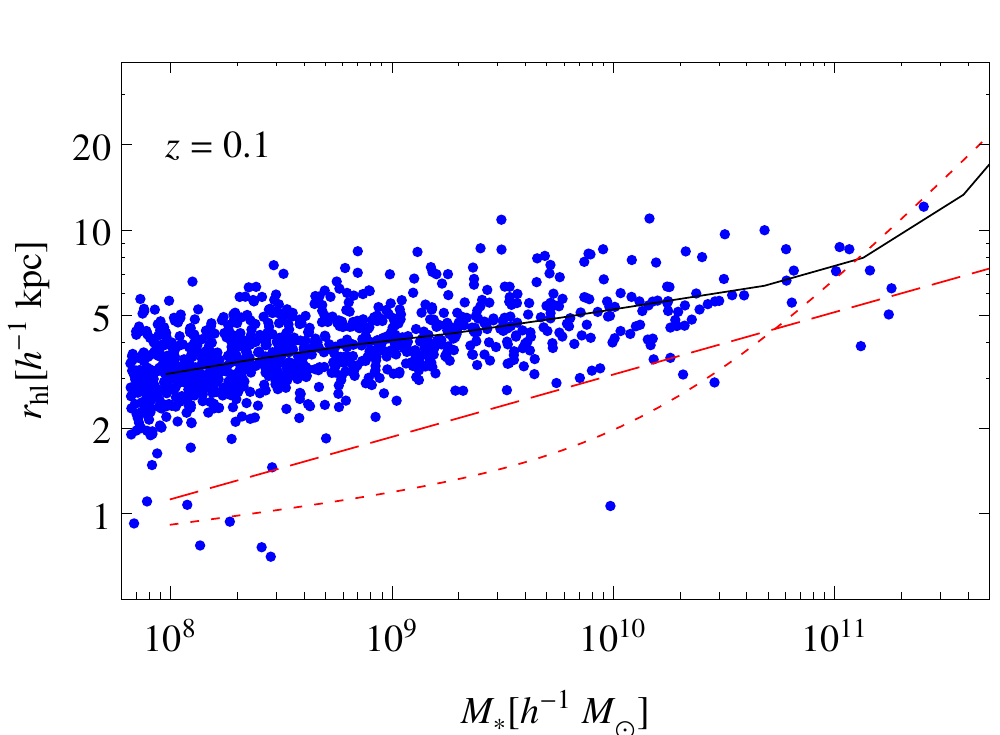}}
\caption{
\label{fig:stellar_mass_size_relation}
Stellar mass-size relation: Shown are the stellar mass $M_*$ and half-light radius $r_{\text{hl}}$ for a randomly selected sample of 1000 galaxies at redshift $z=0.1$. Also shown are the median  $r_{\text{hl}}$ as a function of  $M_*$ (solid line, computed using all galaxies in the simulation), as well as fits to the stellar mass-size relation by \citet{LangeEtal2015} for early-type galaxies (dotted line) and late-type  galaxies (dashed line) in the Galaxy And Mass Assembly (GAMA) survey.
}
\end{figure}

A fundamental property of galaxy image morphology is the apparent size of galaxy images. Fig.~\ref{fig:stellar_mass_size_relation} shows the distribution of galaxy sizes, expressed as half-light radii, as a function of total stellar mass. The stellar mass-size relation is roughly compatible with observed galaxy properties \citep{Shen2003,LangeEtal2015} for larger stellar masses $M_* \gtrsim 10^{11} h^{-1}\Msolar$. For smaller stellar masses, however, the simulated galaxies appear to have larger half-light radii than observed galaxies. This may be in part an artifact of how the stellar components of simulated galaxies are identified and how their light distribution (which also includes very low surface brightness regions at large radii) is quantified. Another possible reason is the limited spatial resolution. However, the larger radii are most likely, at least in part, due to shortcomings of the star formation and feedback model \citep[][]{ScannapiecoEtal2012}.

\subsection{Density correlations}
\label{sec:tests:density_correlations}

Galaxy density correlations are of interest for intrinsic alignment and cosmic shear studies for several reasons. They feature in the expectation value for the common cosmic shear correlation function estimator~\eqref{eq:obsell_correlation_estimator}. They allow one to estimate the galaxy bias required to infer matter density-intrinsic ellipticity correlations from galaxy density-intrinsic ellipticity correlations. Particular features in the density correlations are likely also visible in intrinsic ellipticity correlations through terms $(1+\deltagalS{}{})$ in these correlations.

\begin{figure}
\centerline{\includegraphics[width=\linewidth]{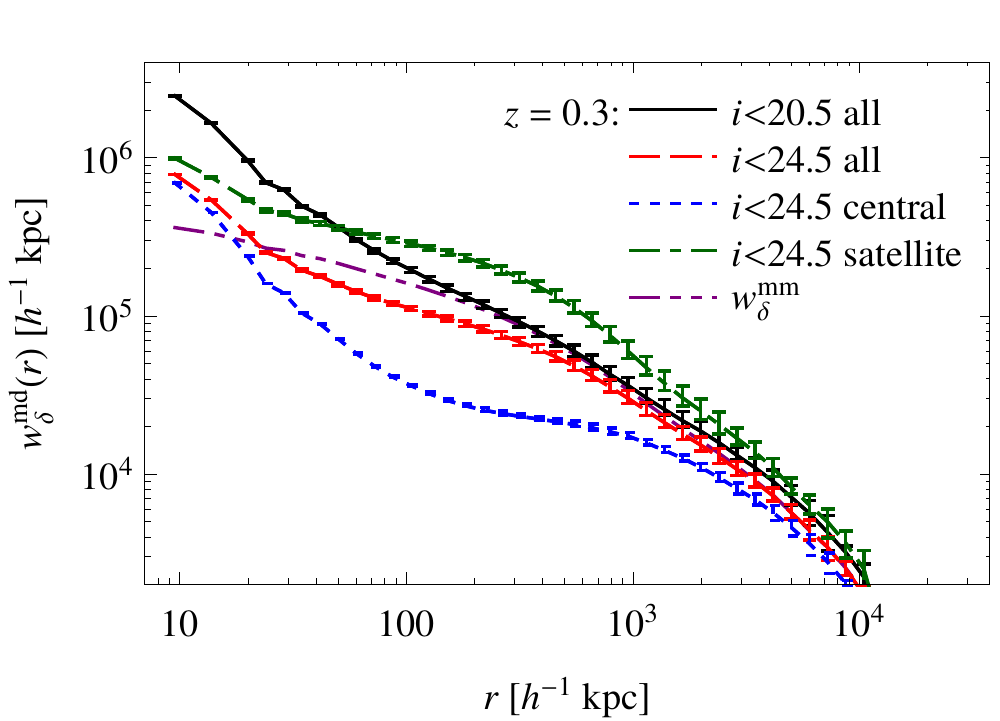}}
\centerline{\includegraphics[width=\linewidth]{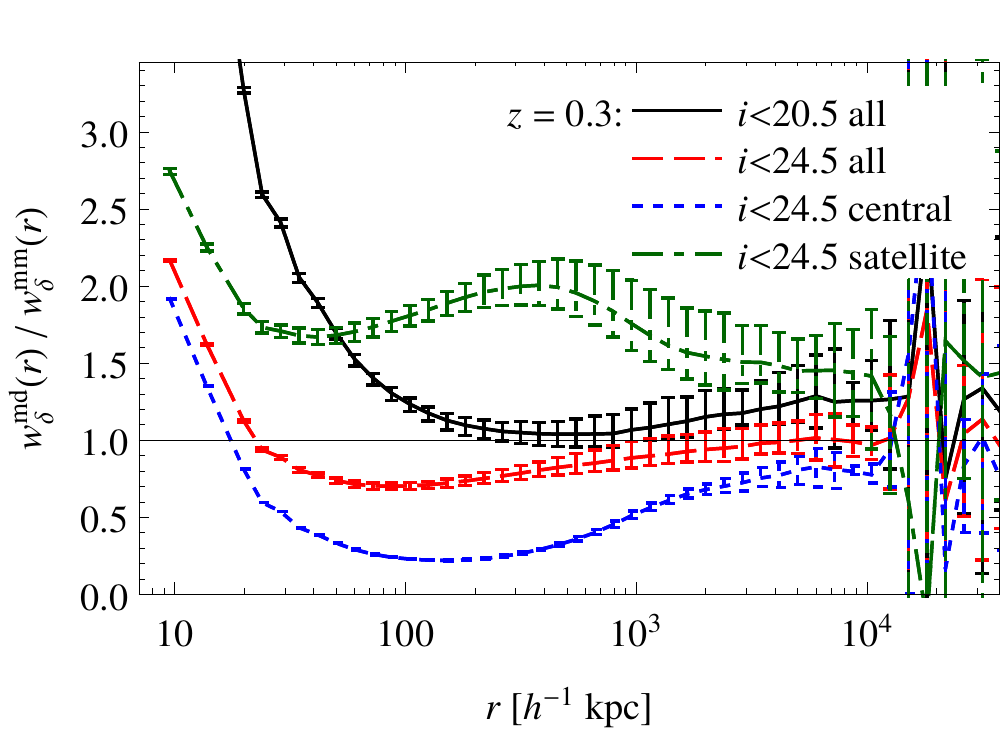}}
\caption{
\label{fig:w_delta_md}
Projected matter density-galaxy density correlations $\pcorrmd{}(r)$ as a function of the separation $r$ (upper panel) for galaxy samples with different apparent $i$-band magnitude limit, and also selected by positional type. The projected matter density correlation $\pcorrmm{}$ is also shown for comparison (upper panel), as well as the ratios $\pcorrmd{}(r)/\pcorrmm{}(r)$  (lower panel).
}
\end{figure}

Here, we present results for the projected auto- and cross-correlations of the matter and galaxy density fields that we measure in the Illustris-1 simulation. Note that due to the finite simulation box size of $75h^{-1}\,\Mpc$, any correlation measurements become unreliable for separations $r\gtrsim 10h^{-1}\,\Mpc$. We thus generally restrict our discussion to separations  $r \leq 10h^{-1}\,\Mpc$.

Fig.~\ref{fig:w_delta_md} compares the projected matter density-galaxy density correlations $\pcorrmd{}$ for various galaxy samples and the projected matter density correlation $\pcorrmm{}$. The cross-correlation of the matter and the bright sample defined by an apparent magnitude $i<20.5$ is noticeably larger than the matter auto-correlation. The excess indicates the galaxy density field is biased with respect to the matter density field by a bias factor $\biasgal{}{} \approx 1.1 - 1.3$ (simply estimated as the ratio of the cross- and auto-correlation) on scales $r\gtrsim 0.1h^{-1}\,\Mpc$. The fainter sample with $i<24.5$ shows a smaller cross-correlation indicative of a bias $\biasgal{}{} \approx 0.8-1$. Samples of central and satellite galaxies show opposite trends, in particular on scales $0.1h^{-1}\,\Mpc \lesssim r \lesssim 1h^{-1}\,\Mpc$, a weaker cross-correlation for centrals and a stronger cross-correlation for satellites.
This is expected, since on average, galaxies in the satellite sample reside in more massive and more extended halos than those in the central sample.

\section{Results}
\label{sec:results}

Here, we present the main results of our study. First, we discuss how the choice of ellipticity estimator affects the resulting intrinsic ellipticity distributions and IA correlations.
We then discuss the matter density-intrinsic ellipticity (mI) and galaxy density-intrinsic ellipticity (dI) correlations, including a comparison to a tidal field model.
These parts are followed by a discussion of the shear-intrinsic ellipticity (GI) correlations and their relation to the mI and dI correlations, and by a discussion of intrinsic ellipticity-intrinsic ellipticity correlations.
In the final parts of this Section, the results for the intrinsic ellipticity-intrinsic ellipticity and shear-intrinsic ellipticity correlations are used to assess the GI and II contributions to the observed ellipticity correlations in tomographic cosmic shear surveys.
Note that (unless explicitly stated otherwise) values for length-like quantities such as line-of-sight distances, transverse separations, projected correlations, etc. are stated in comoving units.

\subsection{Intrinsic ellipticity distributions}
\label{sec:tests:intrinsic_ellipticity_distributions}

\begin{figure}
\centerline{\includegraphics[width=\linewidth]{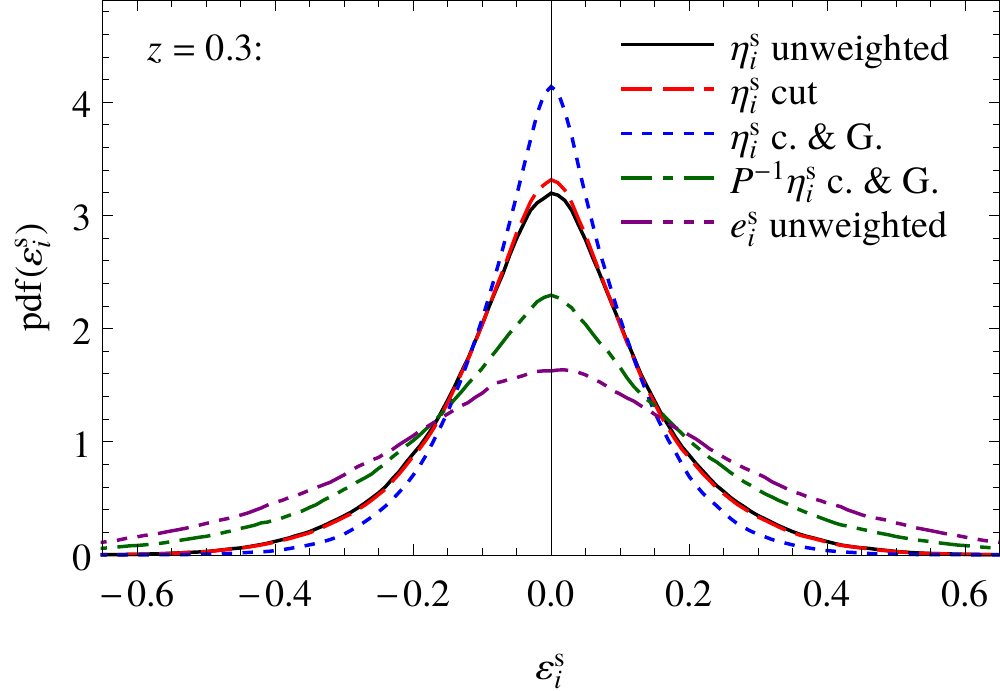}}
\caption{
\label{fig:ellipticity_distribution}
Distribution of the intrinsic ellipticity components $\intell{}{i}$ for galaxies with apparent $i$-band magnitude $i\leq 24.5$ at redshift $z=0.3$: \lq{}unweighted\rq{} ellipticities $\inteta{}{}$ derived from moments employing a flat weight~\eqref{eq:uniform_weight} (solid line), \lq{}weighted\rq{} ellipticities from moments employing a sharp radial cutoff~\eqref{eq:cut_weight} (dashed line), or a sharp cutoff combined with a Gaussian~\eqref{eq:cut_and_Gauss_weight} (dotted line), ellipticities based on weight function~\eqref{eq:cut_and_Gauss_weight} corrected by the inverse shear polarizability (dash-dotted line),
and \lq{}unweighted\rq{} ellipticities $\inte{}{}$ (dash-dot-dotted line).
}
\end{figure}

\begin{figure}
\centerline{\includegraphics[width=\linewidth]{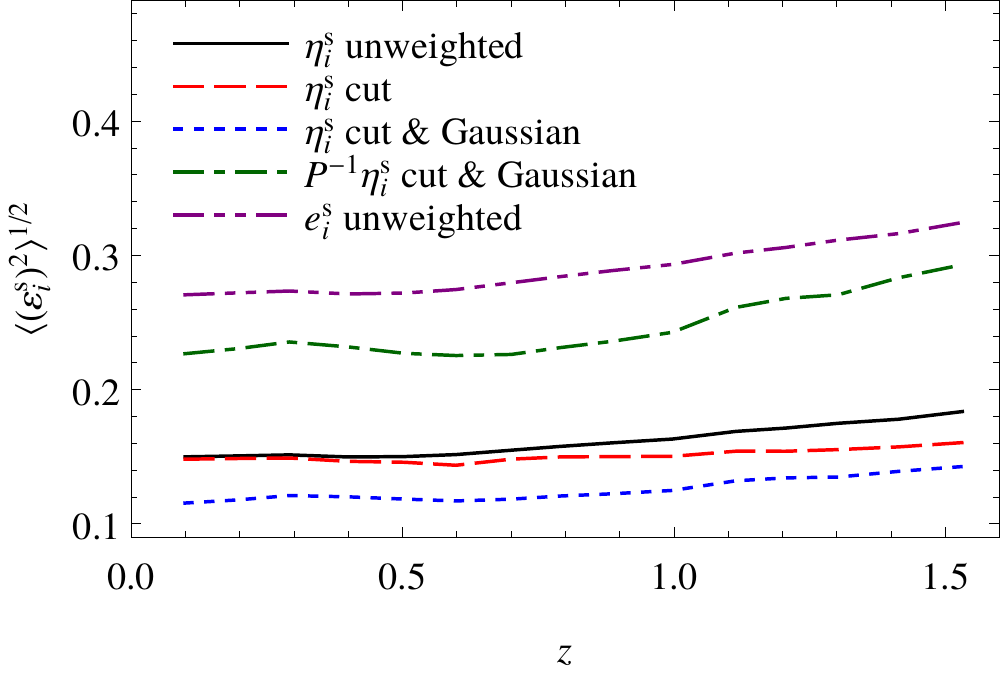}}
\caption{
\label{fig:ellipticity_redshift_evolution}
Standard deviation $\EV{(\intell{}{i})^2}^{1/2}$ of the intrinsic ellipticity components $\intell{}{i}$ as a function of galaxy redshift $z$ for galaxies with apparent $i$-band magnitude $i\leq 24.5$.
}
\end{figure}

The fundamental properties of galaxy image morphology central to our study are the intrinsic ellipticities of galaxy images. 
There are various ellipticity estimators that can be constructed from  second-order brightness moments with different choices for the radial and flux weight function $\RadialFluxWeight(\theta, \ImageSurfaceBrightness)$, ellipticity definition ($\obseta{}{}$ or $\obse{}{}$), and shear response correction (mean vs. individual shear polarizability).  Here, we consider ellipticities $\inteta{}{}$ derived from moments using a uniform weight~\eqref{eq:uniform_weight}, a weight~\eqref{eq:cut_weight} with a sharp cutoff at a radius $\theta_{\text{cut}}$ corresponding to a comoving projected radius of $r_{\text{cut}}=50\,\kpc$ to exclude contributions from intra-cluster light (see Section~\ref{sec:methods:galaxy_shapes}), and a sharp radial cutoff combined with Gaussian weight~\eqref{eq:cut_and_Gauss_weight} with width given by the image's half-light radius $\HalfLightRadius$.
Distributions for ellipticities $\inte{}{}$ can be derived from the distributions of $\inteta{}{}$ exploiting relation~\eqref{eq:e_from_eps}.

We confirm that all considered ellipticities have a mean that is compatible with zero. Moreover, the joint distribution $\pdf(\intell{}{1},\intell{}{2})$ of the ellipticity components $\intell{}{1}$ and $\intell{}{2}$ is compatible with a rotationally symmetric two-dimensional distribution. Thus there is no indication of an artificially preferred orientation of the galaxy images. 

The distributions of ellipticity components are compared in Fig.~\ref{fig:ellipticity_distribution} for galaxies with apparent magnitude $i\leq 24.5$ at redshift $z = 0.3$. Introducing a sharp cutoff at large radii for the moment calculation does hardly affect the ellipticity distribution. In contrast, down-weighting outer parts of the galaxy image by a Gaussian adjusted to the galaxy size does make the galaxies appear significantly rounder. The ellipticity distribution becomes much broader again when the individual shear polarizabilities are also taken into account to obtain an ellipticity that can serve as estimator for gravitational shear. 

As Fig.~\ref{fig:ellipticity_redshift_evolution} indicates, the ellipticity distributions gently broaden with increasing redshifts. Standard deviations rise from $0.15$ per component at $z=0.1$ to $0.18$ at $z=1.5$ for ellipticities $\inteta{}{}$ computed from unweighted moments. 
The same trend with redshift is visible for the other ellipticities.
For example, the unweighted ellipticities for the alternative definition~\eqref{eq:observed_alt_ellipticity} have a standard deviation of $\bEV{(\inte{}{i})^2}^{1/2} = 0.27$ at redshifts $z \leq 0.5$, rising to $0.30$ at $z=1$.
These values are very close to those \citet{TennetiEtal2014} found in the MassiveBlack-II simulation, i.e. $\bEV{(\inte{}{i})^2}^{1/2} \approx 0.26$ at $z=0.06$ rising to $\approx0.30$ at $z = 1$ for host subhalo masses $M_\text{sub} \approx 10^{10} h^{-1}\,\Msolar$.

\begin{figure}
\centerline{\includegraphics[width=\linewidth]{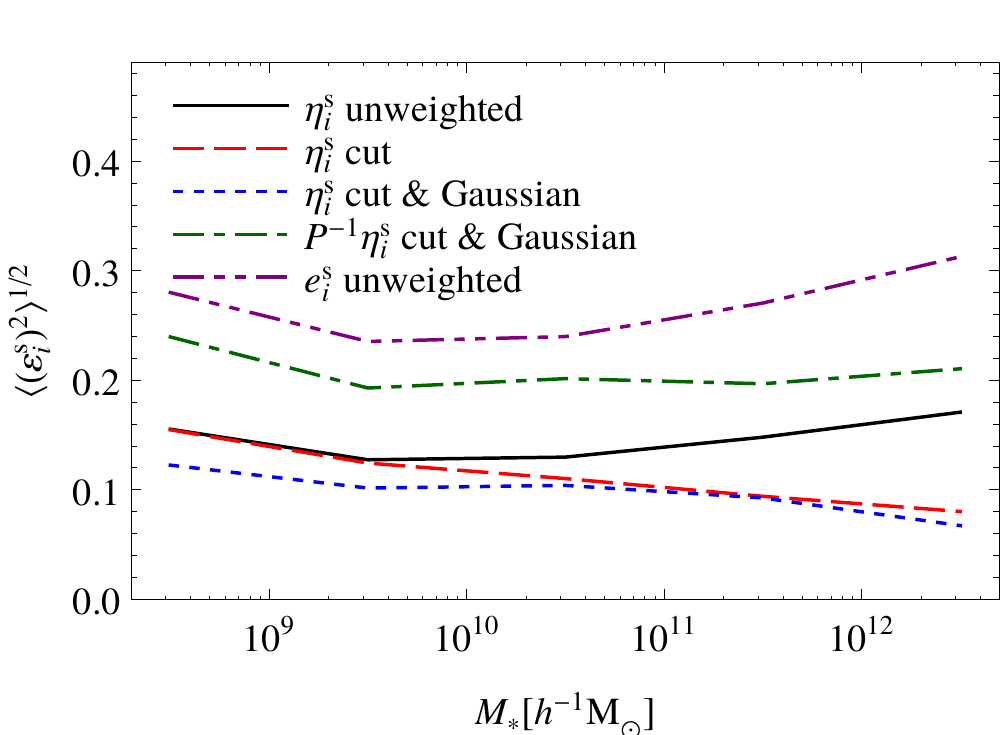}}
\caption{
\label{fig:ellipticity_stellar_mass_dependence}
Standard deviation $\EV{(\intell{}{i})^2}^{1/2}$ of the intrinsic ellipticity components $\intell{}{i}$ as a function of galaxy stellar mass $M_*$ at redshift $z=0.3$.
}
\end{figure}

The stellar-mass dependence of standard deviation of the ellipticity distribution is shown in Fig.~\ref{fig:ellipticity_stellar_mass_dependence}. The standard deviation for the weighted ellipticities slightly decreases with increasing stellar mass due to the increasing impact of the hard cutoff at fixed radius on the measured brightness moments.
In contrast, the standard deviation for the unweighted ellipticities is smallest for galaxies with stellar masses $M_* \sim 10^{10} h^{-1}\,\Msolar$, and slightly increases for larger and smaller stellar masses.
For example,  $\bEV{(\inte{}{i})^2}^{1/2} = 0.24$ for unweighted ellipticity components $\inte{}{i}$ of galaxies with $M_* \approx 10^{10} h^{-1}\,\Msolar$, and $\bEV{(\inte{}{i})^2}^{1/2} = 0.32$ for galaxies with $M_* \approx 10^{12} h^{-1} \Msolar$.
These values agree well with those by \citet{VelliscigEtal2015a} computed from the EAGLE and cosmo-OWLS simulations, i.e. $\bEV{(\inte{}{i})^2}^{1/2} \approx 0.25$ for host stellar masses $M_* \approx 10^8h^{-1}\Msolar$ to $10^{11}h^{-1},\Msolar$, and $\bEV{(\inte{}{i})^2}^{1/2}\approx 0.32$ for $M_* \approx 10^{12}h^{-1}\,\Msolar$ at $z=0$.

The ellipticity distributions for simulated galaxies show slightly less dispersion than the ellipticity distributions of observed galaxy samples. For example, \citet{ReyesEtal2012} obtained $\bEV{(\inte{}{i})^2}^{1/2}= 0.36$ for a source galaxy catalog based on the Sloan Digital Sky Survey\footnote{\href{http://www.sdss.org/}{\texttt{http://www.sdss.org}}} (SDSS), in agreement with the estimate $\bEV{(\inte{}{i})^2}^{1/2}= 0.36$ by \citet{MandelbaumEtal2012} and $\bEV{(\inte{}{i})^2}^{1/2}\approx 0.38$ by \citet{JoachimiEtal2013_IA_I} for galaxies in the Cosmic Evolution Survey\footnote{\href{http://cosmos.astro.caltech.edu/}{\texttt{http://cosmos.astro.caltech.edu/}}} (COSMOS). However, there are various observational effects (imperfect seeing and PSF correction, sky noise, etc.) that may broaden the observed ellipticity distributions, but which have not been accounted for in the simulations.

\subsection{Impact of ellipticity estimator on correlations}
\label{sec:results:shear_estimator}

\begin{figure}
\centerline{\includegraphics[width=\linewidth]{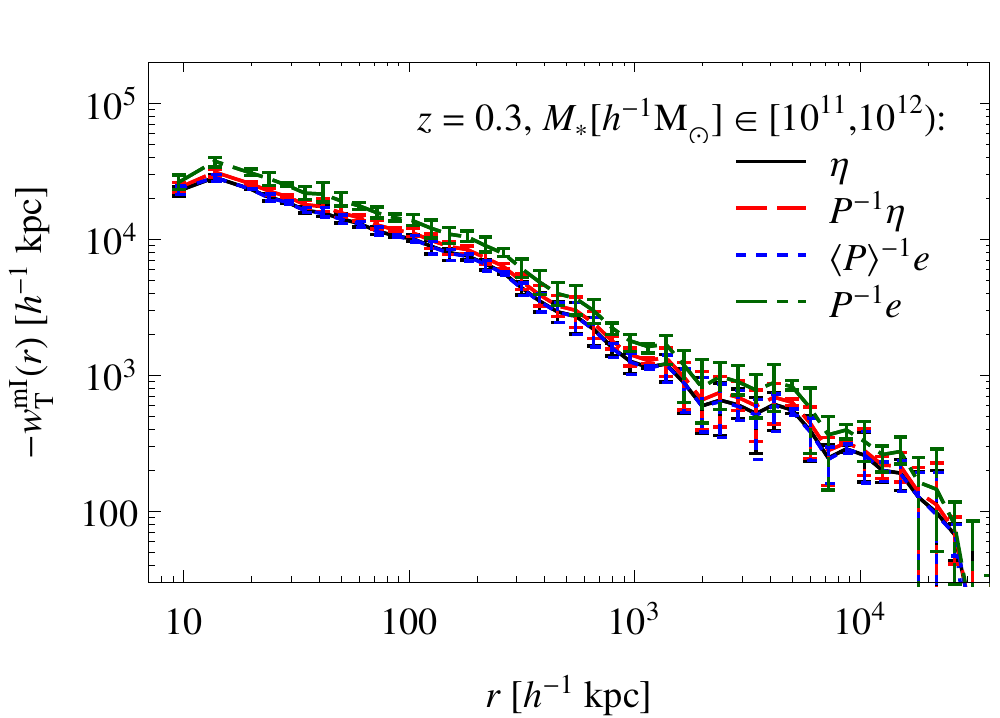}}
\caption{
\label{fig:w_t_mI_estimator_comparison_unweighted}
Projected matter density-intrinsic ellipticity correlations $\pcorrtmI{}(r)$ as a function of the separation $r$ for galaxies with stellar mass $10^{11}h^{-1}\Msolar \leq M_* < 10^{12}h^{-1}\Msolar$ at redshift $z=0.3$ for different shear estimators based on moments computed with uniform weight function~\eqref{eq:uniform_weight}.
}
\end{figure}

In this Section, we study how the choice of ellipticity estimator impacts two-point correlations involving galaxy ellipticities.
For a given choice of weight function $\RadialFluxWeight$ and ellipticity definitions~\eqref{eq:observed_ellipticity} or \eqref{eq:observed_alt_ellipticity}, the estimator~\eqref{eq:shear_estimator_based_on_response} employing individual shear polarizabilities yields similar matter density-intrinsic ellipticity correlations to those computed with estimator~\eqref{eq:shear_estimator_based_on_mean_response} employing mean polarizabilities. We also do not find large differences in the correlations between choosing $\obseta{}{}$ and $\obse{}{}$. Indeed the vast majority of measured ellipticities is small (see Section~\ref{sec:tests:intrinsic_ellipticity_distributions}), where $\obse{}{} \approx 2\obseta{}{}$ and $(\partial \obse{}{}/ \partial g) \approx 2(\partial \obseta{}{}/ \partial g)$ are very good approximations. 

In general, we find that, for a given choice of weight function, differences in the correlation $\pcorrtmI{}(r)$ due to choosing a different estimator $\obsell{}{}$ can be roughly described by an amplitude scaling as the ratio of the estimator's standard deviation $\EV{\obsell{2}{}}^{1/2}$. As Fig.~\ref{fig:w_t_mI_estimator_comparison_unweighted} illustrates, this leads to a very similar scale dependence of $\pcorrtmI{}(r)$. For the correlations shown there with ellipticities based on unweighted moments, the estimators $\obseta{}{}$ and $\EV{P}^{-1}\obse{}{}$ yield almost identical values (differences $\lesssim 2\%$), the estimator $P^{-1}\obseta{}{}$ yields about $10\%$ larger values, and the estimator  $P^{-1}\obse{}{}$ yields $\approx40\%$ larger values.

\begin{figure}
\centerline{\includegraphics[width=\linewidth]{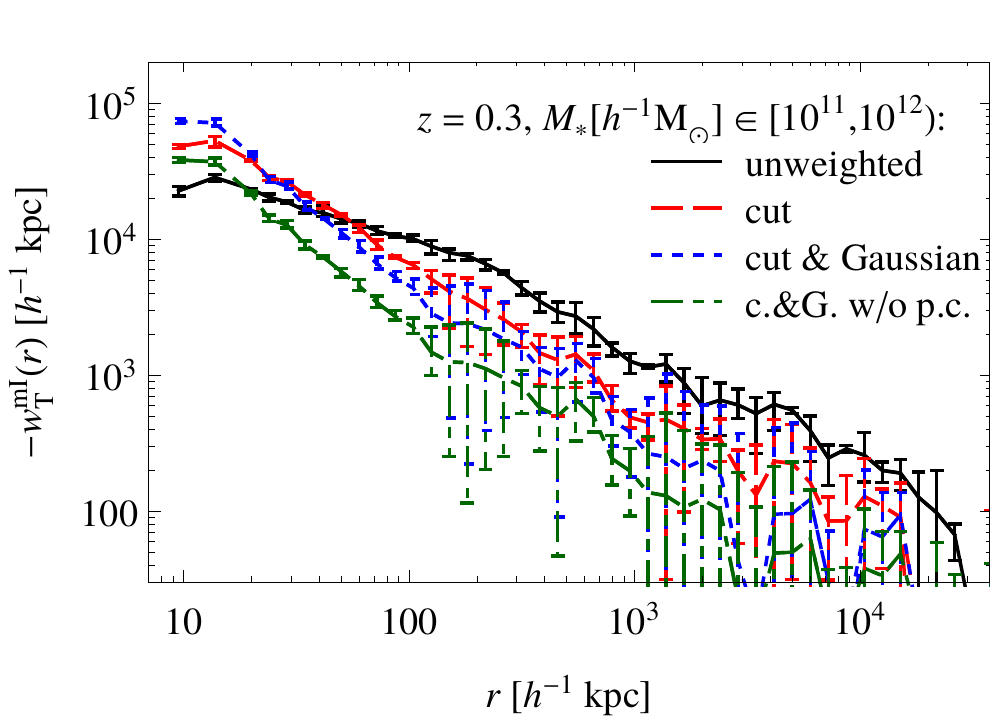}}
\centerline{\includegraphics[width=\linewidth]{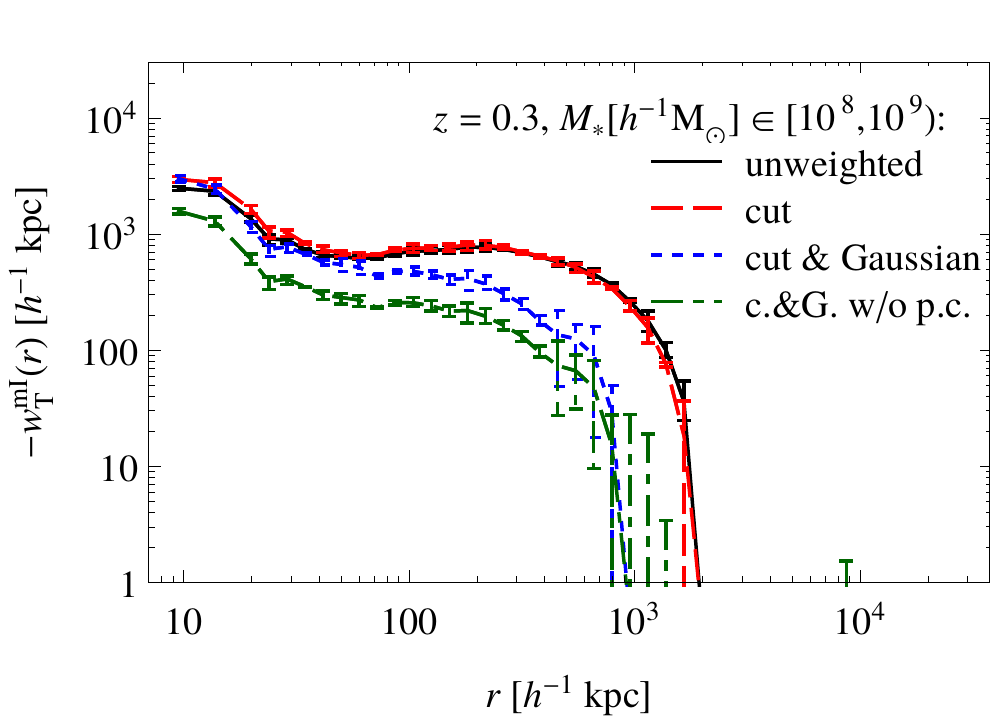}}
\caption{
\label{fig:w_t_mI_estimator_comparison}
Projected matter density-intrinsic ellipticity correlations $\pcorrtmI{}(r)$ as a function of the separation $r$ for more (upper panel) and less (lower panel) massive galaxies at redshift $z=0.3$ for different weight functions~\eqref{eq:uniform_weight} (solid lines), \eqref{eq:cut_weight} (dashed lines), and \eqref{eq:cut_and_Gauss_weight} (dotted lines) with polarizability correction, and for weight function~\eqref{eq:cut_and_Gauss_weight} but without polarizability correction (dash-dotted lines).
}
\end{figure}

In contrast, we find large differences both in amplitude and separation dependence for different choices of the weight function. We consider $\obsell{}{} = \obseta{}{}$ with a uniform weight~\eqref{eq:uniform_weight}, $\obsell{}{} = P^{-1} \obseta{}{}$ with a weight~\eqref{eq:cut_weight} featuring a sharp cutoff at a radius $\theta_{\text{cut}}$ corresponding to a comoving projected radius of $r_{\text{cut}}=50\,\kpc$, and $\obsell{}{} = P^{-1} \obseta{}{}$ with a weight~\eqref{eq:cut_and_Gauss_weight} that combines the sharp cutoff with a Gaussian weight with width given by the image's half-light radius $\HalfLightRadius$.
Figure~\ref{fig:w_t_mI_estimator_comparison} compares the resulting projected matter density-intrinsic ellipticity correlations $\pcorrtmI{}(r)$ for the different weight functions for galaxy samples with stellar masses $M_* \in [10^{11}, 10^{12})h^{-1}\,\Msolar$ and $M_* \in [10^{8}, 10^{9}) h^{-1}\,\Msolar$.
For the less massive (and thus small) galaxies, a sharp radial cutoff~\eqref{eq:cut_weight} at $50h^{-1}\,\kpc$ hardly changes the moments, ellipticities, and correlations compared to a uniform weight (only 6\% of these galaxies  have half-light radii exceeding $10h^{-1}\,\kpc$). The combination~\eqref{eq:cut_and_Gauss_weight} of the sharp cutoff with a Gaussian weight, however, significantly reduces the correlations on larger scales. For the more massive (and thus larger) galaxies, the sharp cutoff also lowers the amplitude of the correlations on larger scales (over 50\% of these galaxies have  half-light radii $>10h^{-1}\,\kpc$). The correlation amplitude on very small scales is, however, increased in the case of non-uniform weights.

The larger amplitude of the correlations for weighted moments compared to unweighted moments at very small separations may be related to the larger variance of the ellipticities based on weighted moments when their lower shear polarizability is accounted for. A possible explanation for the weaker correlations on large scales for weighted moments is that processes inducing galaxy shape alignments (e.g. large-scale tidal fields) on large scales more strongly affect the outskirts of galaxies than their inner regions. Shape estimators that down-weight the galaxies' outer regions are thus less susceptible to the large-scale shape alignments imprinted in these outer regions. A similar dependence of the alignment signal on the shape measurement method has also been found recently in observations \citep[][]{SinghMandelbaum2016}. 
This is also in qualitative agreement with the findings of \citet{VelliscigEtal2015b} that galaxy shape position angles in simulations show less alignments when only the inner regions of galaxies are considered.

A lower correlation amplitude compared to the case of unweighted moments has been noted in simulations, e.g., by \citet{TennetiEtal2015} when reduced moments are used, and by \citet{VelliscigEtal2015b} when moments are computed with a radial cutoff at the half-light radius. However, those studies do not take into account a (correct) polarizability correction, which is essential for unbiased shear estimators for weak lensing. Thus, it is not clear from those studies to what extent these smaller correlation amplitudes are merely due to the smaller galaxy ellipticity estimates resulting from the employed non-uniform radial weight functions (and which require larger polarizability corrections to turn them into shear estimators, which then also increases the correlation amplitudes).

Among the cases shown in Fig.~\ref{fig:w_t_mI_estimator_comparison}, the correlation amplitude is smallest when ellipticities are computed using the radial weight function~\eqref{eq:cut_and_Gauss_weight}, but without applying a polarizability correction. As stated before, such ellipticities are not suitable as shear estimates, and therefore the correlations based on these ellipticities are only of very limited use for constraining  models of IA for weak lensing. However, even when the polarisability correction is included to account for the non-uniform weights (which increases the correlation amplitude by a factor two in the illustrated case), the resulting correlation is still smaller (except on very small scales) than the correlation for flat weights.

Since the correlations are very different depending on whether uniform or centrally concentrated weights are used to compute image moments and ellipticities, we will present results for both \lq{}unweighted\rq{} ellipticities $\intell{}{}=\inteta{}{}$ computed from moments employing a uniform weight~\eqref{eq:uniform_weight}, and \lq{}weighted\rq{} ellipticities $\intell{}{}=(\partial \obseta{}{}/ \partial g)^{-1}\inteta{}{}$ computed from moments employing the weight function~\eqref{eq:cut_and_Gauss_weight}.   

\subsection{Matter density-intrinsic ellipticity correlations}
\label{sec:results:matter_density_ellipticity_correlations}

Here, we present results for the correlations of the matter density field (taking into account \emph{all} matter including baryons and CDM) and the intrinsic galaxy image ellipticities that we measure in the simulation.

\subsubsection{Halo mass dependence}
\label{sec:results:matter_density_ellipticity_correlations:halo_mass}

\begin{figure}
\centerline{\includegraphics[width=\linewidth]{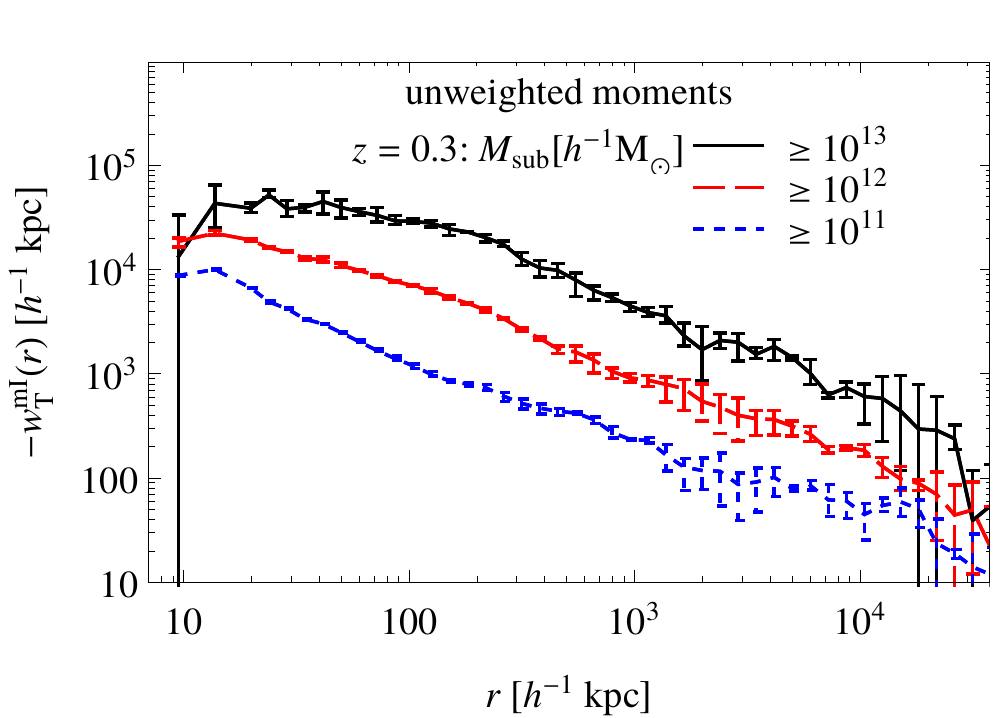}}
\centerline{\includegraphics[width=\linewidth]{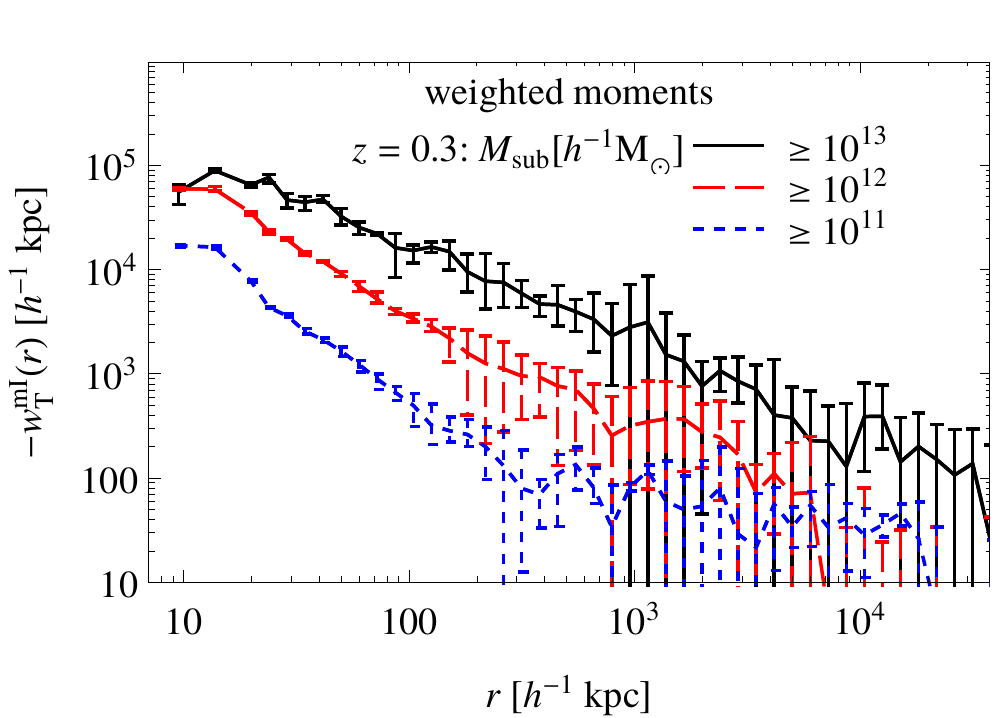}}
\caption{
\label{fig:w_t_mI_subhalo_mass_selected}
Projected matter density-intrinsic ellipticity correlations $\pcorrtmI{}(r)$ as a function of the separation $r$ for galaxies with subhalo masses $\Msub$ above different minimum subhalo masses at redshift $z=0.3$, employing either unweighted (upper panel) or weighted ellipticities (lower panel).
}
\end{figure}

The (gravitationally bound) subhalo mass of galaxies is not easily accessible in observations, but it can be measured in simulations (we use the mass as measured by \softwarename{subfind}).
Fig.~\ref{fig:w_t_mI_subhalo_mass_selected} shows the matter density-intrinsic ellipticity correlation $\pcorrtmI{}(r)$ as a function of the separation $r$ for galaxy samples with subhalo masses $\Msub$ above different lower subhalo mass limit $\Msubmin$. Generally, the correlations are stronger for higher minimum mass limits, with factors of a few increase per decade in minimum subhalo mass, suggesting $\pcorrtmI{} \propto \Msubmin^{0.5 - 0.8}$. The dependence of the correlations on separation $r \gtrsim 200 h^{-1}\,\kpc$ can be loosely described by a power law $\propto r^{-0.8}$ for unweighted ellipticities, and $\propto r^{-1.2}$ for weighted ellipticities.

The correlations $\pcorrtmI{}$ we measure for the subhalo mass-limited samples using unweighted moments are in rough agreement (i.e. within a factor two) with the results of \citet{TennetiEtal2015} from the Massive Black-II simulations. For example, we obtain $\pcorrtmI{}(1 h^{-1}\,\Mpc) = (250 \pm 50) h^{-1}\,\kpc$, $(900 \pm 200) h^{-1}\,\kpc$, and $(4000 \pm 500) h^{-1}\,\kpc$, compared to their results of  $250 h^{-1}\,\kpc$, $600 h^{-1}\,\kpc$, and $2000 h^{-1}\,\kpc$,  for $\Msub \geq 10^{11} h^{-1}\,\Msolar$, $\Msub \geq 10^{12} h^{-1}\,\Msolar$, and $\Msub \geq 10^{13} h^{-1}\,\Msolar$, resp.

\subsubsection{Stellar mass dependence}
\label{sec:results:matter_density_ellipticity_correlations:stellar_mass}

\begin{figure}
\centerline{\includegraphics[width=\linewidth]{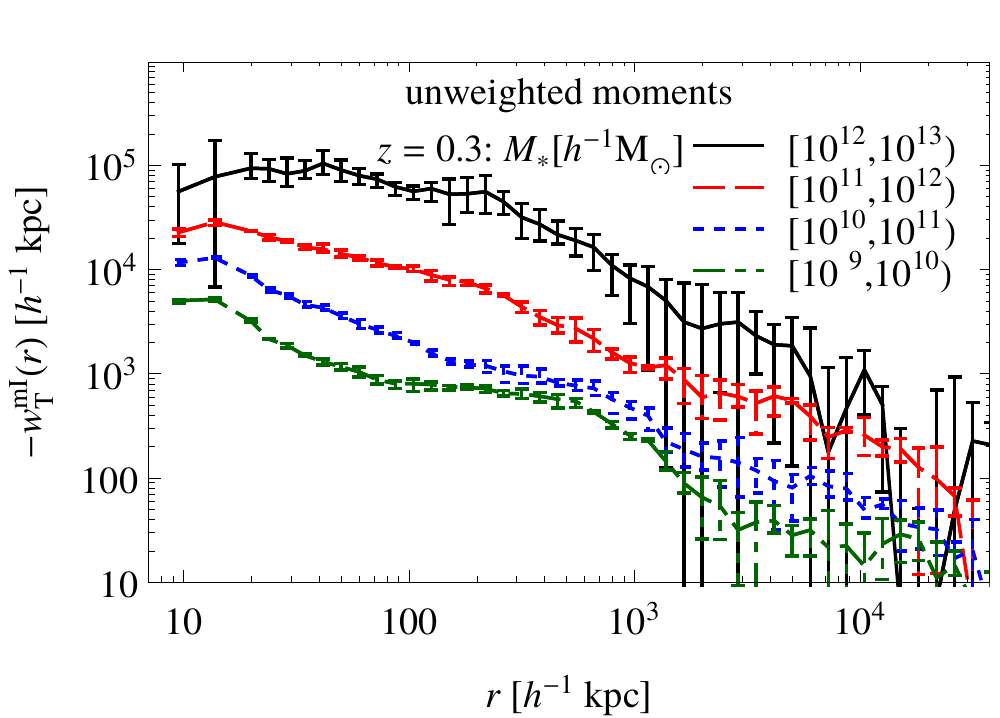}}
\centerline{\includegraphics[width=\linewidth]{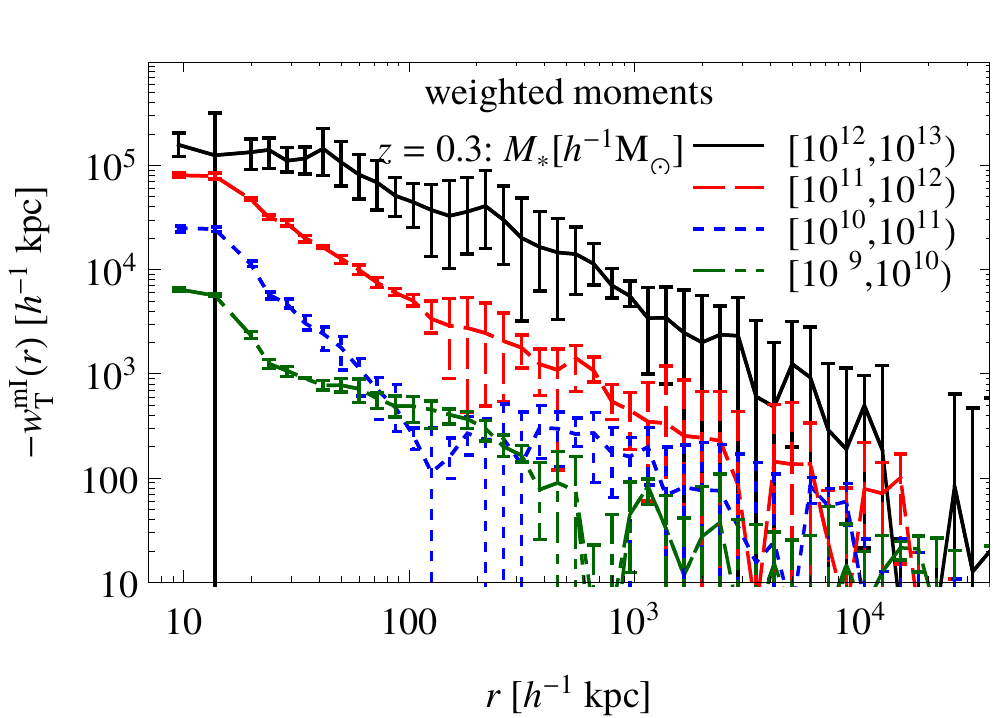}}
\caption{
\label{fig:w_t_mI_mass_selected}
Projected matter density-intrinsic ellipticity correlations $\pcorrtmI{}(r)$ as a function of the separation $r$ for galaxy samples with different stellar mass $\Mstellar$ at redshift $z=0.3$, employing either unweighted (upper panel) or weighted ellipticities (lower panel).
}
\end{figure}

The stellar mass of galaxies is straightforward to measure in simulations, too, but has a much closer connection to readily observable galaxy properties (such as magnitudes) than the subhalo mass. Results for the density-intrinsic ellipticity correlations of stellar mass-selected samples are shown in Fig.~\ref{fig:w_t_mI_mass_selected}. As for the subhalo mass-selected samples, the dependence of the correlations on separation $r$ can be loosely described by a power law $\propto r^{-0.8}$ for unweighted ellipticities, and $\propto r^{-1.2}$ for weighted ellipticities. The magnitude of correlation increases with increasing stellar mass.

The results for unweighted ellipticities are consistent with the findings of \citet{ChisariEtal2015} on the stellar-mass dependence of the density-ellipticity signal in the Horizon-AGN simulation, though there is an indication that the correlations we measure in the Illustris simulation are slightly stronger than those in the Horizon-AGN simulation (see, e.g., their figures~11 and 12).
Our results are also in agreement with the measurements of \citet{TennetiMandelbaumDiMatteo2016} from MassiveBlack-II and Illustris simulations, with  MassiveBlack-II showing somewhat stronger correlations than Illustris. In the considered redshift range $0 < z < 1$, we find no significant redshift dependence of the density-ellipticity signal of the stellar-mass selected galaxy samples.

\subsubsection{Magnitude dependence}
\label{sec:results:matter_density_ellipticity_correlations:magnitude}

\begin{figure}
\centerline{\includegraphics[width=\linewidth]{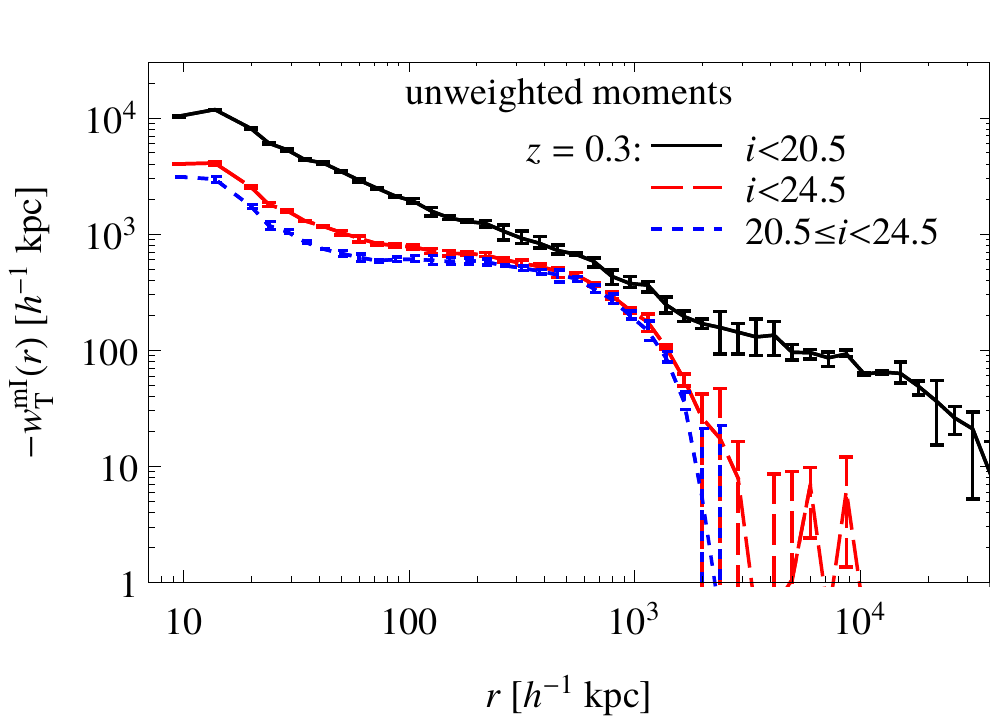}}
\centerline{\includegraphics[width=\linewidth]{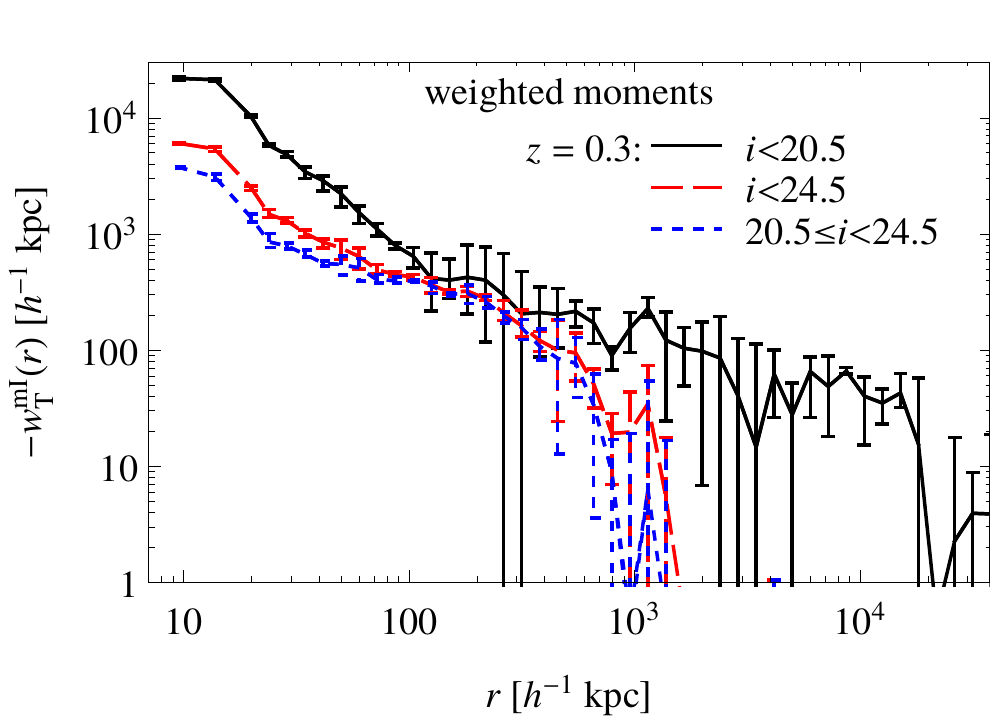}}
\caption{
\label{fig:w_t_mI_mag_selected}
Projected matter density-intrinsic ellipticity correlations $\pcorrtmI{}(r)$ as a function of the separation $r$ for galaxy samples with different magnitude limits at redshift $z=0.3$ employing unweighted moments (upper panel) or weighted moments (lower panel).
}
\end{figure}

Apparent magnitudes can be much more easily inferred from actual galaxy observations than stellar masses. The matter density-intrinsic ellipticity correlation $\pcorrtmI{}$ for galaxy samples selected at redshift $z = 0.3$ depending on their apparent $i$-band magnitude are shown in Fig.~\ref{fig:w_t_mI_mag_selected}. As for the mass-selected samples, correlations computed employing weighted moments generally show a stronger decrease in amplitude with increasing separation. 

The density-ellipticity correlation for bright galaxies with $i < 20.5$ shows a separation dependence $\sim r^{-0.8}$ like the mass selected samples in the case of unweighted moments. The fainter sample with $i < 24.5$ shows a lower correlation amplitude, and also stronger departure from a simple power-law separation dependence with a relatively larger matter density-intrinsic ellipticity correlation $\pcorrtmI{}(r)$ in the projected separation range $0.1h^{-1}\,\Mpc \lesssim r \lesssim 1h^{-1}\,\Mpc$. This shoulder-like feature is also present in the  matter density-galaxy density correlation $\pcorrmd{}(r)$, and is likely due to the fact that many of the galaxies in this sample are satellite galaxies in extended halos.
At projected separations $r >2h^{-1}\,\Mpc$, we do not detect a significant density-ellipticity correlation for the fainter sample.

Brighter galaxies are less numerous than fainter galaxies, but show much stronger alignment than fainter galaxies. Thus removing the brighter galaxies from a galaxy sample could substantially reduce the sample's intrinsic alignment correlations at the justifiable expense of a slightly reduced galaxy number density. However, as Fig.~\ref{fig:w_t_mI_mag_selected} shows, selecting only the galaxies with $20.5 \leq i < 24.5$ does not significantly reduce the alignment compared to the sample with just the faint limit $i<24.5$ imposed.

\subsubsection{Redshift dependence}
\label{sec:results:matter_density_ellipticity_correlations:redshift}

\begin{figure}
\centerline{\includegraphics[width=\linewidth]{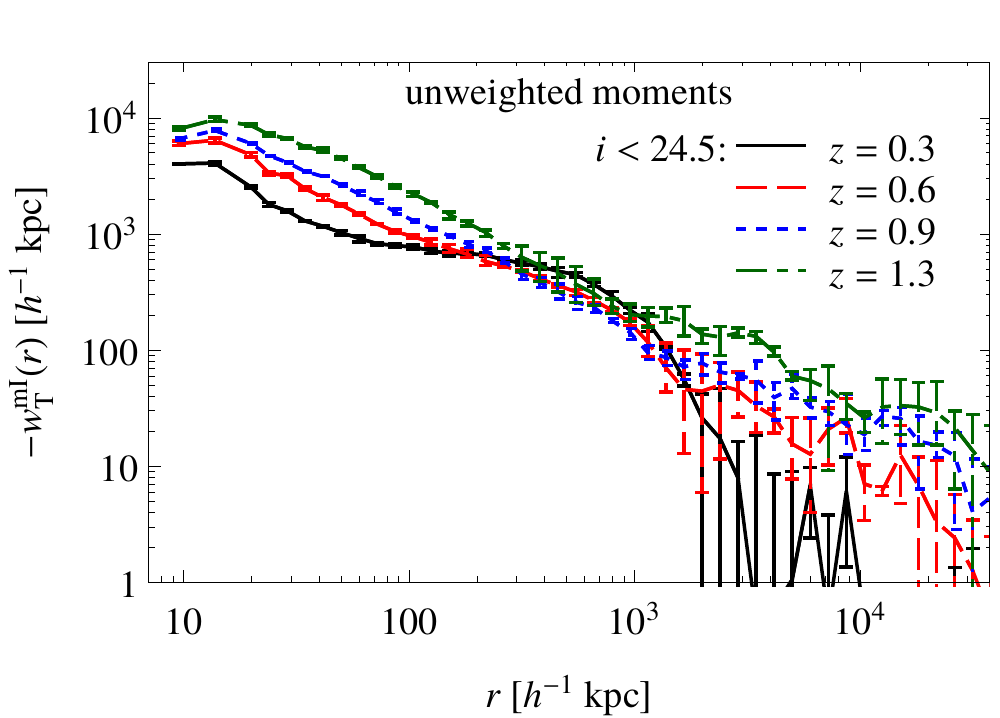}}
\centerline{\includegraphics[width=\linewidth]{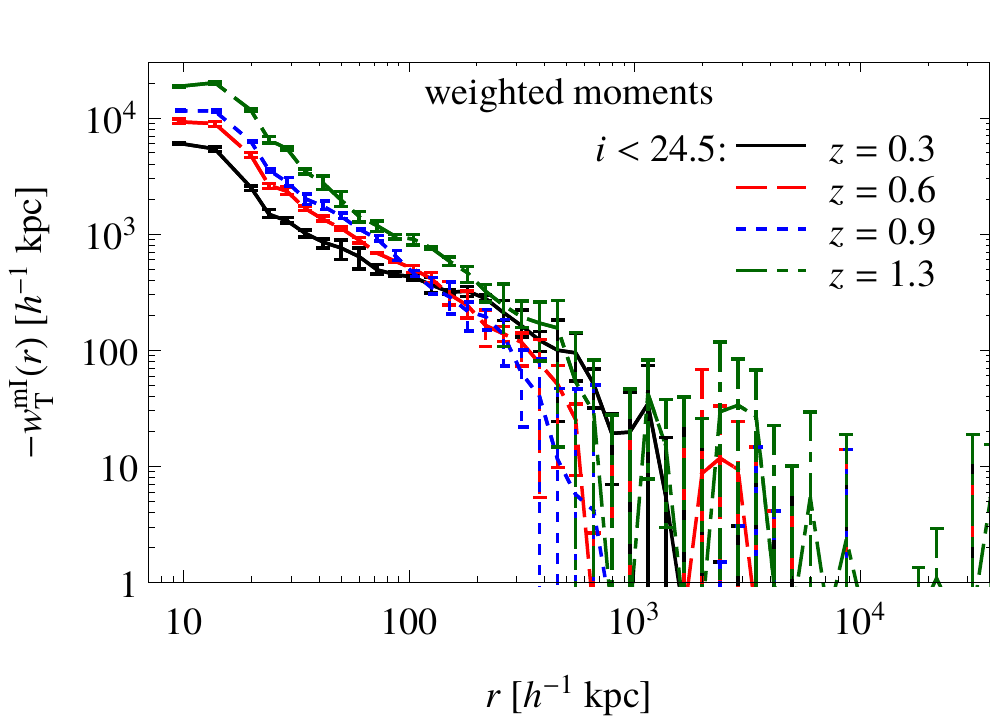}}
\caption{
\label{fig:w_t_mI_redshift_dependence}
Projected matter density-intrinsic ellipticity correlations $\pcorrtmI{}(r)$ as a function of the separation $r$ employing unweighted moments (upper panel) or weighted moments (lower panel) for galaxy samples with apparent magnitude $i<24.5$ at different redshifts.
}
\end{figure}

The redshift dependence of the projected matter density-intrinsic ellipticity correlations $\pcorrtmI{}(r)$  for a magnitude selected sample is shown in Fig.~\ref{fig:w_t_mI_redshift_dependence}. With increasing redshift, the correlation amplitude increases for small and, in particular, for large separations $r > 2h^{-1}\,\Mpc$. This effect is primarily due to the fact that the apparent magnitude limit tends to select more massive galaxies with increasing redshift.

\subsubsection{Dependence on positional type}
\label{sec:results:matter_density_ellipticity_correlations:positional_type}

\begin{figure}
\centerline{\includegraphics[width=\linewidth]{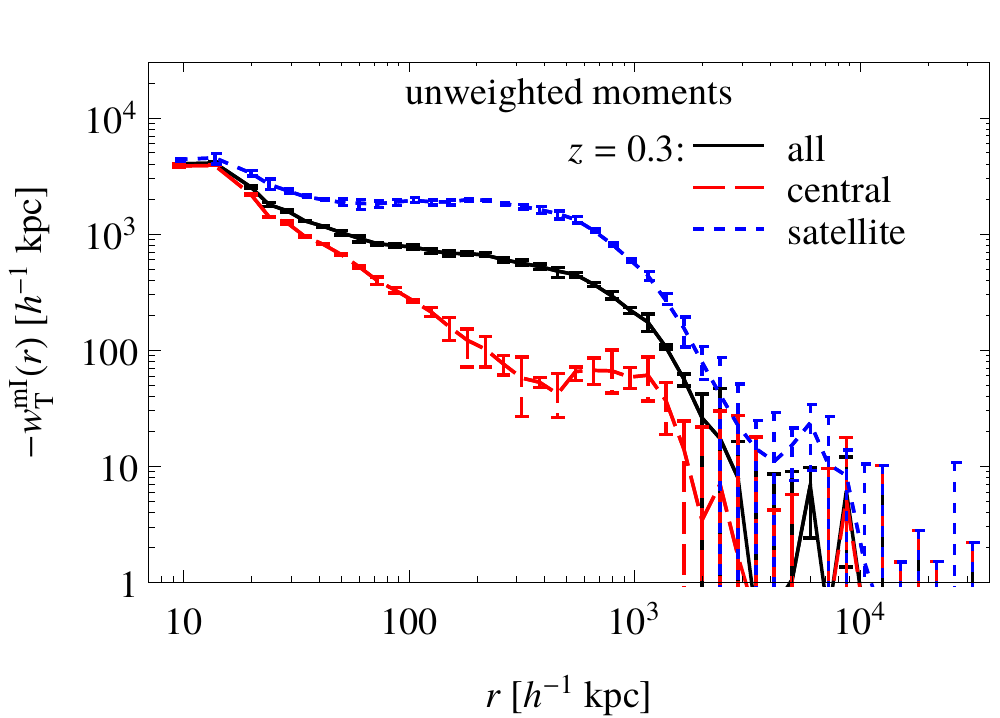}}
\centerline{\includegraphics[width=\linewidth]{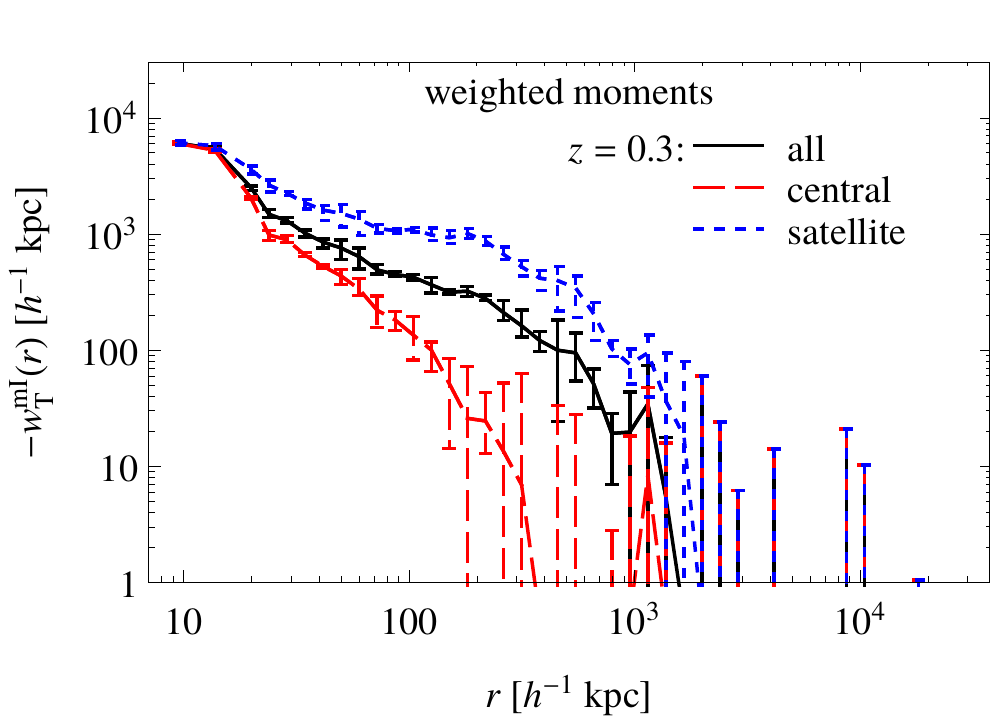}}
\caption{
\label{fig:w_t_mI_central_vs_satellite}
Projected matter density-intrinsic ellipticity correlations $\pcorrtmI{}(r)$ as a function of the separation $r$ for central and satellite galaxies with apparent magnitude $i < 24.5$ at redshift $z=0.3$, employing unweighted moments (upper panel) or weighted moments (lower panel).
}
\end{figure}

Physical processes influencing the shape of a galaxy differ depending on the position of the galaxy within its group. Moreover, the spatial distribution of central and satellite galaxies is different, and appreciable numbers of satellite galaxies are only found in more massive groups. These aspects impact the density-ellipticity correlation $\pcorrtmI{}$, resulting in possibly different correlations for central and satellite galaxies.

As shown in Fig.~\ref{fig:w_t_mI_central_vs_satellite}, the density-ellipticity correlation $\pcorrtmI{}$ for satellites is noticeably stronger on separations $20h^{-1}\,\kpc \lesssim r \lesssim 1h^{-1}\,\Mpc$. This is in accord with the expectation from Fig.~\ref{fig:w_delta_md}, which shows stronger correlations between the matter density and the galaxy density for satellites than for centrals.

\subsubsection{Dependence on photometric type}
\label{sec:results:matter_density_ellipticity_correlations:photometric_type}

\begin{figure}
\centerline{\includegraphics[width=\linewidth]{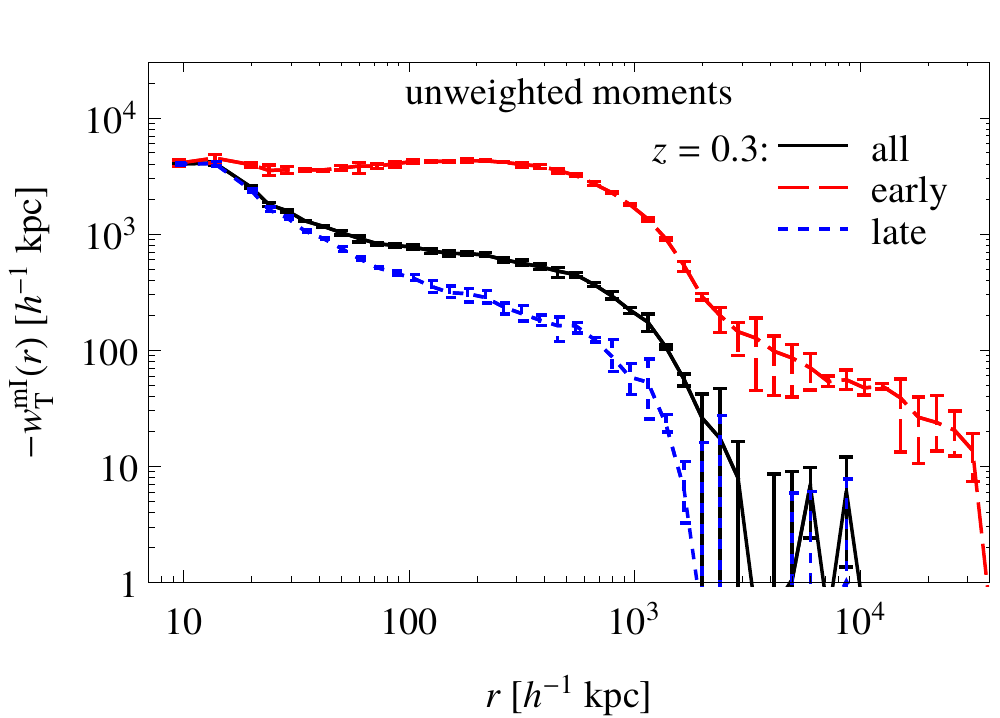}}
\centerline{\includegraphics[width=\linewidth]{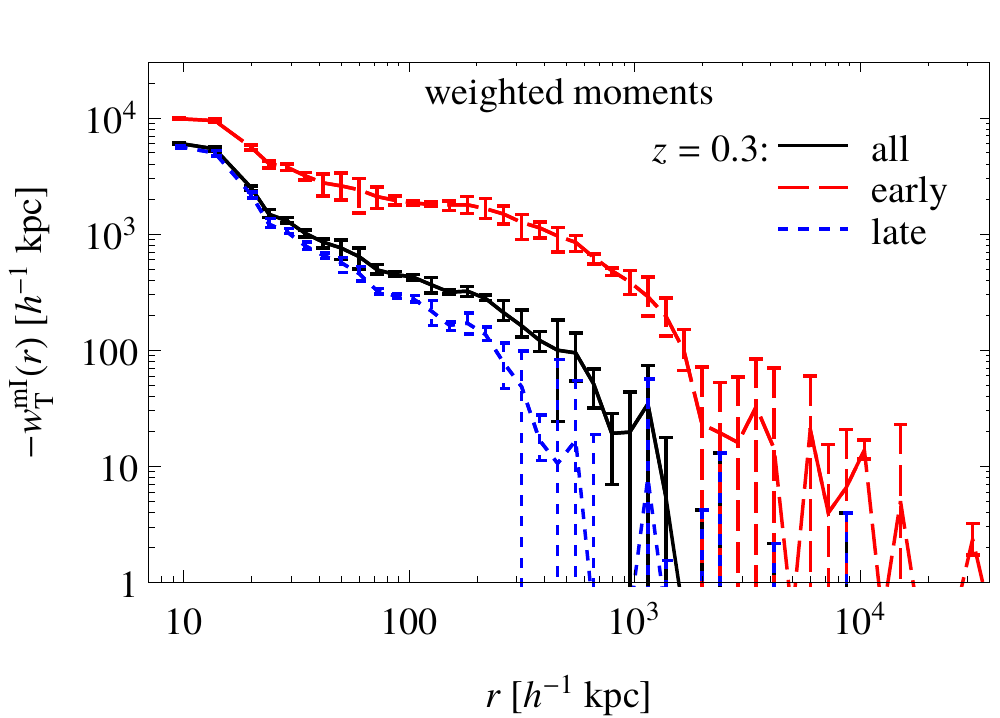}}
\caption{
\label{fig:w_t_mI_early_vs_late}
Projected matter density-intrinsic ellipticity correlations $\pcorrtmI{}(r)$ as a function of the separation $r$ for early-type (\softwarename{BPZ} type $\leq 2$) and late-type (bpz type $>2$) galaxies with apparent magnitude $i < 24.5$ at redshift $z=0.3$, employing unweighted moments (upper panel) or weighted moments (lower panel).
}
\end{figure}

The way a galaxy's shape reacts to its surrounding depend on its morphological type. The morphology, however, is difficult to determine for very distant galaxies. More easily accessible are a galaxy's colors and photometric type, which correlate with it's morphology. The photometric type correlates also with properties of the galaxy's environment such as galaxy density and matter density, which may have an impact on the strength of IA correlations, too.

Several observational studies \citep[e.g.][]{HirataEtal2007,HeymansEtal2013} indicate that red or photometric early-type galaxies show a stronger intrinsic alignment than blue or photometric late-type galaxies. This is also the case in the Illustris simulation. Fig.~\ref{fig:w_t_mI_early_vs_late} shows that galaxies classified as early type (\softwarename{BPZ} type $\leq 2$) show a much higher density-ellipticity correlation amplitude than galaxies classified as late type (\softwarename{BPZ} type $> 2$). Moreover, the alignment signal for the early-type galaxies extends to large separations $r>10h^{-1}\,\Mpc$, whereas the signal for the late-type galaxies becomes too weak to be detectable in the simulation for separations  $r>2h^{-1}\,\Mpc$.

The stronger alignment for the photometric early-type sample we measure in the Illustris simulation is also in agreement with other recent studies of IA in simulations. For example, \citet{ChisariEtal2015} found that very red (i.e. $u-r \gtrsim 1.7$) galaxies in the Horizon-AGN simulation have a much stronger mI signal than less red galaxies.
They also measured a weaker, but positive mI correlation $\pcorrtmI{}(r) > 0$ for samples of blue galaxies ($u - r \approx 1.15$), which have a larger fraction of disk-like galaxies (i.e. galaxies with a high ratio of stellar rotation to stellar velocity dispersion). Certain models of disk galaxy spin alignment indeed predict satellite galaxy spins to align toward the main halo center resulting in a positive $\pcorrtmI{}$ for these galaxies \citep[e.g.][]{JoachimiEtal2013_IA_II}.
We however do not detect a positive mI correlation for photometric late-type galaxies, in concordance with the results of \citet{TennetiMandelbaumDiMatteo2016} for disk-like galaxies in the Illustris and MassiveBlack-II simulations.

\subsection{Comparison to tidal field model}
\label{sec:results:tidal_field_model}

\begin{figure}
\centerline{\includegraphics[width=\linewidth]{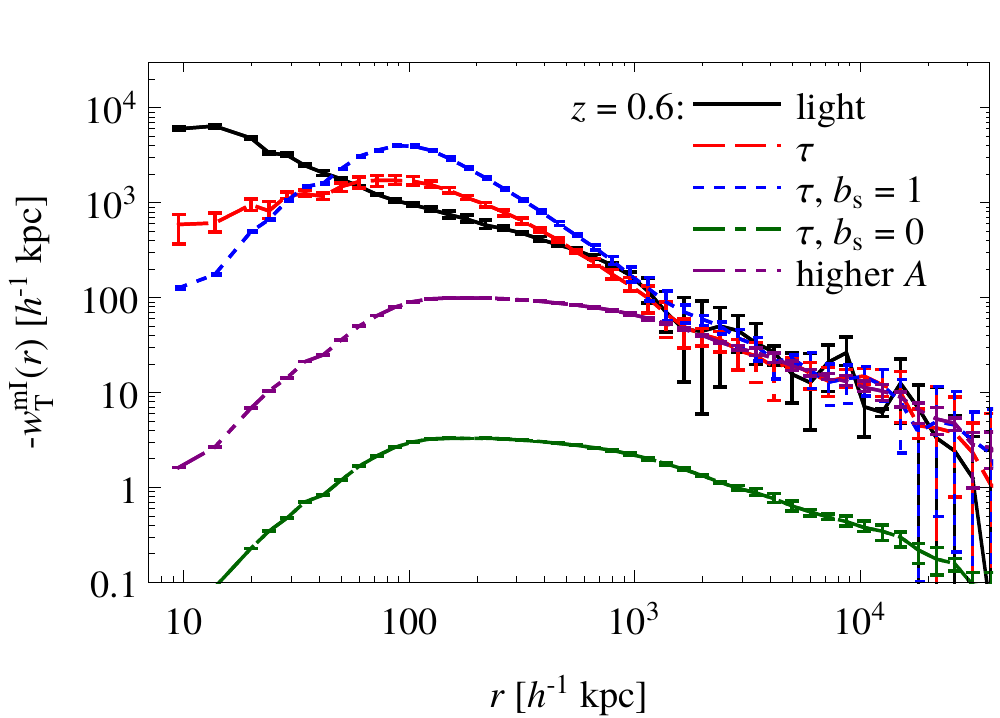}}
\caption{
\label{fig:w_t_mI_tft}
Projected matter shear-intrinsic ellipticity correlation $\pcorrtmI{}(r)$ as a function of separation $r$ for galaxies with apparent magnitude $i < 24.5$ at redshift $z=0.6$ based on ellipticities measured from the galaxy light (using unweighted moments, solid line), or ellipticities predicted by a tidal-field model of galaxy alignment (dashed line). Also shown are tidal field model predictions for a galaxy sample with linear bias $\biasgalS{}{}=1$ exactly tracing the matter density field, i.e. $\deltagalS{}{}=\deltaMatter$ (dotted line), for galaxy sample with uniform density and $\biasgalS{}{}=0$, i.e. $\deltagalS{}{}=0$, using the same amplitude $A$ as for the "$\biasgalS{}{}=1$" model (dash-dotted line), and using a higher amplitude (dash-dot-dotted line).
Note that the tidal field model-based correlations for $r \lesssim 100h^{-1}\,\kpc$ suffer from lack of resolution in the tidal field computation.
}
\end{figure}

Here, we compare results we obtain from ellipticities of the galaxy stellar light in the simulation to predictions from simple tidal field models of galaxy alignment. For this comparison, we compute the tidal field~\eqref{eq:tidal_field} and the tidal field prediction~\eqref{eq:tidal_field_ellipticity} for the intrinsic galaxy ellipticities in the simulation as described in Section~\ref{sec:methods:densities_and_tidal_fields}.

Figure~\ref{fig:w_t_mI_tft} compares the matter density-intrinsic ellipticity correlation $\pcorrtmI{}(r)$ for galaxy ellipticities measured from the galaxy light to the correlation $\pcorrtmI{}(r)$ for galaxies at the same position, but with ellipticities given by the tidal field prediction~\eqref{eq:tidal_field_ellipticity} with an amplitude parameter $A = 0.03$ and assuming the same galaxy positions.\footnote{
We consider a higher redshift $z=0.6$ here, so our magnitude-limited sample carries a robust $\pcorrtmI{}(r)$ signal on separations $r \sim 10h^{-1}\,\Mpc$, where we match the amplitude for the tidal field model.
}
The amplitude was chosen to match the correlations on scales $r\gtrsim 1h^{-1}\,\Mpc$. The tidal field prediction follows the measured correlation also at smaller scales $100h^{-1}\,\kpc \lesssim r \lesssim 1h^{-1}\,\Mpc$ well into the non-linear regime of structure formation, with deviations $\lesssim 50\%$. At even smaller scales, the tidal field prediction suffers from finite-resolution effects (the mesh used to compute the tidal field only has a resolution of $75h^{-1}\,\kpc$).

For the magnitude-limited galaxy shape sample in our comparison, the galaxy bias $\biasgalS{}{}\approx 1$. Fig.~\ref{fig:w_t_mI_tft} also shows the prediction for $\pcorrpGI{}(r)$ from the tidal field model in combination with a linear deterministic galaxy bias model with galaxy bias $\biasgalS{}{}=1$, where the galaxy density exactly follows the matter density. That prediction also roughly traces the measured correlation $\pcorrpGI{}(r)$, albeit with a noticeable excess in amplitude at scales $100h^{-1}\,\kpc \lesssim r \lesssim 400h^{-1}\,\kpc$. At separations $r<100h^{-1}\,\kpc$, our \lq{}tidal field plus linear bias $\biasgalS{}{}=1$\rq{} model prediction also suffers from lack of resolution in the computation of the tidal field.

Neglecting the galaxy density weighting in the computation of the matter density-intrinsic ellipticity correlation $\pcorrpGI{}(r)$ is equivalent to the assumption that the density of the galaxy shape sample is uniform, which in turn is equivalent to a linear deterministic galaxy bias model with galaxy bias $\biasgalS{}{}=0$. As  Fig.~\ref{fig:w_t_mI_tft} illustrates, such a model severely under-predicts the correlation amplitude. On large scales $r\gtrsim 5h^{-1}\,\Mpc$, the prediction is too small by a factor $\sim 30$, which indicates $\sigma_S^2\sim60$ in the IA model of \citet{BlazekVlahSeljak2015}. At smaller separations, the difference is orders of magnitude larger.

The discrepancy at large scales can be remedied by choosing a larger amplitude $A=0.9$, but this still leads to a severe underprediction of the IA correlations on scales $\lesssim 1 h^{-1}\,\Mpc$. Moreover, since galaxies cluster in regions of larger tidal fields in our Universe, interpreting IA observations with a linear-response tidal field model that neglects density weighting (i.e. assumes $\biasgalS{}{}=0$ and $\pcorrtmI{}\propto A \EV{\deltaMatter \tau}$ instead of  $\pcorrtmI{}\propto A \EV{\deltaMatter(1 + \deltagalS{}{})\tau}$) leads to an overestimation of the susceptibility~\eqref{eq:tidalfieldtoellipticityfactor} of galaxy ellipticities to tidal fields.

\subsection{Galaxy density-intrinsic ellipticity correlations}
\label{sec:results:galaxy_density_ellipticity_correlations}

\begin{figure}
\centerline{\includegraphics[width=\linewidth]{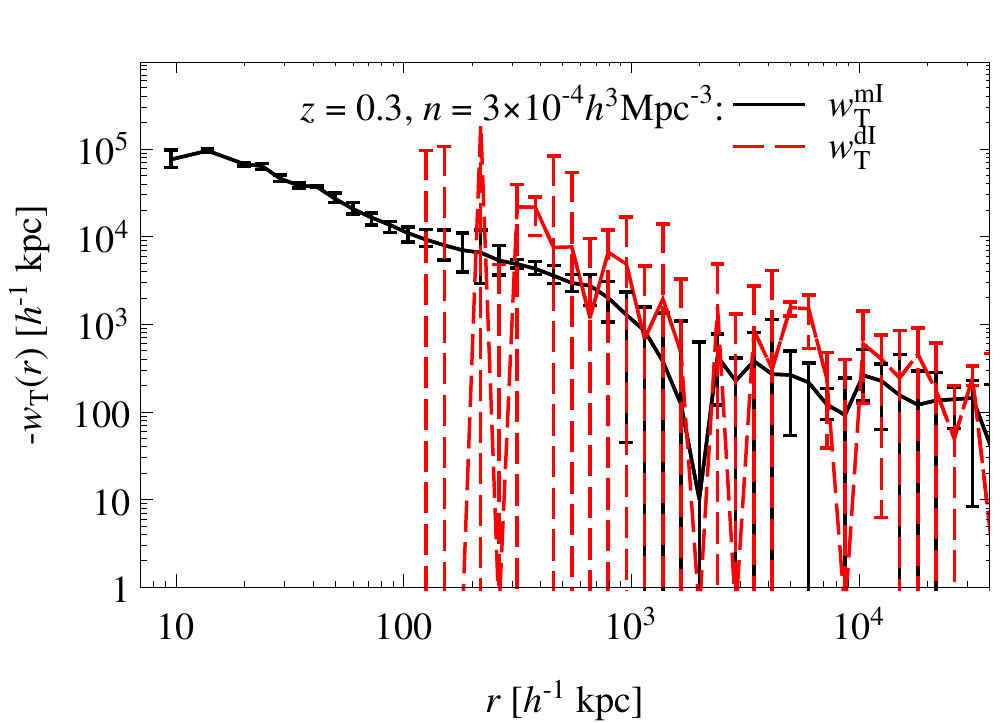}}
\centerline{\includegraphics[width=\linewidth]{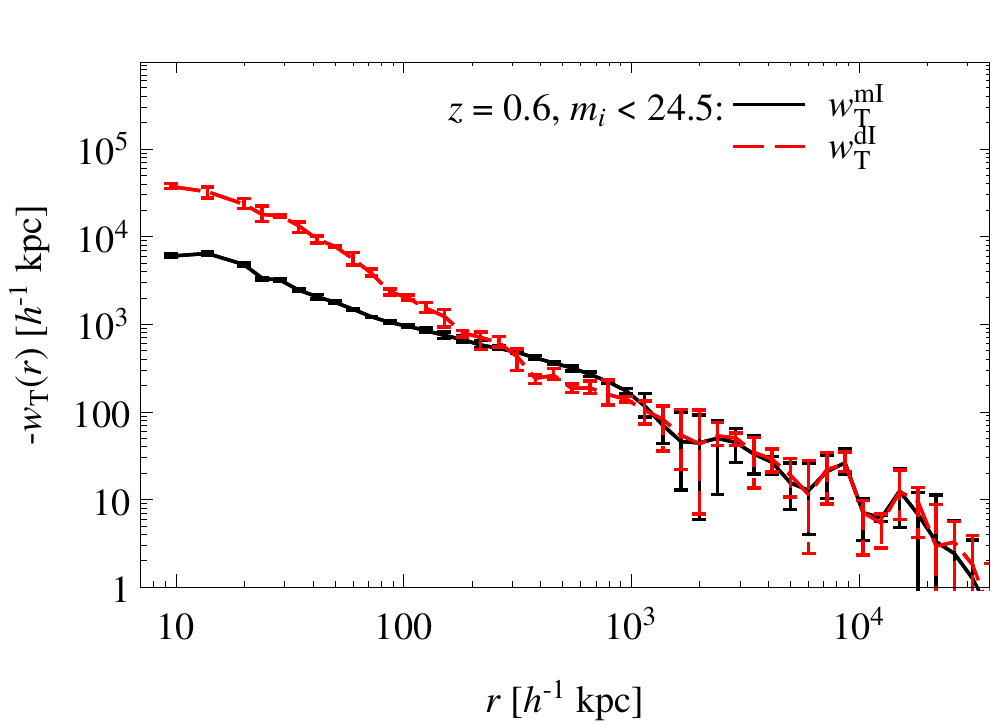}}
\caption{
\label{fig:w_t_mI_vs_w_dI}
Projected matter density-intrinsic ellipticity correlations $\pcorrtmI{}(r)$ as a function of the separation $r$ compared to the projected galaxy density-intrinsic ellipticity correlations $\pcorrtdI{}$ 
using weighted moments for stellar mass-selected galaxies (used both as density sample and shape sample) at redshift $z=0.3$ with comoving number density $n= 3\times10^{-4}h^{3}\,Mpc^{-3}$ (upper panel), and using unweighted moments for galaxies at $z=0.6$ with apparent magnitude $i < 24.5$ (lower panel).
}
\end{figure}

The total matter density on cosmological scales appears to be dominated by dark matter, whose density cannot be easily estimated. 
Galaxies can be employed as tracers of the total matter distribution. However, galaxies are usually biased and noisy tracers.
This may then also cause differences between the matter density-intrinsic ellipticity correlation $\pcorrtmI{}(r)$, which is what various IA models \citep[e.g.][]{HirataSeljak2004} yield (at least as intermediate result), and the galaxy density-intrinsic ellipticity correlation $\pcorrtdI{}(r)$, which is more directly accessible in observations \citep[e.g.][]{SinghMandelbaumMore2015}.

In the upper panel of Fig.~\ref{fig:w_t_mI_vs_w_dI}, we compare the projected matter density-intrinsic ellipticity correlation $\pcorrtmI{}(r)$ and the galaxy density-intrinsic ellipticity correlation $\pcorrtdI{}(r)$ for a stellar mass-selected sample of galaxies with a comoving number density $n= 3\times10^{-4}h^{3}\,\Mpc^{-3}$ to roughly match the SDSS-III BOSS DR11 LOWZ galaxy sample considered by \citet{SinghMandelbaumMore2015} serving both as shape and density sample.
For the Illustris simulation with a comoving volume of $(75 h^{-1}\,\Mpc)^3$, this number density corresponds to a (very small) sample of 127 galaxies with stellar mass $\Mstellar \geq 2.5\times10^{11}h^{-1}\,\Msolar$ (and roughly $\Msub \geq 10^{13}h^{-1}\,\Msolar$). The resulting dI signal (which is very noisy due to the low number of galaxies) appears roughly twice as large as the corresponding mI signal. This is consistent with the assumption that the galaxies trace the matter with a galaxy bias $\biasgalD{}{}\approx 2$, which is the value estimated from the galaxy density correlation.
A similar galaxy bias $\biasgalD{}{}=1.8$ has been measured for the SDSS LOWZ sample. However, the dI correlation of the LOWZ sample measured by \citet{SinghMandelbaumMore2015} is roughly a factor 2-4 weaker than the dI correlation we obtain for our \lq{}mock LOWZ\rq{} sample in Illustris. This discrepancy might indicate shortcomings of the simulation itself, but could also be due to our too simple modelling of the LOWZ selection or the shape measurement method.


For a more detailed comparison of $\pcorrtmI{}(r)$ and $\pcorrtdI{}(r)$, we require a galaxy sample large enough to allow us to measure IA correlations with high signal-to-noise ratio. Since we intend to use the insights gained here to interpret the relation between the dI and GI correlations in Section~\ref{sec:results:results:cosmic_shear}, we consider a magnitude-limited sample at a redshift typical of an intermediate-redshift tomographic bin in a deep weak lensing survey.
The lower panel of Fig.~\ref{fig:w_t_mI_vs_w_dI} shows a comparison of $\pcorrtmI{}(r)$ and $\pcorrtdI{}(r)$ for galaxies with $i<24.5$ at redshift $z=0.6$. 
The two correlations agree very well for projected separations $r\gtrsim 1h^{-1}\,\Mpc$. For small separations $r \lesssim 100h^{-1}\,\kpc$, the galaxy density-intrinsic ellipticity correlation $\pcorrtdI{}(r)$ is stronger by factors of a few than the matter density-intrinsic ellipticity correlation $\pcorrtmI{}(r)$. 
A similar effect has been found in other simulations \citep[e.g.][]{TennetiEtal2015,ChisariEtal2016}.
This feature cannot be explained by a large galaxy bias. The galaxy bias, as e.g. estimated by the ratio $\pcorrsymbol_{\delta}^{\mathrm{md}}(r)/\pcorrsymbol_{\delta}^{\mathrm{mm}}(r)$, is close to unity ($\biasgalD{}{} \approx 1\pm0.2$) for separations $r\gtrsim 30h^{-1}\,\kpc$. It remains to be investigated how this enhancement could be explained, e.g., by short-range interactions of physically close galaxies.\footnote{Since we measure the shapes of galaxies in isolation, there is no blending of galaxy images merely close in projection.}

\subsection{Shear-intrinsic ellipticity correlations}
\label{sec:results:shear_ellipticity_correlations}

\begin{figure}
\centerline{\includegraphics[width=\linewidth]{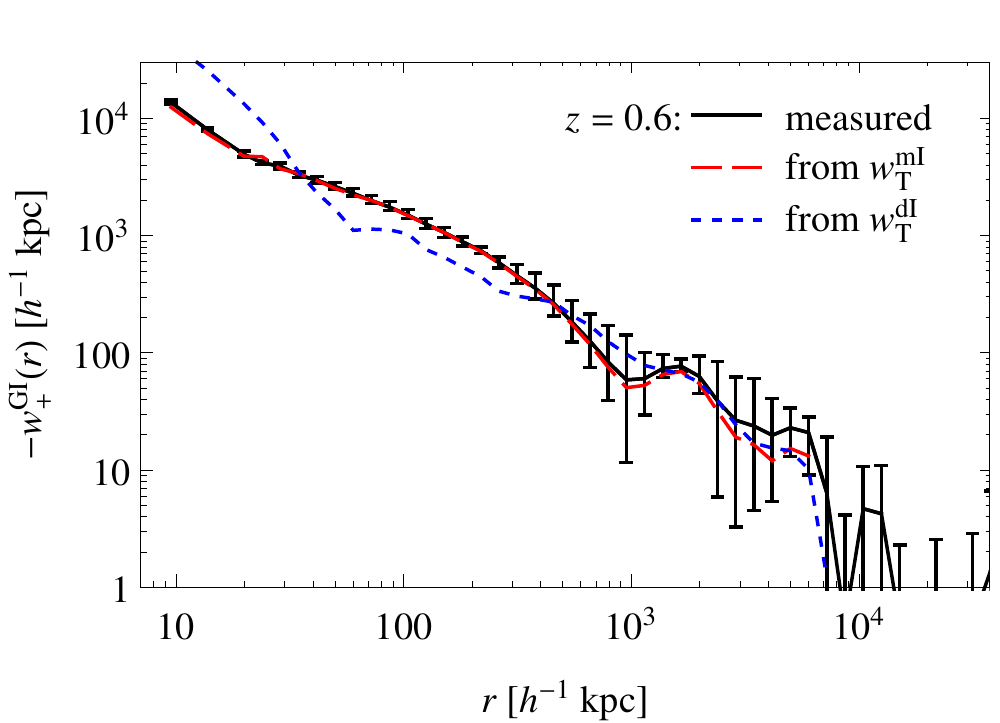}}
\caption{
\label{fig:w_p_GI_direct_vs_prediction}
Projected matter shear-intrinsic ellipticity correlation $\pcorrpGI{}(r)$ as a function of separation $r$ for galaxies with apparent magnitude $i < 24.5$ at redshift $z=0.6$, directly measured in the simulation (solid line), and as inferred from the measured matter density-intrinsic ellipticity correlation (dashed line), or the galaxy density-intrinsic ellipticity correlation $\pcorrtdI{}$ (dotted line).
}
\end{figure}

The gravitational shear-intrinsic ellipticity correlations $\pcorrpmGI{}$ determine the GI contribution to the observed correlation functions. Thus it is very desirable to obtain predictions for these, either directly from theory or simulations, or from additional measurements in observations.

The correlation $\pcorrpGI{}$ we measure in the Illustris simulation for a magnitude-limited galaxy sample at redshift $z=0.6$ is shown in Fig.~\ref{fig:w_p_GI_direct_vs_prediction}. Also shown are values for that correlation inferred from applying the integral transformation~\eqref{eq:relation_GI_mI} to the measured density-intrinsic ellipticity correlations $\pcorrtmI{}$ and $\pcorrtdI{}$ (using the galaxy shape sample also as galaxy density sample). The correlation inferred from the matter density-intrinsic ellipticity correlation $\pcorrtmI{}$ agrees well with the directly measured gravitational shear-intrinsic ellipticity correlation $\pcorrpGI{}$, in particular for separations $r \lesssim 2 h^{-1}\,\Mpc$, where the difference is $\lesssim5\%$. The correlation inferred from the galaxy density-intrinsic ellipticity correlation $\pcorrpGI{}$ also roughly follows the directly measured $\pcorrpGI{}$, but the difference is generally larger, often $\sim 50\%$.

\subsection{Intrinsic ellipticity correlations}
\label{sec:results:ellipticity_correlations}

\begin{figure}
\centerline{\includegraphics[width=\linewidth]{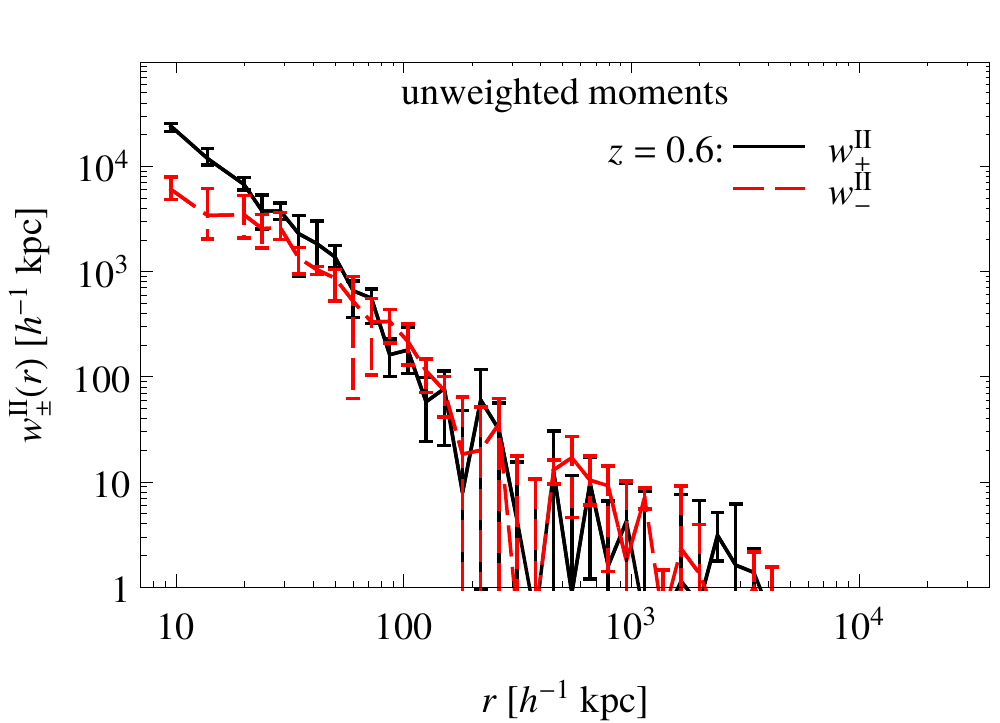}}
\centerline{\includegraphics[width=\linewidth]{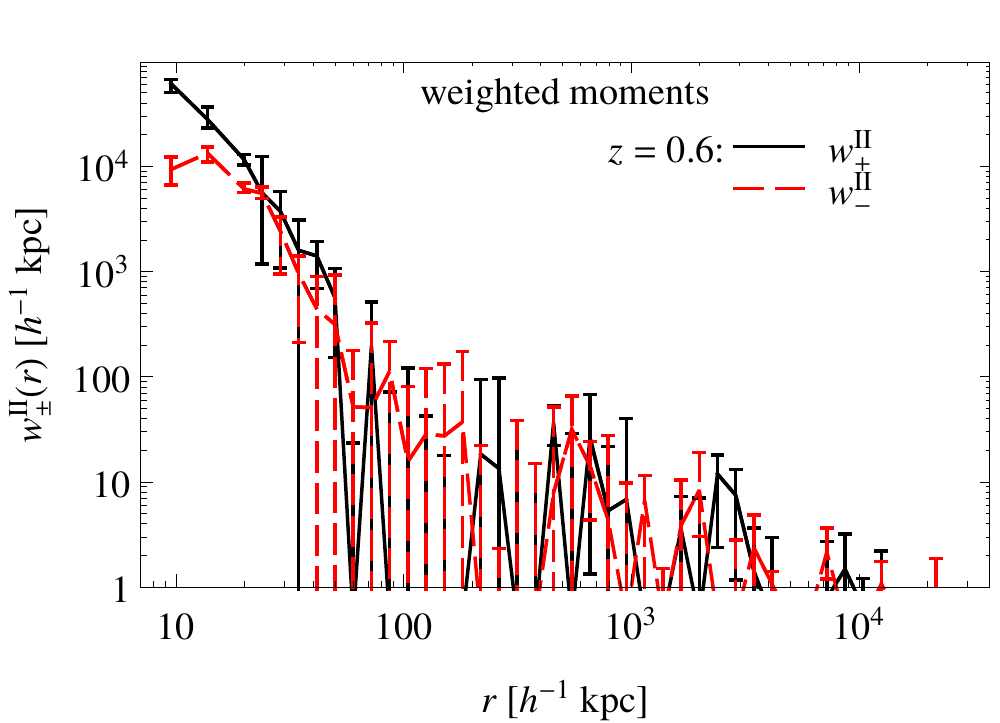}}
\caption{
\label{fig:w_pm_II}
Projected intrinsic ellipticity correlations $\pcorrpmII{}(r)$ as a function of the separation $r$ for galaxies with apparent magnitude $i < 24.5$ at redshift $z=0.6$ for ellipticities computed from unweighted moments (upper panel) and weighed moments (lower panel).
}
\end{figure}

Intrinsic ellipticity correlations contribute to the observed ellipticity correlation when the galaxy shape samples overlap in redshift. Figure~\ref{fig:w_pm_II} shows the correlations we measure in the simulation for galaxies with $i<24.5$ at $z=0.6$. The measurements are generally noisy, and we obtain a statistically significant measurement only for very small separations $r \lesssim 100h^{-1}\,\kpc$.

\subsection{Cosmic shear}
\label{sec:results:results:cosmic_shear}

\begin{figure}
\centerline{\includegraphics[width=\linewidth]{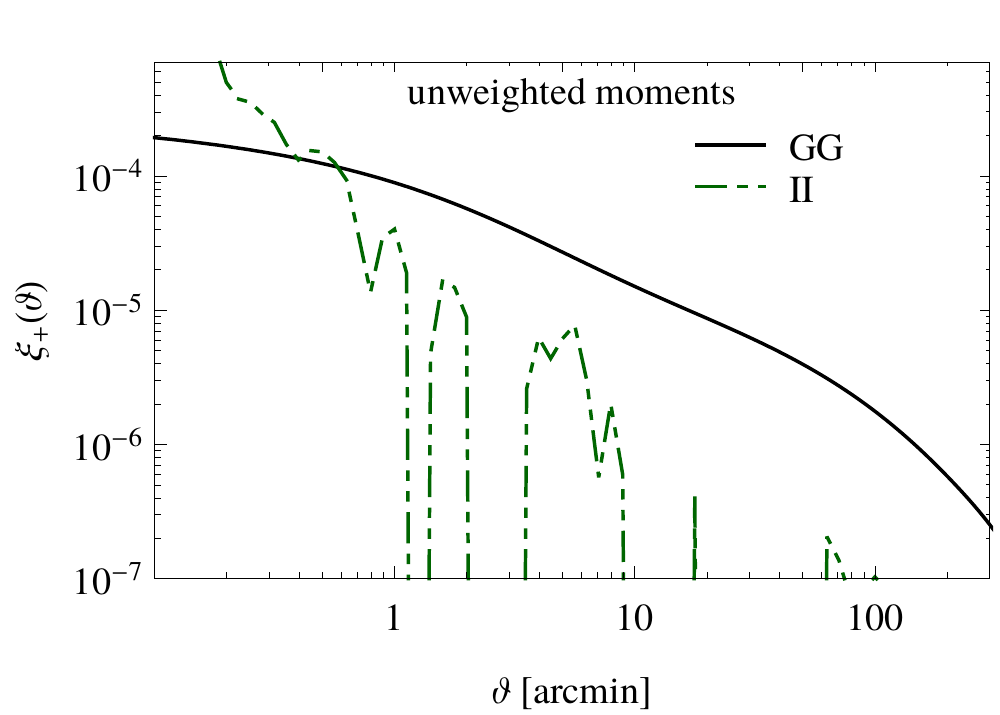}}
\caption{
\label{fig:cosmic_shear_GG_vs_II}
II contribution to cosmic shear: Comparison of the GG (solid line) and II (dash-dotted line) contribution to the observed ellipticity correlation $\xi_+(\vartheta)$ as a function of angular separation $\vartheta$ for galaxies with a redshift distribution  similar to that of tomographic bin 3 (median redshift $z_\text{median} \approx 0.65$) of \citet{HeymansEtal2013}.
}
\end{figure}

The observed ellipticity auto-correlation of a galaxy sample contains contributions from intrinsic ellipticity correlations (II) besides contributions from gravitational shear (GG), and possibly contributions from gravitational shear-intrinsic ellipticity correlations (GI). As an example, we consider the auto-correlation of a galaxy sample with apparent magnitude $i\leq24.5$ and a redshift distribution given by tomographic bin 3 (with a median redshift $z_\text{median} \approx 0.65$) used by \citet{HeymansEtal2013} in their tomographic analysis of CFHTLenS. The expected GG and II contributions to the cosmic shear correlation function estimator $\estacorrpm{\obsell{}{}}{\obsell{}{}}^{(3|3)}(\vartheta)$ are compared in Fig.~\ref{fig:cosmic_shear_GG_vs_II}. The measured $\pcorrpmII{}{}$ and thus also the predicted II contribution are very noisy (for simplicity, we refrain from estimating errors here). There is indication that the II signal surpasses the GG signal on arc-second scales. On scales $\vartheta \geq 1\,\arcmint$, the II contribution is $\lesssim 10 \%$ of the GG signal.

\begin{figure*}
\centerline{\includegraphics[width=0.47\linewidth]{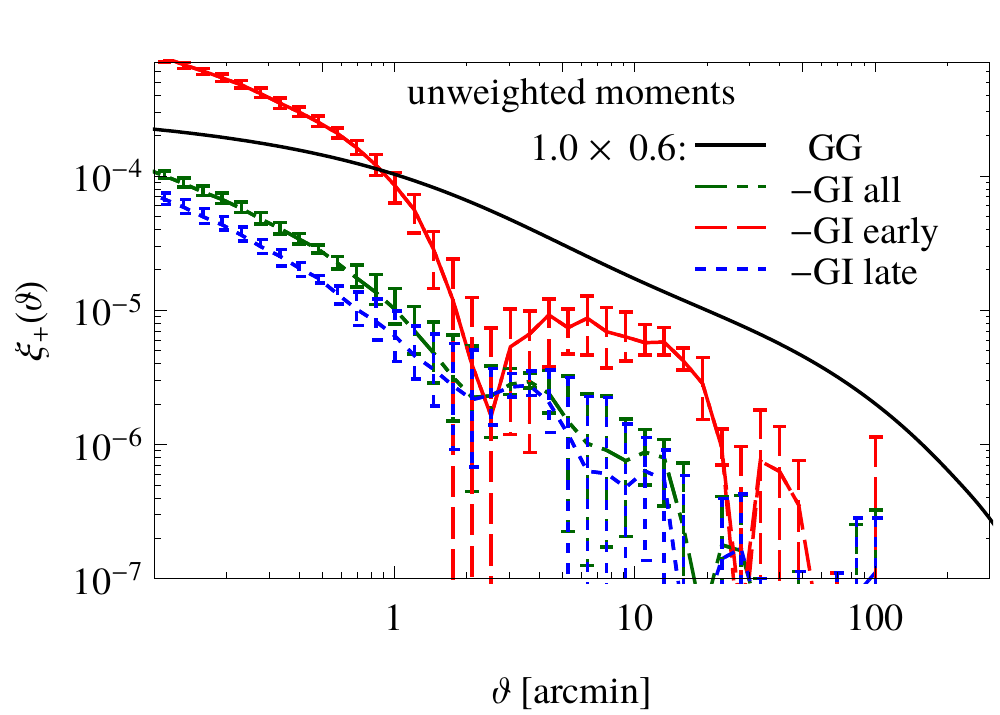} \hfill \includegraphics[width=0.47\linewidth]{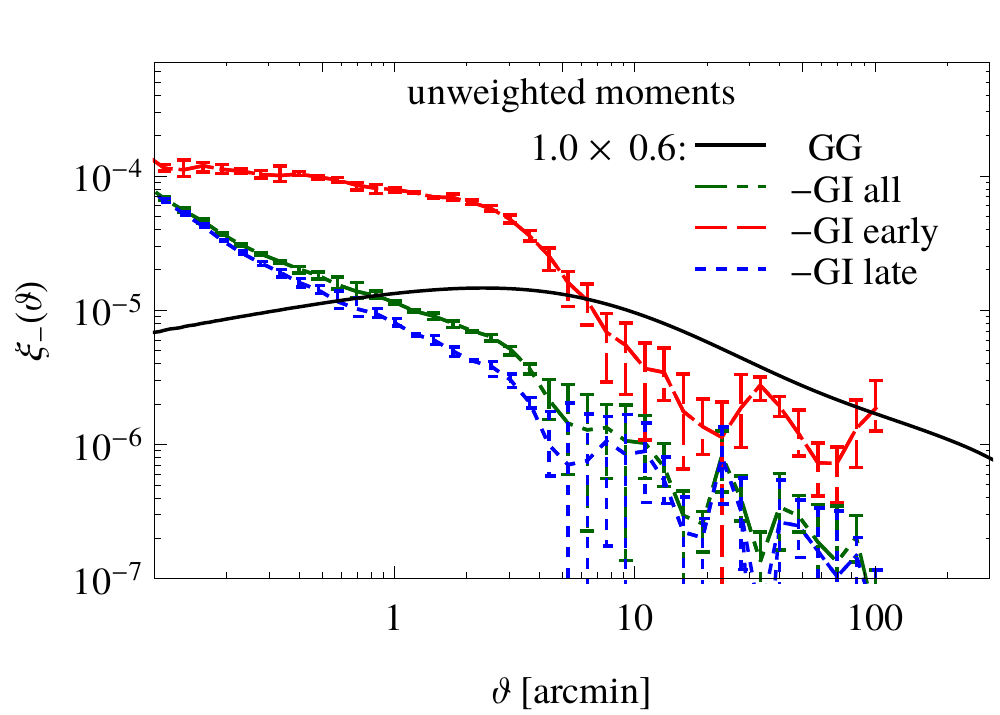}}
\centerline{\includegraphics[width=0.47\linewidth]{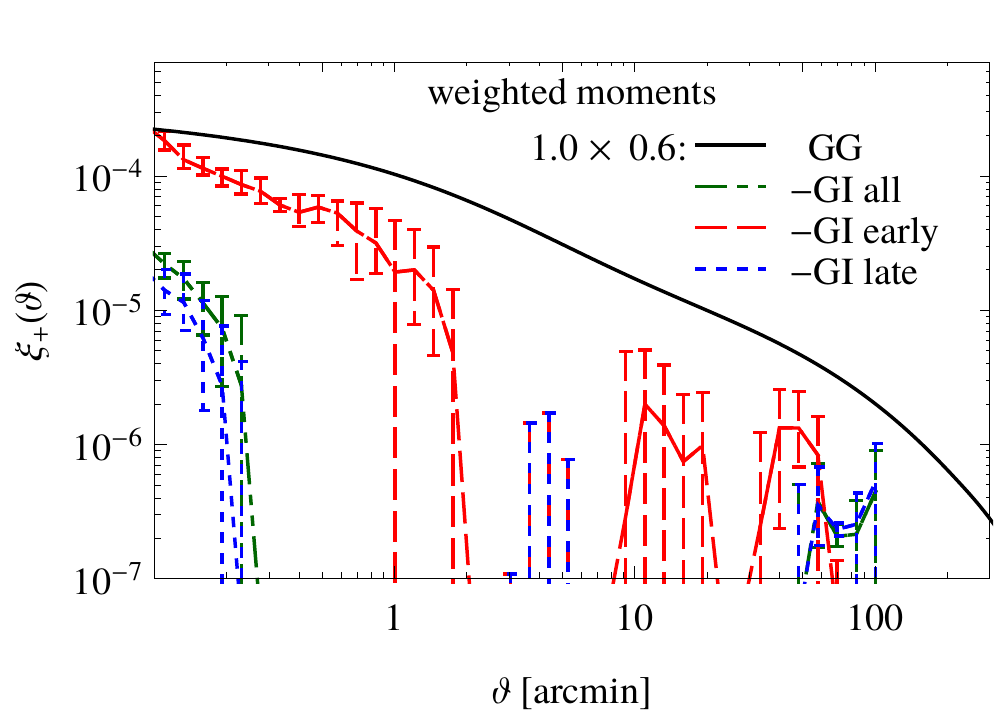}   \hfill \includegraphics[width=0.47\linewidth]{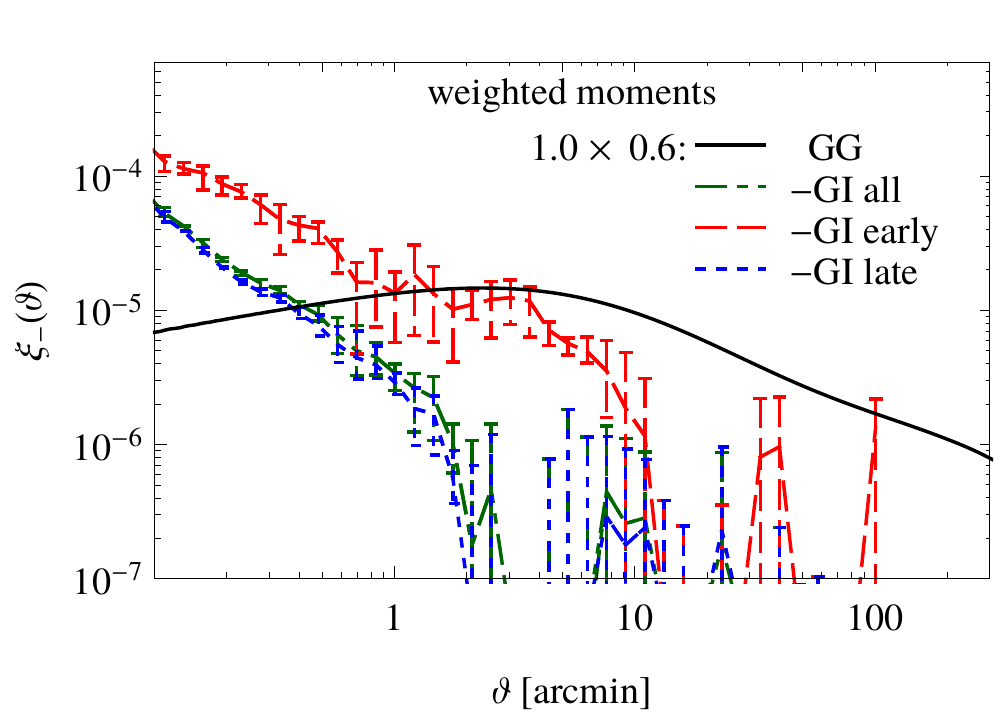}  }
\caption{
\label{fig:cosmic_shear_GG_vs_GI}
GI contribution to cosmic shear: Comparison of the GG (solid lines) and GI contribution to the observed ellipticity correlation $\xi_+(\vartheta)$ (left panels) and $\xi_-(\vartheta)$ (right panels) as a function of angular separation $\vartheta$ for galaxies at redshift $z=0.6$ and $z=1$. Shown are the GI contributions for all galaxies (dash-dotted lines) with apparent magnitude $i < 24.5$, early-type galaxies (dashed lines), and late-type galaxies (dotted lines), either employing unweighted moments (upper panels) or weighted moments (lower panels).
}
\end{figure*}

Contributions from intrinsic ellipticity correlations (II) can be avoided by cross-correlating observed ellipticities of galaxy samples that do not overlap in redshift. The cross-correlations then only contain contributions from gravitational shear (GG) and from gravitational shear-intrinsic ellipticity (GI/IG) correlations. The expected GG and GI contributions to the observed ellipticity correlations $\estacorrpm{\obsell{}{}}{\obsell{}{}}^{(1|2)}(\vartheta)$ are compared in Fig.~\ref{fig:cosmic_shear_GG_vs_GI} for the cross correlation of galaxy samples at two different redshifts $z_1=1.0$ and $z_2=0.6$. 
For a sample of galaxies with apparent $i$-band magnitude limit $i<24.5$ and shapes measured from unweighted moments, the amplitude of the GI contribution to $\xi_+(\vartheta)$ is $10 - 50 \%$ of the GG signal for separations $\vartheta < 1\,\arcmint$, and drops to $\sim5\%$ for  $\vartheta > 1 \,\arcmint$. The GI amplitude for the subsample of early-type galaxies appears much higher, exceeding the GG signal on small scales and reaching $\sim20\%$ of the GG amplitude for $\vartheta > 1 \,\arcmint$. Just considering late-type galaxies decreases the GI amplitude slightly compared to the GI signal for the galaxy sample including all types. However, since the fraction of early-type galaxies is small, the effect of removing these galaxies is not very large.

On small scales, the relative impact of the GI contribution is larger for $\xi_-$ than for $\xi_+$. As Fig.~\ref{fig:cosmic_shear_GG_vs_GI} shows, the GI signal estimated from the simulation for galaxies with $i<24.5$ exceeds $5\%$ of the GG signal for separations $\vartheta < 10\,\arcmint$, and may be larger than the GG signal on scales $\vartheta \lesssim 10\,\arcsect$.

For shapes measured from weighted moments, there is a noticeable GI contribution for separations $\vartheta \lesssim 1 \,\arcmint$. However, in contrast to shapes based on unweighted moments, we do not measure a significant GI contribution (i.e. $<1\%$ of the GG signal) for larger scales except for early-type galaxies.

\begin{figure}
\centerline{\includegraphics[width=\linewidth]{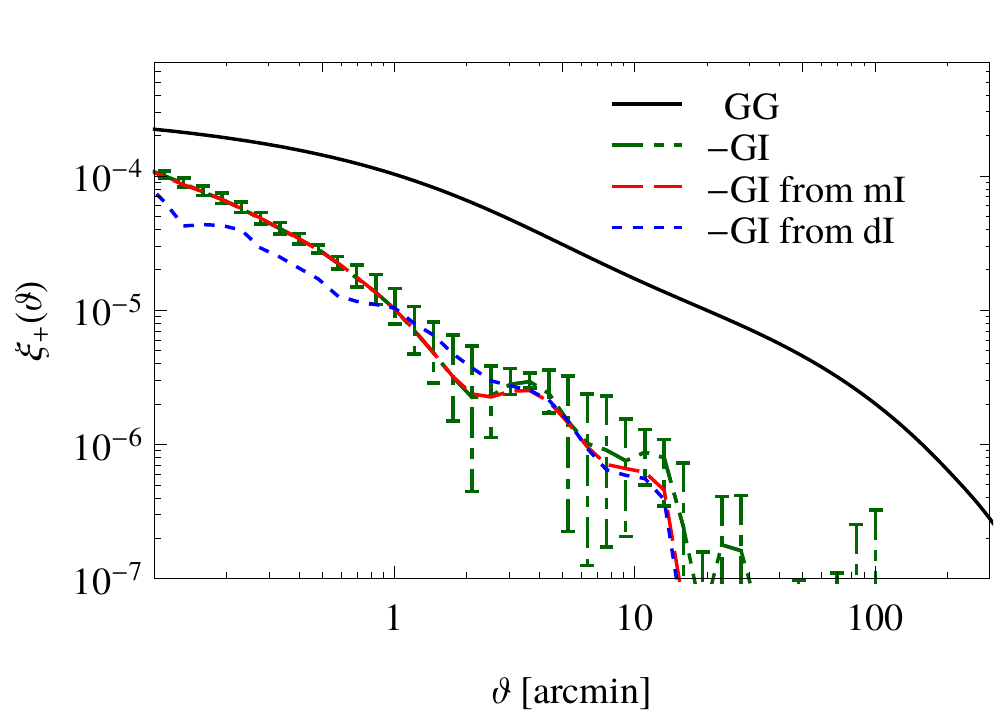}}
\caption{
\label{fig:cosmic_shear_GI_prediction}
Comparison of directly measured GI contribution to the observed ellipticity correlation $\xi_+(\vartheta)$ for galaxies at redshift $z=0.6$ and $z=1$ (dash-dotted line) and the contribution predicted from either the matter density-intrinsic ellipticity correlation $\pcorrtmI{}{}$ (dashed line) or the  galaxy density-intrinsic ellipticity correlation $\pcorrtdI{}{}$ (dotted line). The GG contribution (solid line) is also shown for comparison.
}
\end{figure}

As indicated in Section~\ref{sec:results:shear_ellipticity_correlations}, one may try to infer the GI contribution to the observed ellipticity correlations from measurements of the galaxy density-intrinsic ellipticity correlations. As an example, we consider the GI contribution to the observed ellipticity cross-correlation $\xi_+(\vartheta)$ between a higher-redshift ($z=1$) and a lower-redshift ($z=0.6$) galaxy sample. The GI contribution is estimated from the galaxy density-ellipticity (dI) correlation $\pcorrtdI{}{}$ of the lower-redshift galaxy sample (which has a galaxy-matter bias $b \approx 1$).
The results are shown in Fig.~\ref{fig:cosmic_shear_GI_prediction}. For separations $\vartheta \sim 10\,\arcmint$, the GI contribution estimated from $\pcorrtdI{}{}$ reproduces the \lq{}true\rq{} GI signal (i.e. directly measured from the simulation) within $\approx 30\%$, with differences at least in part due to our large measurement uncertainties because of the small simulation volume. On sub-arcminute scales, the differences between the directly measured GI contribution and the one inferred from the dI signal are about $50\%$, which is mainly due to the systematic differences between the dI and gI signal.
The discrepancies between predicted and actual GI contribution may be smaller for a density sample that very closely traces the matter also on small scales. If for example, the matter density-intrinsic ellipticity (mI) can be used to estimate the GI signal, differences between predicted and actual GI signal vanish on small scales, as shown in Fig.~\ref{fig:cosmic_shear_GI_prediction}.

\section{Summary and discussion}
\label{sec:summary}

In this work we presented our study of intrinsic alignments within the Illustris-1 cosmic structure formation simulation \citep[][]{VogelsbergerEtal2014_Illustris}.
We considered the scale and redshift dependence of various correlations involving the ellipticities of the projected light distribution of galaxies selected according to their properties.
The correlations considered include the matter density-intrinsic ellipticity (mI), galaxy density-intrinsic ellipticity (dI), gravitational shear-intrinsic ellipticity (GI), and intrinsic ellipticity-intrinsic ellipticity (II) correlations. The galaxy properties considered include stellar mass, apparent magnitude, positional type, and photometric type.

In accordance with other works \citep[e.g.][]{TennetiMandelbaumDiMatteo2016}, we find that the amplitude of the intrinsic alignment correlations is larger for galaxy samples with larger stellar mass. 
Similarly, intrinsic alignment correlations are stronger for more luminous galaxies. As a result, galaxy samples at fixed redshift with a bright apparent magnitude limit have stronger intrinsic alignment correlations than samples with a fainter magnitude limit. Thus going deeper in cosmic shear surveys may reduce the intrinsic alignment contribution to the observed ellipticity correlations.

The IA correlations in the simulation also depend on the photometric type as classified by photometric redshift codes. Early-type galaxy samples have much stronger correlations than late-type galaxies at the same redshift and magnitude limit, which is in accordance with observations \citep[e.g.][]{HeymansEtal2013}. Thus removing early-type galaxies from a galaxy sample reduces the IA correlations.

We also examined the impact of the ellipticity definition and radial weighting in the shape measurement method. 
Compared to a shape estimation using flat weights, a method down-weighting the outer parts of the galaxy images results in much lower intrinsic alignment correlations on intermediate and large scales. This is in agreement with recent findings of \citet[][]{SinghMandelbaum2016} in observations. This implies that the intrinsic alignment contribution to the observed ellipticity correlations in cosmic shear surveys may strongly depend on the details of the employed shape estimator. The aim to keep the IA contribution small might thus favor shape measurement methods that down-weight the outer parts of galaxy images.

Furthermore, we compared the correlations based on the measured intrinsic shapes of galaxies to predictions from a simple tidal field model of intrinsic alignment. For that model, we assumed that the galaxy image shapes are proportional to the tidal field at the galaxy positions. This simple model reproduces the IA correlations well on scales $\gtrsim 100h^{-1}\,\kpc$, provided that the galaxy density weighting (either as given by the actual galaxy distribution or by the matter density and a linear bias model) has been taken into account. If the galaxy density weighting is neglected, the tidal field model fails badly.

The matter density-intrinsic ellipticity (mI) and the galaxy density-intrinsic ellipticity (dI) correlations we measure in the simulation agree well on scales $>1h^{-1}\,\Mpc$ for a suitably chosen galaxy sample with galaxy bias $b\approx1$. On smaller scales, however, the mI and dI correlations may differ substantially, even if the galaxy bias is close to unity on these scales. This may hamper the use of the dI correlations as a proxy for the mI correlations on small scales.

Our results indicate that the GI contribution to the observed ellipticity correlation in a tomographic cosmic shear survey can be inferred from the corresponding mI correlation without the need for a parametric model if the mI signal can be measured accurately enough. If only measurements of the dI correlation are available (as in most actual observations), the GI contribution can still be predicted well on larger scales.

The GI and II contributions to the observed ellipticity correlations in a tomographic cosmic shear survey depend on survey parameters such as the survey depth and the redshift distributions of the tomographic redshift bins. As mentioned above, the choice of shape measurement method may also be significant factor. For example, the estimated GI contribution to the observed ellipticity correlation between galaxy samples with apparent magnitude $i<24.5$ at redshift $z=0.6$ and $z=1$ is about $5\%$ for angular separations $\vartheta > 1 \,\arcmint$ if flat moments are used. In contrast, the simulation shows no significant GI contribution on these scales if radially weighted moments are used.

The underlying Illustris simulation exhibits several shortcomings that indicate the need for an improved astrophysical description. For example, the fraction of photometric early-type galaxies in the simulation is lower than we observe in our Universe. This could affect the estimation of the relative importance of early-type galaxies in the IA correlations of magnitude-limited surveys. Furthermore, lower-mass galaxies appear systematically larger in the simulation than observed. This likely affects their susceptibility to environmental influences on their shapes.

A simulation with higher mass and spatial resolution will be required to reliably measure galaxy shapes and intrinsic alignments for deeper surveys. A larger simulation volume would reduce the statistical errors on the still very noisy measurements of the IA correlations. A larger box size would also reduce the suppression of correlations on scales $r\gtrsim 10h^{-1}\,\Mpc$ due to the lack of large-scale modes in the simulation.

\section*{Acknowledgments}
We thank Elisa Chisari, Benjamin Joachimi, and Rachel Mandelbaum for useful discussions.
We also thank the reviewer for helpful comments.
SH acknowledges support by the DFG cluster of excellence \lq{}Origin and Structure of the Universe\rq{} (\href{http://www.universe-cluster.de}{\texttt{www.universe-cluster.de}}).
DX acknowledges support by the Alexander von Humboldt Foundation.
PS and VS acknowledge support by the Transregional Collaborative Research Center TR33 \lq{}The Dark Universe\rq{} of the German Deutsche Forschungsgemeinschaft, DFG.
VS also acknowledges support by the European Research Council through ERC-StG grant EXAGAL-308037. 
LH acknowledges support from NASA grant NNX12AC67G and NSF grant AST-1312095.
SH thanks the Max Planck Institute for Astrophysics where part of the work has been conducted.
DX and VS would like to thank the Klaus Tschira Foundation.

\bibliographystyle{mnras}
\bibliography{\bibpath}

\onecolumn

\appendix

\section{Correlations of projected fields}
\label{app:projected_correlations}

In the computation of expectation values of the estimators involving observed galaxy ellipticities, there arise several terms that are conveniently interpreted as correlations of projected fields.
Here, we briefly review how such correlations of projected fields can be related to projected correlation functions.

Consider two statistically homogeneous and isotropic fields $\phi{(i)}(\vx, \chi, z)$, $i=1,2$ that are function of transverse comoving position $\vx$, l.o.s. comoving distance $\chi$, and redshift $z$. 
Their general two-point correlation reads:
\begin{equation}
  \EV{\phi^{(1)}(\vx, \chi, z) \phi^{(2)}(\vx', \chi', z')} = \ccorr{\phi^{(1)}}{\phi^{(2)}} \bigl(\lvert \fK(\chi')\vtheta' -  \fK(\chi)\vtheta \rvert, \chi' - \chi, z(\chi), z(\chi') \bigr).
\end{equation} 
Define projected fields
\begin{equation}
  \pi^{(i)}(\vtheta) = \int\idiff[]{\chi}\, q^{(i)}(\chi) \, \phi^{(i)}\bigl(\fK(\chi)\vtheta, \chi, z(\chi) \bigr)
\end{equation}
by radial projections of the fields $\phi^{(i)}$ along the lightcone with l.o.s. projection weights $q^{(i)}(\chi)$. The two-point correlation of the projected fields can then be expressed as an integral over the correlation $\ccorr{\phi^{(1)}}{\phi^{(2)}}$:
\begin{equation}
\label{eq:app:correlation_of_projected_fields}
 \EV{\pi^{(1)}(\vtheta) \pi^{(2)}(\vtheta')} 
 = \int\idiff[]{\chi}\, q^{(1)}(\chi) \int\idiff[]{\chi'}\, q^{(2)}(\chi')\, \ccorr{\phi^{(1)}}{\phi^{(2)}} \bigl(\lvert \fK(\chi')\vtheta' -  \fK(\chi)\vtheta \rvert, \chi' - \chi, z(\chi), z(\chi') \bigr).
\end{equation} 
The following approximations all assume that $\ccorr{\phi^{(1)}}{\phi^{(2)}}$ falls off fast enough with increasing l.o.s. separation such that substantial contributions to the integral arise only for $\chi \approx \chi'$ The approximations then replace $\chi'$ by $\chi$  in the integrand except for the second argument of $\ccorr{\phi^{(1)}}{\phi^{(2)}}$. 

Let  $\Pi_{\mathrm{c}}$ denote the effective l.o.s. range of the correlation such that 
\begin{equation}
	\ccorr{\phi^{(1)}}{\phi^{(2)}} \bigl(r, \chi' - \chi, z(\chi), z(\chi') \bigr) \approx 0 \quad\text{for}\quad \lvert \chi' - \chi \rvert \geq \Pi_{\mathrm{c}},
\end{equation}
for the purpose of evaluating the integral in Eq.~\eqref{eq:app:correlation_of_projected_fields}.
The first step assumes slowly varying weights, i.e. $q^{(2)}(\chi') \approx q^{(2)}(\chi)$ for $\lvert \chi' - \chi \rvert < \Pi_{\mathrm{c}}$:
\begin{equation}
 \EV{\pi^{(1)}(\vtheta) \pi^{(2)}(\vtheta')} 
 \approx \int\idiff[]{\chi}\, q^{(1)}(\chi)\, q^{(2)}(\chi) \int\idiff[]{\chi'}\, \ccorr{\phi^{(1)}}{\phi^{(2)}} \bigl(\lvert \fK(\chi')\vtheta' -  \fK(\chi)\vtheta \rvert, \chi' - \chi, z(\chi), z(\chi') \bigr).
\end{equation}
This approximation constitutes the core of a number of approximations frequently called the \lq{}Limber approximation\rq{}. It turns the correlation of projected fields into an integral over a product of weights $q^{(1)}(\chi) q^{(2)}(\chi)$ and a projection
\begin{equation}
	 \int\idiff[]{\chi'}\, \ccorr{\phi^{(1)}}{\phi^{(2)}} \bigl(\lvert \fK(\chi')\vtheta' -  \fK(\chi)\vtheta \rvert, \chi' - \chi, z(\chi), z(\chi') \bigr)
\end{equation}
of the correlation function $\ccorr{\phi^{(1)}}{\phi^{(2)}}$.

The following steps are frequently employed to simplify the computation of the projected correlations.
The second step assumes slow time evolution, i.e. $\ccorr{\phi^{(1)}}{\phi^{(2)}} \bigl(r, \chi' - \chi, z(\chi), z(\chi') \bigr) \approx \ccorr{\phi^{(1)}}{\phi^{(2)}} \bigl(r, \chi' - \chi, z(\chi), z(\chi) \bigr)$ for $\lvert \chi' - \chi \rvert < \Pi_{\mathrm{c}}$, and replaces $z(\chi')$ by $z(\chi)$ in the last argument of $\ccorr{\phi^{(1)}}{\phi^{(2)}}$:
\begin{equation}
 \EV{\pi^{(1)}(\vtheta) \pi^{(2)}(\vtheta')} 
 \approx \int\idiff[]{\chi}\, q^{(1)}(\chi) \, q^{(2)}(\chi) \int\idiff[]{\chi'}\,  \ccorr{\phi^{(1)}}{\phi^{(2)}} \bigl(\lvert \fK(\chi') \vtheta' -  \fK(\chi) \vtheta \rvert, \chi' - \chi, z(\chi), z(\chi) \bigr).
\end{equation}
This avoids the need to consider correlations involving different cosmological epochs.

The third step assumes slowly varying angular-diameter distances, i.e. $\fK(\chi') \approx \fK(\chi)$ for $\lvert \chi' - \chi \rvert < \Pi_{\mathrm{c}}$, and resulting transverse separations to replace the radial projection by a parallel projection (this is sometimes called distant-observer approximation):
\begin{equation}
\label{eq:app:distant_observer_approximation}
 \EV{\pi^{(1)}(\vtheta) \pi^{(2)}(\vtheta')} 
 \approx \int\idiff[]{\chi}\, q^{(1)}(\chi)\, q^{(2)}(\chi) \int\idiff[]{\chi'}\, \ccorr{\phi^{(1)}}{\phi^{(2)}} \bigl(\fK(\chi) \lvert \vtheta' - \vtheta \rvert, \chi' - \chi, z(\chi), z(\chi) \bigr).
\end{equation}
Defining projected correlations by [cf. Eq.~\eqref{eq:df_projected_correlation}]
\begin{equation}
 \pcorr{\phi^{(1)}}{\phi^{(2)}} (r , z) =  \int\idiff[]{\Pi}\,  \ccorr{\phi^{(1)}}{\phi^{(2)}} \bigl(r , \Pi, z, z \bigr),
\end{equation} 
on obtains for the second integral in Eq.~\eqref{eq:app:distant_observer_approximation}:
\begin{equation}
	 \int\idiff[]{\chi'}\, \ccorr{\phi^{(1)}}{\phi^{(2)}} \bigl(\fK(\chi) \lvert \vtheta' - \vtheta \rvert, \chi' - \chi, z(\chi), z(\chi) \bigr)
	= 
\pcorr{\phi^{(1)}}{\phi^{(2)}} \bigl(\fK(\chi) \lvert \vtheta' - \vtheta \rvert, z(\chi)\bigr).	
\end{equation}
The correlation of the projected fields can then be written as:
\begin{equation}
\label{eq:app:Limber_radial_lightcone}
 \EV{\pi^{(1)}(\vtheta) \pi^{(2)}(\vtheta')} 
 \approx \int\idiff[]{\chi}\, q^{(1)}(\chi) \, q^{(2)}(\chi) \, \pcorr{\phi^{(1)}}{\phi^{(2)}} \bigl(\fK(\chi) \lvert \vtheta' - \vtheta \rvert, z(\chi)\bigr).
\end{equation}
Thus, one may compute predictions for these correlations of projected fields (e.g. for the quantitative interpretation of observations) if one knows the involved l.o.s. projection weights $q^{(i)}$ and the projected correlation functions $\pcorrsymbol$. 
Conversely, Eq.~\eqref{eq:app:Limber_radial_lightcone} shows that the projected correlation functions $\pcorrsymbol$ can be estimated directly (i.e. without estimating 3D correlations first) by correlating suitably defined projected fields, e.g. traced by different galaxy populations, which avoids the need of individual l.o.s. distance estimates for the galaxies.\footnote{With sufficiently accurate l.o.s. distances, the signal-to-noise ratio of the correlation estimates may be improved by introducing a l.o.s. distance weighting function that suppresses pairs with large l.o.s. separations (which contribute primarily noise).}

For projected fields defined by
\begin{equation}
  \pi^{(i)}(\vx) = \int\idiff[]{\chi}\, q^{(i)}(\chi) \, \phi^{(i)}\bigl(\vx, \chi, z(\chi) \bigr),
\end{equation}
one obtains
\begin{equation}
 \EV{\pi^{(1)}(\vx) \pi^{(2)}(\vx')} 
 \approx \int\idiff[]{\chi}\, q^{(1)}(\chi) \, q^{(2)}(\chi)  \pcorr{\phi^{(1)}}{\phi^{(2)}} \bigl( \lvert \vx' - \vx \rvert, z(\chi) \bigr).
\end{equation} 
The distant-observer approximation step is not needed here.

In the analysis of the simulation snapshots, knowledge about fields is not confined to their value on the lightcone. One may thus employ a slightly different definition for projected fields:
\begin{equation}
  \pi^{(i)}(\vx) = \int \idiff []{\chi} \, \phi^{(i)}\bigl(\vx, \chi, z \bigr).
\end{equation}
Here, the projection is over the full depth $L$ along any of the principal directions of the cubic simulation volume. Exploiting the periodic boundary conditions of the simulation, one obtains for the correlation of the projected fields:
\begin{equation}
\label{eq:app:Limber_simulation_snapshot}
\begin{split}
  \EV{\pi^{(1)}(\vx) \pi^{(2)}(\vx')} 
 &= \int_0^L\idiff[]{\chi}  \int_0^L\idiff[]{\chi'}\, \ccorr{\phi^{(1)}}{\phi^{(2)}} \bigl(\lvert \vtheta' -  \vtheta \rvert, \chi' - \chi, z, z \bigr)
 \\&=
  \int_0^L\idiff[]{\chi}  \int_0^L\idiff[]{\Pi}\, \ccorr{\phi^{(1)}}{\phi^{(2)}} \bigl(\lvert \vtheta' -  \vtheta \rvert, \Pi, z, z \bigr)
\\&=
 L  \pcorr{\phi^{(1)}}{\phi^{(2)}} (\lvert \vtheta' -  \vtheta \rvert , z) 
.
\end{split}
\end{equation}
This relation between the correlation of the projected fields and the projected correlation is exact.
We exploit this relation, when estimating projected correlations from the simulation data (see Section~\ref{sec:methods:correlations} and Appendix~\ref{app:projected_correlations_from_simulations}).

\section{Observed ellipticity-density correlations}
\label{app:density_ellipticity_correlations}

Here we discuss in more detail the relation between observed density-ellipticity correlations and correlations involving the intrinsic ellipticities and the galaxy and matter density fields.

\subsection{The statistical model}
\label{app:density_ellipticity_correlations:statistical_model}

Consider two samples of observed galaxies.
The first sample, used for tracing the matter density field, comprises $\NgalD{}{}$ galaxies with indices $i = 1,\ldots,\NgalD{}{}$, observed angular positions $\thetagalD{(i)}{}$, and redshifts $\zgalD{(i)}{}$.
The second sample is used as a probe of the ellipticity field and comprises $\NgalS{}{}$ galaxies with observed angular positions $\thetagalS{(j)}{}$, observed redshifts $\zgalS{(j)}{}$, and observed ellipticities $\obsell{(j)}{}$. For these ellipticities, we assume
\begin{equation} 
\label{app:eq:obsell_from_intell_and_gamma}
  \obsell{(j)}{} = \gamma^{(j)} + \intell{(j)}{},
\end{equation}
with $\gamma^{(j)} = \gamma(\thetagalS{(j)}{}, \zgalS{(j)}{})$ and $\intell{(j)}{} = \intell{}{}(\fS^{(j)} \thetagalS{(j)}{}, \chiS^{(j)} , \zgalS{(j)}{})$, where $\fS^{(j)} = \fK(\chiS^{(j)})$ , and $\chiS^{(j)} = \chi( \zgalS{(j)}{})$. For a pair of galaxies, one galaxy $i$ from the density sample and one galaxy $j$ from the ellipticity sample, we write $\obsellt{(j|i)}{}$, $\intellt{(j|i)}{}$, and $\gammat^{(j|i)}{}$ for the tangential component of the observed ellipticity, intrinsic ellipticity, and shear, resp., of galaxy $j$ relative to the line joining the angular positions of galaxy $i$ and $j$.

For simplicity, the angular positions of the galaxies are confined to a common survey field $\FOV$ with area $\AFOV$. Furthermore, we assume that the redshifts $\zgalD{(i)}{}$ of the density sample are sampled from a common redshift distribution $\pzgalD{}{}(\zgalD{(i)}{})$. The positions $\thetagalD{(i)}{}$ are sampled from a common galaxy density field with overdensity $\deltagalD{}{}\bigl(\fD^{(i)} \thetagalD{(i)}{},\chiD^{(i)}, \zgalD{(i)}{}\bigr)$. Similarly, we assume that the galaxy redshifts $\zgalS{(j)}{}$ of the ellipticity sample follow a common redshift distribution $\pzgalS{}{}$, and their positions $\thetagalS{(j)}{}$ follow a common galaxy density field with overdensity  $\deltagalS{}{}$.

For all the density and ellipticity fields and their correlations, we assume statistical homogeneity and isotropy. We also neglect any effects that might introduce correlations between the observed galaxy and matter densities at very different redshifts \citep[see, e.g.,][]{HilbertEtal2009_RT, HartlapEtal2011}.

Expectation values $\bEV{\est{f}}$ of observables $\est{f}(\thetagalD{(1)}{},\ldots)$ are computed by the following (formal) averages:
\begin{equation}
 \bEV{\est{f}} =\bEV{ \bEV{ \bEV{\est{f}}_{\thetagal{}{}} }_{\zgal{}{}{}} }_{\delta, \intell{}{}}.
\end{equation}
Here, $\bEV{\est{f}}_{\delta,\intell{}{}}$ denotes the ensemble average over the realizations of the galaxy and matter density fields and the intrinsic galaxy ellipticities, 
\begin{equation}
\bEV{\est{f}}_{\zgal{}{}} =
 \int\idiff[]{\zgalD{(1)}{}} \pzgalD{}{}(\zgalD{(1)}{}) 
\cdots
 \int\diff[]{\zgalS{(\NgalS{}{})}{}}\pzgalS{}{}(\zgalS{(\NgalS{}{})}{}) 
\est{f}\bigl(\thetagalD{(1)}{},\ldots\bigr)
\end{equation}
denotes the average over the galaxy redshifts, and
\begin{equation}
\bEV{\est{f}}_{\thetagal{}{}} = 
\frac{1}{\AFOV^{\NgalD{}{} + \NgalS{}{}}}
 \int_{\FOV}\idiff[2]{\thetagalD{(1)}{}}          \left[ 1 + \deltagalD{}{} (\fD^{(1)}          \thetagalD{(1)}{}         ,  \chiD^{(1)}         , \zgalD{(1)}{}         ) \right]
\cdots  
 \int_{\FOV}\idiff[2]{\thetagalS{(\NgalS{}{})}{}} \left[ 1 + \deltagalS{}{} (\fS^{(\NgalS{}{})} \thetagalS{(\NgalS{}{})}{},  \chiS^{(\NgalS{}{})}, \zgalS{(\NgalS{}{})}{}) \right]
\est{f}\bigl(\thetagalD{(1)}{},\ldots\bigr)
\end{equation}
denotes the ensemble average over the galaxy positions.

\subsection{The ellipticity-density correlation estimator}
\label{app:density_ellipticity_correlations:estimator}

We consider the following simple estimator for the ellipticity-density correlation as a function of comoving transverse separation $\sepperp$:
\begin{align}
  \label{app:eq:density_ellipticity_correlation_estimator}
 \estpcorr{\deltagalD{}{} }{\obsellt{}{}} (\sepperp) &= \frac{\EstimatorSum{\deltagalD{}{} }{\obsellt{}{}}(\sepperp)}{\RandomEstimatorSum{1}{1}(\sepperp)}
\text{ with}\\
  \label{app:eq:density_ellipticity_correlation_estimator_signal_sum}
\EstimatorSum{\deltagalD{}{}}{\obsellt{}{}}(\sepperp) &=
  \sum_{i,j=1}^{\NgalD{}{},\NgalS{}{}} \wgalD{(i)}{} \wgalS{(j)}{}  \,\Delta\bigl(\sepperp, \fD^{(i)} \lvert  \thetagalS{(j)}{} -   \thetagalD{(i)}{} \rvert \bigr) \obsellt{(j|i)}{}
\quad\text{and}\\
  \label{app:eq:density_ellipticity_correlation_estimator_norm_sum}
\RandomEstimatorSum{1}{1}(\sepperp) &=
  \sum_{i,j=1}^{\NgalD{}{},\NgalS{}{}} \wgalD{(i)}{} \wgalS{(j)}{} \,\Delta\bigl(\sepperp, \fD^{(i)} \lvert \thetagalSR{(j)}{} - \thetagalDR{(i)}{} \rvert \bigr)
.
\end{align}
Here, $\wgalD{(i)}{}$ and $\wgalS{(i)}{}$ denote weights, assumed statistically independent of galaxy positions, redshifts, or ellipticities.
The bin window function
\begin{equation}
\label{app:eq:comoving_bin_function}
  \Delta\bigl(\sepperp, \sepperp' \bigr)
  = \begin{cases}
  1 & \text{for } \lvert \sepperp' - \sepperp \rvert \leq \varDelta(\sepperp)/2 \text{ and}\\
  0 & \text{otherwise,}
  \end{cases}
\end{equation}
where $\varDelta(\sepperp)$ denotes the bin width (which we assume small compared to scales on which correlations change noticeably). The $\thetagalSR{(j)}{}$ denote positions obtained by randomly distributing the ellipticity sample galaxy positions within the survey area. The $\thetagalDR{(i)}{}$ denote randomized positions of the density sample.

A possible variation of the estimator~\eqref{app:eq:density_ellipticity_correlation_estimator} may use $\fS^{(j)} \lvert  \thetagalS{(j)}{} - \thetagalD{(i)}{} \rvert$ or $\lvert \fS^{(j)} \thetagalS{(j)}{} -  \fD^{(i)} \thetagalD{(i)}{} \rvert$ instead of $\fD^{(i)} \lvert  \thetagalS{(j)}{} - \thetagalD{(i)}{} \rvert$ as comoving transverse separation for binning. Moreover, if the individual galaxy redshifts are very inaccurate (or not known at all), it may be preferable to bin in angular separation $\lvert \thetagalS{(j)}{} - \thetagalD{(i)}{} \rvert$ instead. One could also randomize positions of only one of the two galaxy samples in the denominator. The estimator could also be adapted for galaxy-galaxy lensing to estimate the projected density profiles of galaxies by employing a particular redshift-dependent weighting of the galaxy pairs. That weighting, however, suppresses the intrinsic alignment signal, which we are after here.

\subsection{The normalization}
\label{app:density_ellipticity_correlations:normalization}

To compute the expectation $\bEV{\estpcorr{\deltagalD{}{} }{\obsellt{}{}} (\sepperp)}$ of the estimator~\eqref{app:eq:density_ellipticity_correlation_estimator}, we assume that
\begin{equation}
  \bEV{\estpcorr{\deltagalD{}{} }{\obsellt{}{}} (\sepperp)} \approx \frac{\bEV{\EstimatorSum{\deltagalD{}{}}{\obsellt{}{}}(\sepperp)}}{\bEV{\RandomEstimatorSum{1}{1}(\sepperp)}}
  .
\end{equation}

To compute the expectation $\bEV{\RandomEstimatorSum{1}{1}(\sepperp)}$ of the denominator~\eqref{app:eq:density_ellipticity_correlation_estimator_norm_sum}, we first consider the contribution from a single pair of galaxies:
\begin{equation}
\begin{split}
\EV{\Delta\bigl(\sepperp, \fD^{(i)} \lvert \thetagalSR{(j)}{} - \thetagalDR{(i)}{} \rvert \bigr)}
&=
 \int\idiff[]{\zgalD{(i)}{}} \pzgalD{}{}(\zgalD{(i)}{}) 
 \int\idiff[]{\zgalS{(j)}{}} \pzgalS{}{}(\zgalS{(j)}{}) 
\frac{1}{\AFOV^{2}}
\int_{\FOV}\idiff[2]{\thetagalDR{(i)}{}} 
\int_{\FOV}\idiff[2]{\thetagalSR{(j)}{}} 
\Delta\bigl(\sepperp,  \fD^{(i)} \lvert \thetagalSR{(j)}{} - \thetagalDR{(i)}{} \rvert \bigr)
\\&=
 \int\idiff[]{\zgalD{(i)}{}} \pzgalD{}{}(\zgalD{(i)}{}) 
 \frac{\ABin(\sepperp, \zgalD{(i)}{})}{\AFOV}
.
\end{split}
\end{equation}
Here,
\begin{equation}
  \ABin(\sepperp, \zgalD{}{}) = 
  \frac{1}{\AFOV}\int_{\FOV} \idiff[2]{\thetagalD{}{}} \int_{\FOV}\idiff[2]{\thetagalS{}{}} \Delta\bigl(\sepperp, \fD \lvert \thetagalS{}{} - \thetagalD{}{} \rvert \bigr)
  \leq 2 \pi \sepperp \varDelta(\sepperp) \fD^{-2}
,
\end{equation} 
denotes a redshift-dependent effective bin area, which favors small redshifts $\zgalD{(i)}{}$ for small separation $r$. If the actual galaxy positions $\thetagalS{(i)}{}$ instead of randomized positions $\thetagalSR{(i)}{}$ are used and there is a correlation $\ccorr{\deltagalD{}{}}{\deltagalS{}{}}$ between the galaxy sample positions, then that correlation alters the expected number of galaxy pairs in the denominator.

\subsection{The density-shear contribution}
\label{app:density_ellipticity_correlations:density_shear}

Due to our assumption~\eqref{app:eq:obsell_from_intell_and_gamma}, each observed ellipticity $\obsellt{(j|i)}{}$ in the numerator ~\eqref{app:eq:density_ellipticity_correlation_estimator_signal_sum} contributes a shear term $\gammat^{(j|i)}$ and an intrinsic ellipticity term $\intellt{(j|i)}{}$:
\begin{equation}
  \EV{\Delta\bigl(\sepperp, \fD^{(i)} \lvert \thetagalS{(j)}{} - \thetagalD{(i)}{} \rvert \bigr) \obsellt{(j|i)}{}}=
  \EV{\Delta\bigl(\sepperp, \fD^{(i)} \lvert \thetagalS{(j)}{} - \thetagalD{(i)}{} \rvert \bigr) \gammat^{(j|i)}}  +
  \EV{\Delta\bigl(\sepperp, \fD^{(i)} \lvert \thetagalS{(j)}{} - \thetagalD{(i)}{} \rvert \bigr) \intellt{(j|i)}{}}
  .
\end{equation}

The shear contribution for a single pair of galaxies $i$ and $j$ reads:
\begin{equation}
\begin{split}
&\EV{\Delta\bigl(\sepperp, \fD^{(i)} \lvert \thetagalS{(j)}{} - \thetagalD{(i)}{} \rvert \bigr) \gammat^{(j|i)}}
\\&=
 \int\idiff[]{\zgalD{(i)}{}} \pzgalD{}{}(\zgalD{(i)}{}) 
 \int\idiff[]{\zgalS{(j)}{}} \pzgalS{}{}(\zgalS{(j)}{}) 
\frac{1}{\AFOV^{2}}
\int_{\FOV}\idiff[2]{\thetagalD{(i)}{}} 
\int_{\FOV}\idiff[2]{\thetagalS{(j)}{}} 
\Delta\bigl(\sepperp,  \fD^{(i)} \lvert \thetagalS{(j)}{} - \thetagalD{(i)}{} \rvert \bigr)
\\&\quad\times
\EV{
\left[ 1 + \deltagalD{}{} (\fD^{(i)} \thetagalD{(i)}{}, \chiD^{(i)}, \zgalD{(i)}{} ) \right]
\left[ 1 + \deltagalS{}{} (\fS^{(j)} \thetagalS{(j)}{}, \chiS^{(j)}, \zgalS{(j)}{} ) \right]
\gammat(\thetagalS{(j)}{}, \zgalS{(j)}{}; \thetagalS{(j)}{} - \thetagalD{(i)}{})
}_{\delta, \intell{}{}}
\\&=
 \int\idiff[]{\zgalD{(i)}{}} \pzgalD{}{}(\zgalD{(i)}{}) 
 \int\idiff[]{\zgalS{(j)}{}} \pzgalS{}{}(\zgalS{(j)}{}) 
\frac{1}{\AFOV^{2}}
\int_{\FOV}\idiff[2]{\thetagalD{(i)}{}} 
\int_{\FOV}\idiff[2]{\thetagalS{(j)}{}} 
\Delta\bigl(\sepperp,  \fD^{(i)} \lvert \thetagalS{(j)}{} - \thetagalD{(i)}{} \rvert \bigr)
\int_{\FOV}\idiff[]{\chiL'}  \, \geomweight(\chiL',  \zgalS{(j)}{})
\\&\quad\times
\EV{
\left[ 1 + \deltagalD{}{} (\fD^{(i)} \thetagalD{(i)}{}, \chiD^{(i)}, \zgalD{(i)}{} ) \right]
\left[ 1 + \deltagalS{}{} (\fS^{(j)} \thetagalS{(j)}{}, \chiS^{(j)}, \zgalS{(j)}{} ) \right]
\deltashearMattert(\fL' \thetagalS{(j)}{}, \chiL', \zL'; \thetagalS{(j)}{} - \thetagalD{(i)}{})
}_{\delta, \intell{}{}}
.
\end{split}
\end{equation}
As stated above, we assume that any correlations involving the galaxy overdensities vanish if no other field involved is at a similar redshift. Statistical isotropy implies  $\bEV{\deltashearMattert(\fL' \thetagalS{(j)}{},\ldots)}_{\delta, \intell{}{}}=0$ and $\bEV{\deltagalS{}{} (\fS^{(j)} \thetagalS{(j)}{}, \ldots) \deltashearMattert(\fL' \thetagalS{(j)}{},\ldots)}_{\delta, \intell{}{}}=0$. By definition, $\bEV{\deltagalD{}{}}_{\delta, \intell{}{}}=0$ and $\bEV{\deltagalS{}{}}_{\delta, \intell{}{}} = 0$. Furthermore, the geometric weight $q$ suppresses terms involving the density field $\deltagalS{}{}$ and the matter shear contrast $\deltashearMattert$ at similar redshifts. Thus, above expression reduces to:
\begin{equation}
\begin{split}
\EV{\Delta\bigl(\sepperp, \fD^{(i)} \lvert \thetagalS{(j)}{} - \thetagalD{(i)}{} \rvert \bigr) \gammat^{(j|i)}}
&\approx
 \int\idiff[]{\zgalD{(i)}{}} \pzgalD{}{}(\zgalD{(i)}{}) 
 \int\idiff[]{\zgalS{(j)}{}} \pzgalS{}{}(\zgalS{(j)}{}) 
\frac{1}{\AFOV^{2}}
\!\int_{\FOV}\idiff[2]{\thetagalD{(i)}{}} 
\!\int_{\FOV}\idiff[2]{\thetagalS{(j)}{}} 
\Delta\bigl(\sepperp,  \fD^{(i)} \lvert \thetagalS{(j)}{} - \thetagalD{(i)}{} \rvert \bigr)
\!\int_{\FOV}\idiff[]{\chiL'} \, \geomweight(\chiL',  \zgalS{(j)}{})
\\&\quad\times
\EV{
\deltagalD{}{} (\fD^{(i)} \thetagalD{(i)}{}, \chiD^{(i)}, \zgalD{(i)}{} )
\deltashearMattert(\fL' \thetagalS{(j)}{}, \chiL', \zL'; \thetagalS{(j)}{} - \thetagalD{(i)}{})
}_{\delta, \intell{}{}}
\\&\approx
 \int\idiff[]{\zgalD{(i)}{}} \pzgalD{}{}(\zgalD{(i)}{}) 
  \frac{\ABin(\sepperp, \zgalD{(i)}{})}{\AFOV}
\int_{\FOV}\idiff[]{\chiL'} \, \geomweightS{}{}(\chiL') \,
\ccorr{\deltagalD{}{}}{\deltashearMattert} (r, \chiL' - \chiD^{(i)}, \zgalD{(i)}{}, \zL')
\\&\approx
 \int\idiff[]{\zgalD{(i)}{}} \pzgalD{}{}(\zgalD{(i)}{}) 
  \frac{\ABin(\sepperp, \zgalD{(i)}{})}{\AFOV}
 \geomweightS{}{}(\chiD^{(i)}) \,
\pcorr{\deltagalD{}{}}{\deltashearMattert} (r,  \zgalD{(i)}{})
.
\end{split}
\end{equation}
Here, the source redshift-weighted mean geometric weight
\begin{equation}
  \geomweightS{}{}(\chiL) 
=
 \frac{3 \HubbleConstant^2\OmegaMatter}{2\clight^2}(1 + \zL) \fL
\int_{\zL}^{\infty}\!\!\diff{\zS}\, \pzgalS{}{}(\zS)  \frac{\fSL}{\fS}
.
\end{equation}

\subsection{The density-intrinsic ellipticity contribution}
\label{app:density_ellipticity_correlations:density_intrinsic}

The intrinsic ellipticity contribution for a single pair of galaxies $i$ and $j$ reads:
\begin{equation}
\begin{split}
&\EV{\Delta\bigl(\sepperp, \fD^{(i)} \lvert \thetagalS{(j)}{} - \thetagalD{(i)}{} \rvert \bigr) \intellt{(j|i)}{}}
\\&=
 \int\idiff[]{\zgalD{(i)}{}} \pzgalD{}{}(\zgalD{(i)}{}) 
 \int\idiff[]{\zgalS{(j)}{}} \pzgalS{}{}(\zgalS{(j)}{}) 
\frac{1}{\AFOV^{2}}
\int_{\FOV}\idiff[2]{\thetagalD{(i)}{}} 
\int_{\FOV}\idiff[2]{\thetagalS{(j)}{}} 
\Delta\bigl(\sepperp,  \fD^{(i)} \lvert \thetagalS{(j)}{} - \thetagalD{(i)}{} \rvert \bigr)
\\&\quad\times
\EV{
\left[ 1 + \deltagalD{}{} (\fD^{(i)} \thetagalD{(i)}{}, \chiD^{(i)}, \zgalD{(i)}{} ) \right]
\left[ 1 + \deltagalS{}{} (\fS^{(j)} \thetagalS{(j)}{}, \chiS^{(j)}, \zgalS{(j)}{} ) \right]
\intellt{}{}\bigl(\fS^{(j)} \thetagalS{(j)}{}, \chiS^{(j)}, \zgalS{(j)}{}; \thetagalS{(j)}{} - \thetagalD{(i)}{}\bigr)
}_{\delta, \intell{}{}}
\\&\approx
 \int\idiff[]{\zgalD{(i)}{}} \pzgalD{}{}(\zgalD{(i)}{}) 
 \int\idiff[]{\zgalS{(j)}{}} \pzgalS{}{}(\zgalS{(j)}{}) 
\frac{\ABin(r, \zgalD{(i)}{})}{\AFOV}
\ccorr{\deltagalD{}{}}{(1 + \deltagalS{}{})\intellt{}{}}\bigl(r, \chiS^{(j)} - \chiD^{(i)}, \zgalD{(i)}{}, \zgalS{(j)}{}\bigr)
.
\end{split}
\end{equation}

\subsection{The full result}
\label{app:density_ellipticity_correlations:result}

The above results show that the expectation of the estimator~\eqref{app:eq:density_ellipticity_correlation_estimator} can be expressed as a sum of a density-gravitational shear contribution (dG) and a density-intrinsic ellipticity contribution (dI): 
\begin{equation}
\bEV{\estpcorr{\deltagalD{}{} }{\obsellt{}{}} (\sepperp)} =
\bEV{\estpcorr{\deltagalD{}{} }{\obsellt{}{}} (\sepperp)}^{\text{dG}} +
\bEV{\estpcorr{\deltagalD{}{} }{\obsellt{}{}} (\sepperp)}^{\text{dI}} 
.
\end{equation}
For these contributions,
\begin{align}
  \bEV{\estpcorr{\deltagalD{}{} }{\obsellt{}{}} (\sepperp)}^{\text{dG}}  &\approx \frac{\EstimatorSummandEV{\deltagalD{}{}}{\obsellt{}{}}^{\text{dG}}(\sepperp)}{\RandomEstimatorSummandEV{1}{1}(\sepperp)}
,\quad\text{and}\\
  \bEV{\estpcorr{\deltagalD{}{} }{\obsellt{}{}} (\sepperp)}^{\text{dI}}  &\approx \frac{\EstimatorSummandEV{\deltagalD{}{}}{\obsellt{}{}}^{\text{dI}}(\sepperp)}{\RandomEstimatorSummandEV{1}{1}(\sepperp)} 
,
\end{align}
where
\begin{align}
\label{app:eq:raw_pair_contribution_to_density_ellipticity_normalization}
  \RandomEstimatorSummandEV{1}{1}(\sepperp)
  &\approx
    \int\idiff[]{\zgalD{}{}}\, \pzgalD{}{}(\zgalD{}{}) 
    \frac{\ABin(\sepperp, \zgalD{}{})}{\AFOV}
,\\
\label{app:eq:raw_pair_contribution_to_density_ellipticity_dG_signal}
  \EstimatorSummandEV{\deltagalD{}{}}{\obsellt{}{}}^{\text{dG}}(\sepperp) 
  &\approx
    \int\idiff[]{\zgalD{}{}} \, \pzgalD{}{}(\zgalD{}{}) 
    \frac{\ABin(\sepperp, \zgalD{}{})}{\AFOV}
    \geomweightS{}{}(\chiD^{})\,
    \pcorr{\deltagalD{}{}}{\deltashearMattert} (r,  \zgalD{}{})
,\quad\text{and}\\
\label{app:eq:raw_pair_contribution_to_density_ellipticity_dI_signal}
  \EstimatorSummandEV{\deltagalD{}{}}{\obsellt{}{}}^{\text{dI}}(\sepperp) 
  & \approx
    \int\idiff[]{\zgalD{}{}}\, \pzgalD{}{}(\zgalD{}{}) 
    \frac{\ABin(r, \zgalD{}{})}{\AFOV}
    \int\idiff[]{\zgalS{}{}}\, \pzgalS{}{}(\zgalS{}{}) 
    \ccorr{\deltagalD{}{}}{(1 + \deltagalS{}{})\intellt{}{}}\bigl(r, \chiS - \chiD, \zgalD{}{}, \zgalS{}{}\bigr)
.
\end{align}

If at least one of the redshift distributions varies slowly with redshift compared to the 3D correlation $\ccorr{}{}$, one may substitute that correlation by its projection:
\begin{align}
\label{app:eq:raw_pair_contribution_to_density_ellipticity_dI_signal_Limber}
 \EstimatorSummandEV{\deltagalD{}{}}{\obsellt{}{}}^{\text{dI}}(\sepperp) 
 & \approx
 \int\idiff[]{\zgalD{}{}}\, \pzgalD{}{}(\zgalD{}{}) 
\frac{\ABin(r, \zgalD{}{})}{\AFOV}
 \pzgalS{}{}(\zgalD{}{}) 
\left(\totder{\chiD{}{}}{\zgalD{}{}}\right)^{-1}
\pcorr{\deltagalD{}{}}{(1 + \deltagalS{}{})\intellt{}{}}\bigl(r, \zgalD{}{} \bigr)
.
\end{align} 

If the redshift distributions $\pzgalD{}{}$ and $\pzgalS{}{}$ have no overlap, 
\begin{align}
\label{app:eq:raw_pair_contribution_to_density_ellipticity_dI_signal_no_overlap}
 \EstimatorSummandEV{\deltagalD{}{}}{\obsellt{}{}}^{\text{dI}}(\sepperp) 
 & \approx 0
.
\end{align} 

If there are no galaxies of the density sample at substantially lower redshift than any galaxies of the ellipticity sample,
\begin{equation}
\label{app:eq:raw_pair_contribution_to_density_ellipticity_dG_signal_no_lensing}
  \EstimatorSummandEV{\deltagalD{}{}}{\obsellt{}{}}^{\text{dG}}(\sepperp) \approx 0
  .
\end{equation}

\section{Observed ellipticity correlations}
\label{app:observed_ellipticity_correlations}

Here we discuss in more detail the relation between observed ellipticity-ellipticity correlations and correlations involving the intrinsic ellipticities and the galaxy and matter density fields.
The discussion builds partly upon the results of Appendix~\ref{app:density_ellipticity_correlations}.

\subsection{The statistical model}
\label{app:observed_ellipticity_correlations:statistical_model}

We choose a statistical model similar to that in Section~\ref{app:density_ellipticity_correlations:statistical_model}. 
We consider one or more sets of observed galaxies with position and ellipticity information.
Each set $\alpha \in \{ 1,\ldots,\Nsamples \}$ contains $\NgalS{(\alpha)}{}$ galaxies. These galaxies are labeled by $(\alpha, i)$ where $i \in \{1,\ldots,\NgalS{(\alpha)}{}\}$, and have observed ellipticities $\obsell{(\alpha,i)}{}$, observed angular positions $\thetagalS{(\alpha,i)}{}$ inside a survey field $\FOV$ with area $\AFOV$, and known probability distribution $\pzgalS{(\alpha)}{}(\zgalS{(\alpha,i)}{})$ for their redshifts $\zgalS{(\alpha,i)}{}$. 
The observed ellipticities $\obsell{(\alpha,i)}{} = \gamma^{(\alpha,i)} + \intell{(\alpha,i)}{}$, with $\gamma^{(\alpha,i)} = \gamma(\thetagalS{(\alpha,i)}{}, \zgalS{(\alpha,i)}{})$ and $\intell{(\alpha,i)}{} = \intell{(\alpha)}{}(\fS^{(\alpha,i)} \thetagalS{(\alpha,i)}{}, \chiS^{(\alpha,i)} , \zgalS{(\alpha,i)}{})$, where $\fS^{(\alpha,i)} = \fK(\chiS^{(\alpha,i)})$ , and $\chiS^{(\alpha,i)} = \chi( \zgalS{(\alpha,i)}{})$.
We assume that the positions of all galaxies $i$ within a common galaxy sample $\alpha$ are statistically distributed according to the same underlying galaxy density field with overdensity $\deltagalS{(\alpha)}{}(\fS^{(\alpha,i)} \thetagalS{(\alpha,i)}{}, \chiS^{(\alpha,i)} , \zgalS{(\alpha,i)}{})$.

For a pair of galaxies, one galaxy $i$ from sample $\alpha$ and one galaxy $j$ from sample $\nu$, we write $\obsellt{(\alpha,i|\nu,j)}{}$ and $\obsellx{(\alpha,i|\nu,j)}{}$ for the tangential and cross components of the observed ellipticity of galaxy $i$ relative to the line joining the angular positions of galaxy $i$ and $j$. The same pattern regarding notation of tangential and cross components applies to the intrinsic ellipticities and the shear.

For all the density and ellipticity fields and their correlations, we assume statistical homogeneity and isotropy. Furthermore, we neglect  magnification bias, masking by foreground galaxies, and any other effect that might introduce correlations between the intrinsic ellipticities, observed galaxy densities, or matter densities at very different redshifts.

Expectation values $\bEV{\est{f}}$ of observables $\est{f}(\thetagalS{(1,1)}{},\ldots)$ are computed by the following (formal) averages:
\begin{equation}
 \bEV{\est{f}} =\bEV{ \bEV{ \bEV{\est{f}}_{\thetagal{}{}} }_{\zgal{}{}} }_{\delta, \intell{}{}}.
\end{equation}
Here, $\bEV{\est{f}}_{\delta,\intell{}{}}$ denotes the ensemble average over the realizations of the galaxy and matter density fields and the intrinsic galaxy ellipticities, 
\begin{equation}
\bEV{\est{f}}_{\zgal{}{}} =
\left[
\prod_{\alpha,i} 
\int\idiff[]{\zgalS{(\alpha,i)}{}} \pzgalS{(\alpha)}{}(\zgalS{(\alpha,i)}{}) 
\right]
\est{f}\bigl(\thetagalS{(1,1)}{},\ldots\bigr)
\end{equation}
denotes the average over the galaxy redshifts, and
\begin{equation}
\bEV{\est{f}}_{\thetagal{}{}} = 
\left\{
\prod_{\alpha,i} 
\frac{1}{\AFOV} 
\int_{\FOV}\idiff[2]{\thetagalS{(\alpha,i)}{}}
\left[ 1 + \deltagalS{(\alpha)}{} (\fS^{(\alpha,i)} \thetagalS{(\alpha,i)}{},  \chiS^{(\alpha,i)}, \zgalS{(\alpha,i)}{}) \right]
\right\}
\est{f}\bigl(\thetagalS{(1,1)}{},\ldots\bigr)
\end{equation}
denotes the ensemble average over the galaxy positions.

\subsection{The ellipticity correlation estimator}
\label{app:observed_ellipticity_correlations:estimator}

We consider the following estimator for the observed ellipticity correlation between galaxy sample $\alpha$ and $\nu$ as a function of angular separation $\vartheta$:
\begin{align}
  \label{app:eq:obsell_correlation_estimator}
 \estacorrpm{\obsell{}{}}{\obsell{}{}}^{(\alpha | \nu)} (\vartheta) &= \frac{\EstimatorSumPM{\obsell{}{}}{\obsell{}{}}^{(\alpha | \nu)} (\vartheta)}{\EstimatorSum{1}{1}^{(\alpha | \nu)} (\vartheta)}
\text{ with}\\
  \label{app:eq:obsell_correlation_estimator_signal_sum}
\EstimatorSumPM{\obsell{}{}}{\obsell{}{}}^{(\alpha | \nu)} (\vartheta) &=
  \sum_{i,j=1}^{\NgalS{(\alpha)}{}\!,\NgalS{(\nu)}{}\!} \wgalS{(\alpha,i)}{} \wgalS{(\nu,j)}{} \,\Delta\bigl(\vartheta, \lvert \thetagalS{(\nu,j)}{} - \thetagalS{(\alpha,i)}{} \rvert \bigr)
  \left( \obsellt{(\alpha,i|\nu,j)}{} \obsellt{(\nu,j|\alpha,i)}{} \pm \obsellx{(\alpha,i|\nu,j)}{} \obsellx{(\nu,j|\alpha,i)}{} \right)
\quad\text{and}\\
  \label{app:eq:obsell_correlation_estimator_norm_sum}
\EstimatorSum{1}{1}^{(\alpha | \nu)} (\vartheta) &=
  \sum_{i,j=1}^{\NgalS{(\alpha)}{}\!,\NgalS{(\nu)}{}\!} \wgalS{(\alpha,i)}{} \wgalS{(\nu,j)}{} \,\Delta\bigl(\vartheta, \lvert \thetagalS{(\nu,j)}{} - \thetagalS{(\alpha,i)}{} \rvert \bigr)
.
\end{align}
Here, the $\wgalS{(\alpha,i)}{}$ denote statistical weights. 
The bin window function
\begin{equation}
\label{app:eq:bin_function}
  \Delta\bigl(\vartheta, \theta \bigr)
  = \begin{cases}
  1 & \text{for } \lvert \theta - \vartheta \rvert \leq \varDelta(\vartheta)/2 \text{ and}\\
  0 & \text{otherwise,}
  \end{cases}
\end{equation}
where $\varDelta(\vartheta)$ denotes the bin width (which we assume small compared to scales on which correlations change noticeably).
The separation and the bin width, along with details of the survey geometry, determine the effective bin area,
\begin{equation}
  \ABin(\vartheta) = 
  \frac{1}{\AFOV}\int_{\FOV} \idiff[2]{\vtheta} \int_{\FOV}\idiff[2]{\vtheta'} \Delta\bigl(\vartheta, \lvert \vtheta' - \vtheta \rvert \bigr)
  \leq 2 \pi \vartheta \varDelta(\vartheta)
,
\end{equation}
a main factor for the expected number of galaxy pairs contributing to the estimator.

\subsection{The normalization}
\label{app:observed_ellipticity_correlations:normalization}

To compute the expectation $\bEV{\estacorrpm{\obsell{}{}}{\obsell{}{}}^{(\alpha | \nu)} (\vartheta)}$ of the ellipticity correlation estimator~\eqref{app:eq:obsell_correlation_estimator}, we assume that statistical fluctuations of the denominator $\EstimatorSum{1}{1}^{(\alpha | \nu)}(\vartheta)$ are small such that
\begin{equation}
  \bEV{\estacorrpm{\obsell{}{}}{\obsell{}{}}^{(\alpha | \nu)} (\vartheta)} \approx
  \frac{\bEV{\EstimatorSumPM{\obsell{}{}}{\obsell{}{}}^{(\alpha | \nu)}(\vartheta)}}{\bEV{\EstimatorSum{1}{1}^{(\alpha | \nu)}(\vartheta)}}
  .
\end{equation}

To compute the expectation $\bEV{\EstimatorSum{1}{1}^{(\alpha | \nu)}(\vartheta)}$ of the denominator~\eqref{app:eq:obsell_correlation_estimator_norm_sum}, we first consider the contribution from a single pair of  distinct galaxies:
\begin{equation}
\begin{split}
\label{app:eq:raw_pair_contribution_to_normalization}
&\EV{\Delta\bigl(\vartheta, \lvert \thetagalS{(\nu,j)}{} - \thetagalS{(\alpha,i)}{} \rvert \bigr)}
\\&=
  \int\idiff[]{\zgalS{(\alpha,i)}{}} \pzgalS{(\alpha)}{}(\zgalS{(\alpha,i)}{}) 
  \int\idiff[]{\zgalS{(\nu   ,j)}{}} \pzgalS{(\nu   )}{}(\zgalS{(\nu   ,j)}{}) 
  \frac{1}{\AFOV^{2}}
  \int_{\FOV}\idiff[2]{\thetagalS{(\alpha,i)}{}} 
  \int_{\FOV}\idiff[2]{\thetagalS{(\nu   ,j)}{}} 
  \Delta\bigl(\vartheta, \lvert \thetagalS{(\nu,j)}{} - \thetagalS{(\alpha,i)}{} \rvert \bigr)
  \\&\quad\times
  \EV{
  \left[ 1 + \deltagalS{(\alpha)}{} \bigl(\fS^{(\alpha,i)} \thetagalS{(\alpha,i)}{}, \chiS^{(\alpha,i)}, \zgalS{(\alpha,i)}{} \bigr) \right]
  \left[ 1 + \deltagalS{(\nu   )}{} \bigl(\fS^{(\nu   ,j)} \thetagalS{(\nu   ,j)}{}, \chiS^{(\nu   ,j)}, \zgalS{(\nu   ,j)}{} \bigr) \right]
  }_{\delta, \intell{}{}}
\\&=
  \int\idiff[]{\zgalS{(\alpha,i)}{}} \pzgalS{(\alpha)}{}(\zgalS{(\alpha,i)}{}) 
  \int\idiff[]{\zgalS{(\nu   ,j)}{}} \pzgalS{(\nu   )}{}(\zgalS{(\nu   ,j)}{}) 
  \frac{1}{\AFOV^{2}}
  \int_{\FOV}\idiff[2]{\thetagalS{(\alpha,i)}{}} 
  \int_{\FOV}\idiff[2]{\thetagalS{(\nu   ,j)}{}} 
  \Delta\bigl(\vartheta, \lvert \thetagalS{(\nu,j)}{} - \thetagalS{(\alpha,i)}{} \rvert \bigr)
  \\&\quad\times
  \left[ 1 +
  \ccorr{\deltagalS{(\alpha)}{}}{\deltagalS{(\nu)}{}}
  \bigl( \bigl|\fS^{(\nu,j)} \thetagalS{(\nu,j)}{} - \fS^{(\alpha,i)} \thetagalS{(\alpha,i)}{} \bigr|, \chiS^{(\nu,j)} - \chiS^{(\alpha,i)}, \zgalS{(\alpha,i)}{}, \zgalS{(\nu,j)}{} \bigr) \right]
\\&\approx
  \frac{\ABin(\vartheta)}{\AFOV}
  \int\idiff[]{\zgalS{(\alpha,i)}{}} \pzgalS{(\alpha)}{}(\zgalS{(\alpha,i)}{}) 
  \int\idiff[]{\zgalS{(\nu   ,j)}{}} \pzgalS{(\nu   )}{}(\zgalS{(\nu   ,j)}{}) 
  \left[1 +
  \ccorr{\deltagalS{(\alpha)}{}}{\deltagalS{(\nu)}{}} \bigl(\fS^{(\alpha,i)} \vartheta, \chiS^{(\nu,j)} - \chiS^{(\alpha,i)}, \zgalS{(\alpha,i)}{},  \zgalS{(\nu,j)}{} \bigr) \right]
.
\end{split}
\end{equation}
The last step assumes that $\fS^{(\nu,j)} \approx \fS^{(\alpha,i)}$ unless $\ccorr{\deltagalS{(\alpha)}{}}{\deltagalS{(\nu)}{}} \approx 0$, and moreover narrow bins for $\Delta$ such that $\bigl\lvert\thetagalS{(\nu,j)}{} - \thetagalS{(\alpha,i)}{}\bigr\rvert \approx \vartheta$ for contributing pairs of galaxies.

The expected pair contribution~\eqref{app:eq:raw_pair_contribution_to_normalization} may be subject to a separation-dependent enhancement (or decrement in some cases) due to non-vanishing galaxy density correlations $\ccorr{\deltagalS{(\alpha)}{}}{\deltagalS{(\nu)}{}}$. The enhancement may be negligible if galaxy density correlations are sufficiently weak in general. Moreover, there is no enhancement in the number of expected pairs if the redshift distributions $\pzgalS{(\alpha)}{}$ and $\pzgalS{(\nu)}{}$ ensure that the galaxies of any participating pair are well separated in the line-of-sight direction such that $\ccorr{\deltagalS{(\alpha)}{}}{\deltagalS{(\nu)}{}}$ vanishes.


\subsection{Shear-shear contributions}
\label{app:observed_ellipticity_correlations:shear_shear}

Since we assume $\obsell{}{} = \gamma + \intell{}{}$, each product of observed ellipticities in the numerator ~\eqref{app:eq:obsell_correlation_estimator_signal_sum} can be broken up into four terms,
\begin{equation}
\begin{split}
&\left( \obsellt{(\alpha,i|\mu,j)}{} \obsellt{(\mu,j|\alpha,i)}{} \pm \obsellx{(\alpha,i|\mu,j)}{} \obsellx{(\mu,j|\alpha,i)}{} \right)
\\&=
    \left( \gammat^{(\alpha,i|\mu,j)}   \gammat^{(\mu,j|\alpha,i)}   \pm \gammax^{(\alpha,i|\mu,j)}   \gammax^{(\mu,j|\alpha,i)}   \right)
  + \left( \gammat^{(\alpha,i|\mu,j)}{} \intellt{(\mu,j|\alpha,i)}{} \pm \gammax^{(\alpha,i|\mu,j)}{} \intellx{(\mu,j|\alpha,i)}{} \right) 
\\&\quad
 +  \left( \intellt{(\alpha,i|\mu,j)}{} \gammat^{(\mu,j|\alpha,i)}   \pm \intellx{(\alpha,i|\mu,j)}{} \gammax^{(\mu,j|\alpha,i)}   \right)
  + \left( \intellt{(\alpha,i|\mu,j)}{} \intellt{(\mu,j|\alpha,i)}{} \pm \obsellx{(\alpha,i|\mu,j)}{} \obsellx{(\mu,j|\alpha,i)}{} \right)
.
\end{split}
\end{equation}
These terms are commonly called gravitational lensing shear-shear (GG), shear-intrinsic (GI), intrinsic-shear (IG), and intrinsic-intrinsic (II).

The tangential GG term for a pair of galaxies yields
\begin{equation}
\begin{split}
&\EV{\Delta\bigl(\vartheta, \lvert \thetagalS{(\nu,j)}{} - \thetagalS{(\alpha,i)}{} \rvert \bigr)  \gammat^{(\alpha,i|\mu,j)}   \gammat^{(\mu,j|\alpha,i)} }
\\&=
  \int\idiff[]{\zgalS{(\alpha,i)}{}} \pzgalS{(\alpha)}{}(\zgalS{(\alpha,i)}{}) 
  \int\idiff[]{\zgalS{(\nu   ,j)}{}} \pzgalS{(\nu   )}{}(\zgalS{(\nu   ,j)}{}) 
  \frac{1}{\AFOV^{2}}
  \int_{\FOV}\idiff[2]{\thetagalS{(\alpha,i)}{}} 
  \int_{\FOV}\idiff[2]{\thetagalS{(\nu   ,j)}{}} 
  \Delta\bigl(\vartheta, \lvert \thetagalS{(\nu,j)}{} - \thetagalS{(\alpha,i)}{} \rvert \bigr)
  \\&\quad\times
  \Bigl\langle
  \left[ 1 + \deltagalS{(\alpha)}{} \bigl(\fS^{(\alpha,i)} \thetagalS{(\alpha,i)}{}, \chiS^{(\alpha,i)}, \zgalS{(\alpha,i)}{} \bigr) \right]
  \left[ 1 + \deltagalS{(\nu   )}{} \bigl(\fS^{(\nu   ,j)} \thetagalS{(\nu   ,j)}{}, \chiS^{(\nu   ,j)}, \zgalS{(\nu   ,j)}{} \bigr) \right]
  \\&\quad\qquad\times
  \gammat \bigl(\thetagalS{(\alpha,i)}{}, \zgalS{(\alpha,i)}{}; \thetagalS{(\nu,j)}{} - \thetagalS{(\alpha,i)}{} \bigr)
  \gammat \bigl(\thetagalS{(\nu   ,j)}{}, \zgalS{(\nu   ,j)}{}; \thetagalS{(\nu,j)}{} - \thetagalS{(\alpha,i)}{} \bigr)
  \Bigr\rangle_{\delta, \intell{}{}}
\\&=
  \int\idiff[]{\zgalS{(\alpha,i)}{}} \pzgalS{(\alpha)}{}(\zgalS{(\alpha,i)}{}) 
  \int\idiff[]{\zgalS{(\nu   ,j)}{}} \pzgalS{(\nu   )}{}(\zgalS{(\nu   ,j)}{}) 
  \frac{1}{\AFOV^{2}}
  \int_{\FOV}\idiff[2]{\thetagalS{(\alpha,i)}{}} 
  \int_{\FOV}\idiff[2]{\thetagalS{(\nu   ,j)}{}} 
  \Delta\bigl(\vartheta, \lvert \thetagalS{(\nu,j)}{} - \thetagalS{(\alpha,i)}{} \rvert \bigr)
  \\&\quad\times
  \int\idiff[]{\chiD^{(\alpha)}} \geomweight(\chiD^{(\alpha)}, \zgalS{(\alpha,i)}{}) 
  \int\idiff[]{\chiD^{(\nu   )}} \geomweight(\chiD^{(\nu   )}, \zgalS{(\nu   ,j)}{}) 
  \Bigl\langle
  \left[ 1 + \deltagalS{(\alpha)}{} \bigl(\fS^{(\alpha,i)} \thetagalS{(\alpha,i)}{}, \chiS^{(\alpha,i)}, \zgalS{(\alpha,i)}{} \bigr) \right]
  \left[ 1 + \deltagalS{(\nu   )}{} \bigl(\fS^{(\nu   ,j)} \thetagalS{(\nu   ,j)}{}, \chiS^{(\nu   ,j)}, \zgalS{(\nu   ,j)}{} \bigr) \right]
  \!\!
  \\&\quad\qquad\times
  \deltashearMattert \bigl(\fD^{(\alpha)} \thetagalS{(\alpha,i)}{}, \chiD^{(\alpha)}, \zgalD{(\alpha)}{}; \thetagalS{(\nu,j)}{} - \thetagalS{(\alpha,i)}{} \bigr)
  \deltashearMattert \bigl(\fD^{(\mu   )} \thetagalS{(\nu   ,j)}{}, \chiD^{(\mu   )}, \zgalD{(\mu   )}{}; \thetagalS{(\nu,j)}{} - \thetagalS{(\alpha,i)}{} \bigr)
  \Bigr\rangle_{\delta, \intell{}{}}
.
\end{split}
\end{equation}
If one assumes that any correlation involving the galaxy overdensities is only non-zero when at least one of the other involved fields is at a similar redshift\footnote{
Note that observational effects due to, e.g., magnification bias or masking may violate this assumption.
}, then terms of the form $\EV{\deltagalS{}{}\deltashearMattert \deltashearMattert}$ either vanish themselves or are suppressed by a small geometric weight $q$. Moreover, terms of the form $\EV{\deltagalS{}{}\deltagalS{}{}\deltashearMattert\deltashearMattert}$ factorize into $\ccorr{\deltagalS{}{}}{\deltagalS{}{}}\ccorr{\deltashearMattert}{\deltashearMattert}$. Furthermore, except for very low source redshifts or very wide angular separations, the correlation $\ccorr{\deltashearMattert}{\deltashearMattert}$ may be substituted by the projected correlation $\pcorr{\deltashearMattert}{\deltashearMattert}$ in a Limber-type approximation:
\begin{equation}
\begin{split}
&\EV{\Delta\bigl(\vartheta, \lvert \thetagalS{(\nu,j)}{} - \thetagalS{(\alpha,i)}{} \rvert \bigr)  \gammat^{(\alpha,i|\mu,j)}   \gammat^{(\mu,j|\alpha,i)} }
\\&\approx
  \frac{\ABin(\vartheta)}{\AFOV}
  \int\idiff[]{\zgalS{(\alpha,i)}{}} \pzgalS{(\alpha)}{}(\zgalS{(\alpha,i)}{}) 
  \int\idiff[]{\zgalS{(\nu   ,j)}{}} \pzgalS{(\nu   )}{}(\zgalS{(\nu   ,j)}{}) 
  \left[1 + \ccorr{\deltagalS{(\alpha)}{}}{\deltagalS{(\nu)}{}} \bigl(\fS^{(\alpha,i)} \vartheta, \chiS^{(\nu,j)} - \chiS^{(\alpha,i)}, \zgalS{(\alpha,i)}{},  \zgalS{(\nu,j)}{} \bigr) \right]
  \\&\quad\times
  \int\idiff[]{\chiD^{(\alpha)}} \geomweight(\chiD^{(\alpha)}, \zgalS{(\alpha,i)}{}) 
  \int\idiff[]{\chiD^{(\nu   )}} \geomweight(\chiD^{(\nu   )}, \zgalS{(\nu   ,j)}{}) 
  \ccorr{\deltashearMattert}{\deltashearMattert} \bigl(\fD^{(\alpha)} \vartheta, \chiD^{(\mu)} - \chiD^{(\alpha)}, \zgalD{(\alpha)}{}, \zgalD{(\mu)}{} \bigr) 
\\&\approx
  \frac{\ABin(\vartheta)}{\AFOV}
  \int\idiff[]{\zgalS{(\alpha,i)}{}} \pzgalS{(\alpha)}{}(\zgalS{(\alpha,i)}{}) 
  \int\idiff[]{\zgalS{(\nu   ,j)}{}} \pzgalS{(\nu   )}{}(\zgalS{(\nu   ,j)}{}) 
  \left[1 + \ccorr{\deltagalS{(\alpha)}{}}{\deltagalS{(\nu)}{}} \bigl(\fS^{(\alpha,i)} \vartheta, \chiS^{(\nu,j)} - \chiS^{(\alpha,i)}, \zgalS{(\alpha,i)}{},  \zgalS{(\nu,j)}{} \bigr) \right]
  \\&\quad\times
  \int\idiff[]{\chiD^{(\alpha)}} \geomweight(\chiD^{(\alpha)}, \zgalS{(\alpha,i)}{}) \geomweight(\chiD^{(\alpha)}, \zgalS{(\nu   ,j)}{}) 
  \pcorr{\deltashearMattert}{\deltashearMattert} \bigl(\fD^{(\alpha)} \vartheta, \zgalD{(\alpha)}{} \bigr) 
.
\end{split}
\end{equation}

Combining above results for GG tangential terms with results from a similar computation for the GG cross terms, one obtains:
\begin{equation}
\label{app:eq:raw_pair_contribution_to_GG_signal}
\begin{split}
&\EV{\Delta\bigl(\vartheta, \lvert \thetagalS{(\nu,j)}{} - \thetagalS{(\alpha,i)}{} \rvert \bigr) 
  \left( \gammat^{(\alpha,i|\mu,j)}   \gammat^{(\mu,j|\alpha,i)}   \pm \gammax^{(\alpha,i|\mu,j)}   \gammax^{(\mu,j|\alpha,i)}   \right)
  }
\\&\approx
  \frac{\ABin(\vartheta)}{\AFOV}
  \int\idiff[]{\zgalS{(\alpha,i)}{}} \pzgalS{(\alpha)}{}(\zgalS{(\alpha,i)}{}) 
  \int\idiff[]{\zgalS{(\nu   ,j)}{}} \pzgalS{(\nu   )}{}(\zgalS{(\nu   ,j)}{}) 
  \left[1 + \ccorr{\deltagalS{(\alpha)}{}}{\deltagalS{(\nu)}{}} \bigl(\fS^{(\alpha,i)} \vartheta, \chiS^{(\nu,j)} - \chiS^{(\alpha,i)}, \zgalS{(\alpha,i)}{},  \zgalS{(\nu,j)}{} \bigr) \right]
  \\&\quad\times
  \int\idiff[]{\chiD^{(\alpha)}} \geomweight(\chiD^{(\alpha)}, \zgalS{(\alpha,i)}{}) \geomweight(\chiD^{(\alpha)}, \zgalS{(\nu   ,j)}{}) 
  \pcorrpm{\deltashearMatter}{\deltashearMatter} \bigl(\fD^{(\alpha)} \vartheta, \zgalD{(\alpha)}{} \bigr) 
.
\end{split}
\end{equation}
The term $\ccorr{\deltagalS{(\alpha)}{}}{\deltagalS{(\nu)}{}}$ is again a source of a separation-dependent enhancement or decrement of the expected number of galaxy pairs contributing to the numerator. It also introduces an effective (anti)correlation between the redshifts of the galaxy pairs, which alters the expected number of galaxy pairs at similar redshifts in redshift regions where the distributions $\pzgalS{(\alpha)}{}$ and $\pzgalS{(\nu)}{}$ overlap. This (anti)correlation in the galaxy redshifts increases (decreases) the contribution to the shear signal from matter correlations at higher redshifts. This effect is more pronounced for broader redshift distributions with a larger range of overlap in redshifts.
 

\subsection{Shear-intrinsic contributions}
\label{app:observed_ellipticity_correlations:shear_intrinsic}

For $(\alpha,i)$ and $(\nu,j)$ denoting different galaxies, the tangential GI term becomes
\begin{equation}
\begin{split}
&
\EV{\Delta\bigl(\vartheta, \lvert \thetagalS{(\nu,j)}{} - \thetagalS{(\alpha,i)}{} \rvert \bigr)  \gammat^{(\alpha,i|\mu,j)}   \intellt{(\mu,j|\alpha,i)}{} }
\\&=
  \int\idiff[]{\zgalS{(\alpha,i)}{}} \pzgalS{(\alpha)}{}(\zgalS{(\alpha,i)}{}) 
  \int\idiff[]{\zgalS{(\nu   ,j)}{}} \pzgalS{(\nu   )}{}(\zgalS{(\nu   ,j)}{}) 
  \frac{1}{\AFOV^{2}}
  \int_{\FOV}\idiff[2]{\thetagalS{(\alpha,i)}{}} 
  \int_{\FOV}\idiff[2]{\thetagalS{(\nu   ,j)}{}} 
  \Delta\bigl(\vartheta, \lvert \thetagalS{(\nu,j)}{} - \thetagalS{(\alpha,i)}{} \rvert \bigr)
  \\&\quad\times
  \Bigl\langle
  \left[ 1 + \deltagalS{(\alpha)}{} \bigl(\fS^{(\alpha,i)} \thetagalS{(\alpha,i)}{}, \chiS^{(\alpha,i)}, \zgalS{(\alpha,i)}{} \bigr) \right]
  \left[ 1 + \deltagalS{(\nu   )}{} \bigl(\fS^{(\nu   ,j)} \thetagalS{(\nu   ,j)}{}, \chiS^{(\nu   ,j)}, \zgalS{(\nu   ,j)}{} \bigr) \right]
  \\&\quad\qquad\times
  \gammat          \bigl(\thetagalS{(\alpha,i)}{}, \zgalS{(\alpha,i)}{}; \thetagalS{(\nu,j)}{} - \thetagalS{(\alpha,i)}{}\bigr)
  \intellt{(\nu)}{}\bigl(\fS^{(\nu,j)} \thetagalS{(\nu,j)}{}, \chiS^{(\nu,i)}, \zgalS{(\nu,i)}{}; \thetagalS{(\nu,j)}{} - \thetagalS{(\alpha,i)}{} \bigr)
  \Bigr\rangle_{\delta, \intell{}{}}
\\&=
  \int\idiff[]{\zgalS{(\alpha,i)}{}} \pzgalS{(\alpha)}{}(\zgalS{(\alpha,i)}{}) 
  \int\idiff[]{\zgalS{(\nu   ,j)}{}} \pzgalS{(\nu   )}{}(\zgalS{(\nu   ,j)}{}) 
  \frac{1}{\AFOV^{2}}
  \int_{\FOV}\idiff[2]{\thetagalS{(\alpha,i)}{}} 
  \int_{\FOV}\idiff[2]{\thetagalS{(\nu   ,j)}{}} 
  \Delta\bigl(\vartheta, \lvert \thetagalS{(\nu,j)}{} - \thetagalS{(\alpha,i)}{} \rvert \bigr)
  \\&\quad\times
  \int\idiff[]{\chiD^{(\alpha)}} \geomweight(\chiD^{(\alpha)}, \zgalS{(\alpha,i)}{}) 
  \Bigl\langle
  \left[ 1 + \deltagalS{(\alpha)}{} \bigl( \fS^{(\alpha,i)} \thetagalS{(\alpha,i)}{}, \chiS^{(\alpha,i)}, \zgalS{(\alpha,i)}{} \bigr) \right]
  \left[ 1 + \deltagalS{(\nu   )}{} \bigl( \fS^{(\nu   ,j)} \thetagalS{(\nu   ,j)}{}, \chiS^{(\nu   ,j)}, \zgalS{(\nu   ,j)}{} \bigr) \right]
  \\&\quad\qquad\times
  \deltashearMattert \bigl(\fD^{(\alpha  )} \thetagalS{(\alpha,i)}{}, \chiD^{(\alpha  )}, \zgalD{(\alpha  )}{}; \thetagalS{(\nu,j)}{} - \thetagalS{(\alpha,i)}{} \bigr)
  \intellt{(\nu)}{}  \bigl(\fS^{(\nu   ,j)} \thetagalS{(\nu   ,j)}{}, \chiS^{(\nu   ,j)}, \zgalS{(\nu   ,j)}{}; \thetagalS{(\nu,j)}{} - \thetagalS{(\alpha,i)}{} \bigr)
  \Bigr\rangle_{\delta, \intell{}{}}
.
\end{split}
\end{equation}
Assuming that correlations vanish unless for each involved field there is another involved field at a similar redshift, one obtains:
\begin{equation}
\begin{split}
&
\EV{\Delta\bigl(\vartheta, \lvert \thetagalS{(\nu,j)}{} - \thetagalS{(\alpha,i)}{} \rvert \bigr)  \gammat^{(\alpha,i|\mu,j)}   \intellt{(\mu,j|\alpha,i)}{} }
\\&\approx
  \frac{\ABin(\vartheta)}{\AFOV}
  \int\idiff[]{\zgalS{(\alpha,i)}{}} \pzgalS{(\alpha)}{}(\zgalS{(\alpha,i)}{}) 
  \int\idiff[]{\zgalS{(\nu   ,j)}{}} \pzgalS{(\nu   )}{}(\zgalS{(\nu   ,j)}{}) 
  \int\idiff[]{\chiD^{(\alpha)}} \geomweight(\chiD^{(\alpha)}, \zgalS{(\alpha,i)}{}) \,
  \ccorr{\deltashearMattert}{(1 + \deltagalS{(\nu   )}{}) \intellt{(\nu)}{}} 
  \bigl( \fD^{(\alpha  )} \vartheta, \chiS^{(\nu   ,j)} -  \chiD^{(\alpha  )}, \zgalD{(\alpha  )}, \zgalS{(\nu   ,j)}{} \bigr)
\\&\approx
  \frac{\ABin(\vartheta)}{\AFOV}
  \int\idiff[]{\zgalS{(\nu   ,j)}{}} \pzgalS{(\nu   )}{}(\zgalS{(\nu   ,j)}{}) 
  \geomweightS{(\alpha)}{}( \chiS^{(\nu   ,j)}) \,
  \pcorr{\deltashearMattert}{(1 + \deltagalS{(\nu   )}{}) \intellt{(\nu)}{}} \bigl(\fS^{(\nu   ,j)}  \vartheta, \zgalS{(\nu   ,j)}{} \bigr)
,
\end{split}
\end{equation}
where the source redshift weighted mean geometric weight for source galaxies from sample $\alpha$ is given by:
\begin{equation}
\label{app:eq:geometric_weight_for_source_galaxy_sample}
  \geomweightS{(\alpha)}{}(\chiL) 
=
 \frac{3 \HubbleConstant^2\OmegaMatter}{2\clight^2}(1 + \zL) \fL
\int_{\zL}^{\infty}\!\!\diff{\zS}\, \pzgalS{(\alpha)}{}(\zS)  \frac{\fSL}{\fS}
.
\end{equation}

Repeating the calculation for the cross terms and combining the tangential and cross terms, one obtains:
\begin{equation}
\label{app:eq:raw_pair_contribution_to_GI_signal}
\begin{split}
&
  \EV{
  \Delta\bigl(\vartheta, \lvert \thetagalS{(\nu,j)}{} - \thetagalS{(\alpha,i)}{} \rvert \bigr)  
  \left( \gammat^{(\alpha,i|\mu,j)}{} \intellt{(\mu,j|\alpha,i)}{} \pm \gammax^{(\alpha,i|\mu,j)}{} \intellx{(\mu,j|\alpha,i)}{} \right) 
  }
\\&\approx
  \frac{\ABin(\vartheta)}{\AFOV}
  \int\idiff[]{\zgalS{(\nu   ,j)}{}} \pzgalS{(\nu   )}{}(\zgalS{(\nu   ,j)}{}) 
  \geomweightS{(\alpha)}{}( \chiS^{(\nu   ,j)}) \,
  \pcorrpm{\deltashearMatter}{(1 + \deltagalS{(\nu   )}{}) \intell{(\nu)}{}} \bigl(\fS^{(\nu   ,j)}  \vartheta, \zgalS{(\nu   ,j)}{} \bigr)
.
\end{split}
\end{equation}

\subsection{Intrinsic-intrinsic contributions}
\label{app:observed_ellipticity_correlations:intrinsic_intrinsic}

Building upon the experience gained in the preceding section for the GI and IG terms, we can omit many of the intermediate steps for the II terms. For a single pair of distinct galaxies $(\alpha,i)$ and $(\nu, j)$,
\begin{equation}
\label{app:eq:raw_pair_contribution_to_II_signal}
\begin{split}
&
  \EV{
  \Delta\bigl(\vartheta, \lvert \thetagalS{(\nu,j)}{} - \thetagalS{(\alpha,i)}{} \rvert \bigr)  
  \left( \intellt{(\alpha,i|\mu,j)}{} \intellt{(\mu,j|\alpha,i)}{} \pm \intellx{(\alpha,i|\mu,j)}{} \intellx{(\mu,j|\alpha,i)}{} \right) 
  }
\\&\approx
  \frac{\ABin(\vartheta)}{\AFOV}
  \int\idiff[]{\zgalS{(\alpha,i)}{}} \pzgalS{(\alpha)}{}(\zgalS{(\alpha,i)}{}) 
  \int\idiff[]{\zgalS{(\nu   ,j)}{}} \pzgalS{(\nu   )}{}(\zgalS{(\nu   ,j)}{}) \,
  \ccorrpm{(1 + \deltagalS{(\alpha)}{}) \intell{(\alpha)}{}}{(1 + \deltagalS{(\nu   )}{}) \intell{(\nu)}{}} 
  \bigl( \fS^{(\alpha,i)} \vartheta, \chiS^{(\nu   ,j)} -  \chiS^{(\alpha,i)}, \zgalS{(\alpha,i)}{}, \zgalS{(\nu   ,j)}{} \bigr)
.
\end{split}
\end{equation}

\subsection{The full result}
\label{app:observed_ellipticity_correlations:result}

Combining the results for the different contributions to the expectation value of the ellipticity correlation estimator~\eqref{app:eq:obsell_correlation_estimator}, one obtains
\begin{equation}
   \bEV{\estacorrpm{\obsell{}{}}{\obsell{}{}}^{(\alpha|\nu)} (\vartheta)}
= 
   \bEV{\estacorrpm{\obsell{}{}}{\obsell{}{}}^{(\alpha|\nu)} (\vartheta)}^{\text{GG}}
 + \bEV{\estacorrpm{\obsell{}{}}{\obsell{}{}}^{(\alpha|\nu)} (\vartheta)}^{\text{GI}}
 + \bEV{\estacorrpm{\obsell{}{}}{\obsell{}{}}^{(\alpha|\nu)} (\vartheta)}^{\text{IG}}
 + \bEV{\estacorrpm{\obsell{}{}}{\obsell{}{}}^{(\alpha|\nu)} (\vartheta)}^{\text{II}}
.
\end{equation}
The gravitational shear-shear (GG), gravitational shear-intrinsic ellipticity (GI), intrinsic ellipticity-gravitational shear (IG), and intrinsic ellipticity-intrinsic ellipticity (II) contributions are given by:
\begin{equation}
   \bEV{\estacorrpm{\obsell{}{}}{\obsell{}{}}^{(\alpha|\nu)} (\vartheta)}^{\text{XY}}
\approx
\frac
{ \EstimatorSummandPMEV{\obsell{}{}}{\obsell{}{}}^{(\alpha|\nu)\,\text{XY}}(\vartheta)}
{ \EstimatorSummandEV{1}{1}^{(\alpha|\nu)}(\vartheta)}
,\quad \mathrm{X},\mathrm{Y} \in \{ \mathrm{G}, \mathrm{I} \}, 
\end{equation}
with [see Eqs.~\eqref{app:eq:raw_pair_contribution_to_normalization}, \eqref{app:eq:raw_pair_contribution_to_GG_signal}, \eqref{app:eq:raw_pair_contribution_to_GI_signal}, and \eqref{app:eq:raw_pair_contribution_to_II_signal}]
\begin{subequations}
\begin{align}
\begin{split}
\EstimatorSummandEV{1}{1}^{(\alpha|\nu)}(\vartheta)
  &\approx
  \int\idiff[]{\zgalS{(\alpha)}{}} \pzgalS{(\alpha)}{}(\zgalS{(\alpha)}{}) 
  \int\idiff[]{\zgalS{(\nu   )}{}} \pzgalS{(\nu   )}{}(\zgalS{(\nu   )}{}) 
  \left[1 +
  \ccorr{\deltagalS{(\alpha)}{}}{\deltagalS{(\nu)}{}} \bigl(\fS^{(\alpha)} \vartheta, \chiS^{(\nu)} - \chiS^{(\alpha)}, \zgalS{(\alpha)}{},  \zgalS{(\nu)}{} \bigr) \right]
  ,
\end{split}
\\
\begin{split}
\EstimatorSummandPMEV{\obsell{}{}}{\obsell{}{}}^{(\alpha|\nu)\,\text{GG}}(\vartheta)
  &\approx
  \int\idiff[]{\zgalS{(\alpha)}{}} \pzgalS{(\alpha)}{}(\zgalS{(\alpha)}{}) 
  \int\idiff[]{\zgalS{(\nu   )}{}} \pzgalS{(\nu   )}{}(\zgalS{(\nu   )}{}) 
  \left[1 + \ccorr{\deltagalS{(\alpha)}{}}{\deltagalS{(\nu)}{}} \bigl(\fS^{(\alpha)} \vartheta, \chiS^{(\nu)} - \chiS^{(\alpha)}, \zgalS{(\alpha)}{},  \zgalS{(\nu)}{} \bigr) \right]
  \\&\quad\times
  \int\idiff[]{\chiD} \, \geomweight(\chiD, \zgalS{(\alpha)}{}) \, \geomweight(\chiD, \zgalS{(\nu)}{}) \,
  \pcorrpm{\deltashearMatter}{\deltashearMatter} \bigl(\fD \vartheta, \zgalD{}{} \bigr) 
  ,
\end{split}
\\
\EstimatorSummandPMEV{\obsell{}{}}{\obsell{}{}}^{(\alpha|\nu)\,\text{GI}}(\vartheta)
  &\approx
  \int\idiff[]{\zgalS{}{}} \, \pzgalS{(\nu   )}{}(\zgalS{}{})  \,
  \geomweightS{(\alpha)}{}( \chiS) \,
  \pcorrpm{\deltashearMatter}{(1 + \deltagalS{(\nu   )}{}) \intell{(\nu)}{}} \bigl(\fS \vartheta, \zgalS{}{} \bigr)
  ,
\\
\EstimatorSummandPMEV{\obsell{}{}}{\obsell{}{}}^{(\alpha|\nu)\,\text{IG}}(\vartheta)
  &\approx
  \int\idiff[]{\zgalS{}{}} \, \pzgalS{(\alpha   )}{}(\zgalS{}{}) \,
  \geomweightS{(\nu)}{}( \chiS) \,
  \pcorrpm{\deltashearMatter}{(1 + \deltagalS{(\alpha   )}{}) \intell{(\alpha)}{}} \bigl(\fS \vartheta, \zgalS{}{} \bigr)
  ,\quad\text{and}
\\
\EstimatorSummandPMEV{\obsell{}{}}{\obsell{}{}}^{(\alpha|\nu)\,\text{II}}(\vartheta)
  &\approx
  \int\idiff[]{\zgalS{(\alpha)}{}} \pzgalS{(\alpha)}{}(\zgalS{(\alpha)}{}) 
  \int\idiff[]{\zgalS{(\nu)}{}} \pzgalS{(\nu   )}{}(\zgalS{(\nu)}{}) \,
  \ccorrpm{(1 + \deltagalS{(\alpha)}{}) \intell{(\alpha)}{}}{(1 + \deltagalS{(\nu   )}{}) \intell{(\nu)}{}} 
  \bigl( \fS^{(\alpha)} \vartheta, \chiS^{(\nu)} -  \chiS^{(\alpha)}, \zgalS{(\alpha)}{}, \zgalS{(\nu)}{} \bigr)
  ,
\end{align}
\end{subequations}
with geometric factors $q$ given by Eqs.~\eqref{eq:geom_factor} and \eqref{app:eq:geometric_weight_for_source_galaxy_sample}.

The general expressions above may simplify considerably for certain cases. In the particular case of a higher-redshift source galaxy sample $\alpha$ with all galaxy redshifts well above any source galaxy redshift of a lower-redshift sample $\nu$,
\begin{subequations}
\begin{align}
\EstimatorSummandEV{1}{1}^{(\alpha|\nu)}(\vartheta)
  &=
  1
  ,
\\
\EstimatorSummandPMEV{\obsell{}{}}{\obsell{}{}}^{(\alpha|\nu)\,\text{GG}}(\vartheta)
  &\approx
  \int\idiff[]{\chiD} \, \geomweightS{(\alpha)}{}(\chiD) \, \geomweightS{(\nu)}{}(\chiD) \,
  \pcorrpm{\deltashearMatter}{\deltashearMatter} \bigl(\fD \vartheta, \zgalD{}{} \bigr) 
  ,
\\
\EstimatorSummandPMEV{\obsell{}{}}{\obsell{}{}}^{(\alpha|\nu)\,\text{GI}}(\vartheta)
  &\approx
  \int\idiff[]{\zgalS{}{}} \, \pzgalS{(\nu   )}{}(\zgalS{}{})  \,
  \geomweightS{(\alpha)}{}( \chiS) \,
  \pcorrpm{\deltashearMatter}{(1 + \deltagalS{(\nu   )}{}) \intell{(\nu)}{}} \bigl(\fS \vartheta, \zgalS{}{} \bigr)
  ,
\\
\EstimatorSummandPMEV{\obsell{}{}}{\obsell{}{}}^{(\alpha|\nu)\,\text{IG}}(\vartheta)
  &=
  0
  ,\quad\text{and}
\\
\EstimatorSummandPMEV{\obsell{}{}}{\obsell{}{}}^{(\alpha|\nu)\,\text{II}}(\vartheta)
  &=
  0
  .
\end{align}
\end{subequations}
 
In the particular case of a source galaxy sample $\alpha$ with a redshift distribution that is neither very broad nor narrow,
\begin{subequations}
\begin{align}
\EstimatorSummandEV{1}{1}^{(\alpha|\alpha)}(\vartheta)
  &\approx
  1 +
  \int\idiff[]{\zgalS{}{}} \left[ \pzgalS{(\alpha)}{}(\zgalS{}{}) \right]^2
  \left( \totder{\chiS}{\zgalS{}{}} \right)^{-1}
  \pcorr{\deltagalS{(\alpha)}{}}{\deltagalS{(\alpha)}{}} \bigl(\fS \vartheta,  \zgalS{}{} \bigr)
  ,
\\
\EstimatorSummandPMEV{\obsell{}{}}{\obsell{}{}}^{(\alpha|\alpha)\,\text{GG}}(\vartheta)
  &\approx
  \EstimatorSummandEV{1}{1}^{(\alpha|\alpha)}(\vartheta)
  \int\idiff[]{\chiD} \, \geomweightS{(\alpha)}{}(\chiD) \, \geomweightS{(\nu)}{}(\chiD) \,
  \pcorrpm{\deltashearMatter}{\deltashearMatter} \bigl(\fD \vartheta, \zgalD{}{} \bigr) 
  ,
\\
\EstimatorSummandPMEV{\obsell{}{}}{\obsell{}{}}^{(\alpha|\alpha)\,\text{GI}}(\vartheta)
  &\approx
  0
  ,
\\
\EstimatorSummandPMEV{\obsell{}{}}{\obsell{}{}}^{(\alpha|\alpha)\,\text{IG}}(\vartheta)
  &\approx
  0
  ,\quad\text{and}
\\
\EstimatorSummandPMEV{\obsell{}{}}{\obsell{}{}}^{(\alpha|\alpha)\,\text{II}}(\vartheta)
  &\approx
  \int\idiff[]{\zgalS{}{}} \left[ \pzgalS{(\alpha)}{}(\zgalS{}{}) \right]^2
  \left( \totder{\chiS}{\zgalS{}{}} \right)^{-1}
  \pcorrpm{(1 + \deltagalS{(\alpha)}{}) \intell{(\alpha)}{}}{(1 + \deltagalS{(\nu   )}{}) \intell{(\nu)}{}} 
  \bigl( \fS \vartheta, \zgalS{}{} \bigr)
  .
\end{align}
\end{subequations}

If galaxy density fluctuations and their correlations can all be neglected,
\begin{subequations}
\begin{align}
\EstimatorSummandEV{1}{1}^{(\alpha|\nu)}(\vartheta)
  &\approx 1
  ,
\\
\EstimatorSummandPMEV{\obsell{}{}}{\obsell{}{}}^{(\alpha|\nu)\,\text{GG}}(\vartheta)
  &\approx
  \int\idiff[]{\chiD} \, \geomweightS{(\alpha)}{}(\chiD) \, \geomweightS{(\nu)}{}(\chiD) \,
  \pcorrpm{\deltashearMatter}{\deltashearMatter} \bigl(\fD \vartheta, \zgalD{}{} \bigr) 
  ,
\\
\EstimatorSummandPMEV{\obsell{}{}}{\obsell{}{}}^{(\alpha|\nu)\,\text{GI}}(\vartheta)
  &\approx
  \int\idiff[]{\zgalS{}{}} \, \pzgalS{(\nu   )}{}(\zgalS{}{})  \,
  \geomweightS{(\alpha)}{}( \chiS) \,
  \pcorrpm{\deltashearMatter}{ \intell{(\nu)}{}} \bigl(\fS \vartheta, \zgalS{}{} \bigr)
  ,
\\
\EstimatorSummandPMEV{\obsell{}{}}{\obsell{}{}}^{(\alpha|\nu)\,\text{IG}}(\vartheta)
  &\approx
  \int\idiff[]{\zgalS{}{}} \, \pzgalS{(\alpha   )}{}(\zgalS{}{}) \,
  \geomweightS{(\nu)}{}( \chiS) \,
  \pcorrpm{\deltashearMatter}{\intell{(\alpha)}{}} \bigl(\fS \vartheta, \zgalS{}{} \bigr)
  ,\quad\text{and}
\\
\EstimatorSummandPMEV{\obsell{}{}}{\obsell{}{}}^{(\alpha|\nu)\,\text{II}}(\vartheta)
  &\approx
  \int\idiff[]{\zgalS{(\alpha)}{}} \pzgalS{(\alpha)}{}(\zgalS{(\alpha)}{}) 
  \int\idiff[]{\zgalS{(\nu)}{}} \pzgalS{(\nu   )}{}(\zgalS{(\nu)}{}) \,
  \ccorrpm{\intell{(\alpha)}{}}{\intell{(\nu)}{}} 
  \bigl( \fS^{(\alpha)} \vartheta, \chiS^{(\nu)} -  \chiS^{(\alpha)}, \zgalS{(\alpha)}{}, \zgalS{(\nu)}{} \bigr)
  .
\end{align}
If, in addition, the redshift distributions are slowly varying functions of redshift, the II term may be written as:
\begin{equation}
  \EstimatorSummandPMEV{\obsell{}{}}{\obsell{}{}}^{(\alpha|\nu)\,\text{II}}(\vartheta)
  \approx
  \int\idiff[]{\zgalS{}{}} \pzgalS{(\alpha)}{}(\zgalS{}{}) \,
                           \pzgalS{(\nu   )}{}(\zgalS{}{}) \,
  \left( \totder{\chiS}{\zgalS{}{}} \right)^{-1}
  \pcorrpm{\intell{(\alpha)}{}}{\intell{(\nu)}{}} \bigl( \fS \vartheta, \zgalS{}{} \bigr)
  .
\end{equation}
\end{subequations}

 \section{Integral relations for correlations involving intrinsic ellipticities}
 \label{app:integral_relations}

Here, we briefly outline how one can derive the relation~\eqref{eq:relation_GI_mI} connecting the density-intrinsic ellipticity correlation and the shear-intrinsic ellipticity correlation. 
Consider statistically homogeneous and isotropic real random fields
$\rfX{(i,0)}{}:\R^2 \to \R$, where $i\in\N$ and $\mathrm{X} \in \{\mathrm{E}, \mathrm{B}\}$. Due to statistical homogeneity and isotropy, the two-point correlation function of $\rfX{(i,0)}{}$ and  $\rfY{(j,0)}{}$ can be expressed as:
 \begin{equation}
 \begin{split}
   \EV{\rfX{(i,0)}{} (\vx + \vr) \, \rfY{(j,0)}{} (\vx)} &=
   \frac{1}{2\pi} 
   \int_0^{\infty} \idiff[]{k}\, k\,
   J_0 (k |\vr|) \,
   \PSXY{i}{j}(k)
 .
 \end{split}
 \end{equation} 
 Here, $J$ denotes the Bessel function of the first kind, and $\PSXY{i}{j}$ denotes the (cross-)power spectrum of $\rfX{(i,0)}{}$ and  $\rfY{(j,0)}{}$, satisfying
 \begin{equation}
 \EV{\ftrfX{(i,0)*}{} (\vk) \, \ftrfY{(j,0)}{} (\vk')} = (2\pi)^2 \DiracDelta(\vk - \vk') \PSXY{i}{j}(|\vk|),
 \end{equation}
and $\ft{\rfsymbol}$ denotes the two-dimensional Fourier transform of $\rfsymbol$.

The \lq{}E-mode\rq{} scalar field $\rfE{(i,0)}{}$ and the \lq{}B-mode\rq{} scalar field $\rfB{(i,0)}{}$ may be combined to obtain a complex scalar field $\rf{(i,0)}{} = \rfE{(i,0)}{} + \ii \rfB{(i,0)}{}$.
%
One may then define associated complex spin-2 fields $\rf{(i,2)}{}: \R^2  \to \C$ that are connected to the complex scalar fields $\rf{(i,0)}{}$ via the relation
\begin{equation}
  \ftrf{(i,2)}{}(\vk) = \ee^{ 2 \ii \varphi(\vk)}  \ftrf{(i,0)}{}(\vk),
\end{equation} 
where $\varphi(\vk)$ denotes the polar angle of $\vk=(k_1,k_2)$. Fields $\rfr{(i,2)}{}(\vx; \vr)$, tangential components $\rft{(i,2)}{}(\vx; \vr)$, and cross components $ \rfx{(i,2)}{}(\vx; \vr)$ relative to direction $\vr$ are defined by:
\begin{align}
	 \rfr{(i,2)}{}(\vx; \vr) &=  - \ee^{-2 \ii \varphi(\vr)} \rf{(i,2)}{}(\vx),\\
	 \rft{(i,2)}{}(\vx; \vr) &=  \,\frac{1}{2}\left[ \rfr{(i,2)}{}(\vx; \vr) + \rfr{(i,2)*}{}(\vx; \vr) \right],\\
	 \rfx{(i,2)}{}(\vx; \vr) &=  \frac{1}{2\ii}\left[ \rfr{(i,2)}{}(\vx; \vr) - \rfr{(i,2)*}{}(\vx; \vr) \right].
\end{align}
Then,\footnote{
The fields $\rf{}{}$ will be later identified with spacetime fields also depending on l.o.s. distance $\chi$ and redshift $z$. We thus use the symbol $\ccorrsymbol$ to denote correlations of these fields to maintain consistency with the notation introduced in Section~\ref{sec:theory:intrinsic_correlations}.
}
\begin{equation}
 \begin{split}
\ccorr{\rfE{(i,0)}{}}{\rft{(j,2)}{}} (|\vr|) &= 
   \EV{\rfE{(i,0)}{} (\vx + \vr) \, \rft{(j,2)}{} (\vx; \vr)} 
	= 
   \frac{1}{2\pi}
	 \int_0^{\infty} \idiff[]{k}\, k\,
   J_2 (k |\vr|) \,
 	\PSEE{i}{j}(k)
.
\end{split}
\end{equation}
Conversely,
\begin{equation}
 \begin{split}
 	\PSEE{i}{j}(k)
	= 
  2\pi
	 \int_0^{\infty} \idiff[]{r}\, r\,
		J_2 (k r) \,
\ccorr{\rfE{(i,0)}{}}{\rft{(j,2)}{}} (r)
.
\end{split}
\end{equation}

Moreover,
\begin{equation}
 \begin{split}
	\ccorrp{\rf{(i,2)}{}}{\rf{(j,2)}{}} (|\vr|)
	&= 
   \EV{\rft{(i,2)}{} (\vx + \vr; \vr) \, \rft{(j,2)}{} (\vx; \vr) + \rfx{(i,2)}{} (\vx + \vr; \vr) \, \rfx{(j,2)}{} (\vx; \vr)}
	\\&= 
   \frac{1}{2\pi}
	 \int_0^{\infty} \idiff[]{k}\, k\,
   J_0 (k |\vr|) \,
 	\left[ \PSEE{i}{j}(k)	+  \PSBB{i}{j}(k) \right]
.
\end{split}
\end{equation} 

If $\rfB{(i,0)}{} = 0$, then $  \PSBB{i}{j}(k) = 0$, and thus,
\begin{equation}
	\ccorrp{\rf{(i,2)}{}}{\rf{(j,2)}{}} (r)
		= 
   \frac{1}{2\pi}
	 \int_0^{\infty} \idiff[]{k}\, k\,
   J_0 (k r) \, \PSEE{i}{j}(k)
,
\end{equation} 
By exploiting the properties of the Bessel functions, one obtains the relation: 
\begin{equation}
\begin{split}
	\ccorrp{\rf{(i,2)}{}}{\rf{(j,2)}{}} (r) &=
 			\frac{1}{2\pi}
 	 \int_0^{\infty} \idiff[]{k}\, k\,
    J_0 (k r) \,
   2\pi
 	 \int_0^{\infty} \idiff[]{r'}\, r'\,
 		J_2 (k r') \,
 \ccorr{\rfE{(i,0)}{}}{\rft{(j,2)}{}} (r')
 	\\&=
 	 \int_0^{\infty} \idiff[]{r'}\, r'	\left[
 	 \int_0^{\infty} \idiff[]{k}\, k\,
    J_0 (k r) \,
 		J_2 (k r') \right]
 \ccorr{\rfE{(i,0)}{}}{\rft{(j,2)}{}} (r')	
 	\\&=
	 \int_0^{\infty} \idiff[]{r'}\, r'	 \mathcal{G}_{+, \tang}(r,r')
\ccorr{\rfE{(i,0)}{}}{\rft{(j,2)}{}} (r')	
	.
\end{split}
\end{equation}
with
\begin{equation}
	 \mathcal{G}_{+, \tang}(r,r') =
-\frac{1}{r'} \DiracDelta(r' - r) + 
\frac{2}{r^{\prime 2}} \HeavisideTheta(r' - r).
\end{equation}

Setting
\begin{subequations}
\begin{align}
	\rfE{(i,0)}{}(\vx ) &= \deltaMatter (\vx, \chi', z),\\
   \rf{(j,2)}{}(\vx ) &= \left[1 + \deltagalS{}{}(\vx, \chi, z) \right]  \intell{}{}(\vx, \chi, z),
\end{align}
\end{subequations}
one obtains
\begin{equation}
\begin{split}
	\ccorrp{\deltashearMatter}{(1 + \deltagalS{}{})\intell{}{}} (r, \chi' - \chi, z, z) &=
	 \int_0^{\infty} \idiff[]{r'}\, r'	 \mathcal{G}_{+, \tang}(r,r')
\ccorr{\deltaMatter}{(1 + \deltagalS{}{})\intellt{}{}} (r', \chi' - \chi, z, z)
,
\end{split}
\end{equation}
and after carrying out the line-of-sight projection,
\begin{equation}
\begin{split}
	\pcorrp{\deltashearMatter}{(1 + \deltagalS{}{})\intell{(2)}{}} (r, z) &=
	 \int_0^{\infty} \idiff[]{r'}\, r'	 \mathcal{G}_{+, \tang}(r,r')
\ccorr{\deltaMatter}{(1 + \deltagalS{}{})\intellt{}{}} (r', z)
,
\end{split}
\end{equation}
which is equivalent to Eq.~\eqref{eq:relation_GI_mI}.

\section{Estimating projected correlation functions from simulations}
\label{app:projected_correlations_from_simulations}

We estimate projected correlations \eqref{eq:df_projected_correlation} from the simulation snapshots by correlating projected fields and exploiting the Limber approximation~\eqref{eq:app:Limber_simulation_snapshot}. As discussed in Appendix~\ref{app:projected_correlations}, this approximation becomes exact for our setup, which employs fields at constant redshift, parallel projection, periodic boundary conditions, and a constant projection weight over the full depth $L$ of the simulation box.

We use the simulation particles in a simulation snapshot at redshift $z$ to define the matter overdensity field within the simulation box at that redshift by:\footnote{For simplicity, we omit redshift arguments of the fields, since we only correlative fields of the same snapshot, but never across snapshots.}
\begin{equation}
	\deltaMatter(\vx, \chi) = \frac{1}{\rhoMatterMean}\sum_{i=1}^{\Npart{}{}} \mpart{(i)}{} \DiracDelta (\vx - \xpart{(i)}{}) \DiracDelta (\chi - \chipart{(i)}{}) - 1.
\end{equation}
Here, $\vx$ denotes the two-dimensional comoving position transverse to the viewing direction (chosen as one of the principal directions of the simulation box),  $\chi$ denotes the comoving position along the viewing direction, $\Npart{}{}$ denotes the number of particles in the snapshot, $\rhoMatterMean$ denotes the cosmic mean comoving matter density, $\mpart{(i)}{}$ denotes the mass of the simulation particle $i$, and $\xpart{(i)}{}$ and $\chipart{(i)}{}$, resp., denote its comoving positions perpendicular and along the viewing direction.

We also use the galaxies in each simulation snapshot to define fields within the simulation box. In particular, the galaxy number density of the density sample reads:
\begin{equation}
 \ngalD{}{}(\vx, \chi) = \sum_{i=1}^{\NgalD{}{}} \DiracDelta (\vx - \xgalD{(i)}{}) \DiracDelta (\chi - \chiD^{(i)}),
\end{equation} 
where $\xgalD{(i)}{}$ and $\chiD^{(i)})$, resp., denote the comoving positions of the galaxies perpendicular and along the viewing direction. For the ellipticity sample, we define 
\begin{equation}
 [\ngalS{}{} \intell{}{}] (\vx, \chi) = \sum_{i=1}^{\NgalS{}{}} \intell{(i)}{} \DiracDelta (\vx - \xgalS{(i)}{}) \DiracDelta (\chi - \chiS^{(i)}).
\end{equation} 

For each principal direction of the simulation box as viewing direction, we compute the following two-dimensional projected fields:
\begin{align}
 \pi_{\deltaMatter} (\vx) &= \int_0^{L} \idiff{\chi}\, \deltaMatter(\vx, \chi),
 \\
 \pi_{\deltashearMatter} (\vx) &= \int_0^{L} \idiff{\chi}\, \deltashearMatter(\vx, \chi),
 \\
 \pi_{\ngalD{}{}} (\vx) &= \int_0^{L} \idiff{\chi}\, \ngalD{}{}(\vx, \chi) = \sum_{i=1}^{\NgalD{}{}} \DiracDelta(\vx -\xgalD{(i)}{}),
 \\
 \pi_{\ngalS{}{} \intell{}{}} (\vx) &=  \int_0^{L} \idiff{\chi}\, [\ngalS{}{} \intell{}{}](\vx, \chi) = \sum_{i=1}^{\NgalS{}{}}\intell{(i)}{}  \DiracDelta(\vx -\xgalS{(i)}{}).
\end{align}
The projected matter shear contrast $\pi_{\deltashearMatter} (\vx)$ can be obtained directly from the projected matter overdensity $\pi_{\deltaMatter}$,  e.g. by exploiting their simple relation $\ft{\pi}_{\deltashearMatter}(\vk) = (k_1 + \ii k_2)^2 |\vk|^{-2} \ft{\pi}_{\deltaMatter}(\vk)$ in Fourier space.

Raw correlations of the projected fields are then computed as:
\begin{align}
\RandomEstimatorSum{1}{1} (r) &= \!
 \int_{\FOV}\idiff{\vx} \int_{\FOV}\idiff{\vx'} 
   \Delta\bigl(r, |\vx' - \vx| \bigr)
,\\
  \PEstimatorSum{\deltaMatter}{\ngalS{}{} \intellt{}{}} (r) &= \!
 \int_{\FOV}\idiff{\vx} \! \int_{\FOV}\idiff{\vx'} 
   \Delta\bigl(r, |\vx' - \vx| \bigr)
  \pi_{\deltaMatter} \bigl(\vx \bigr) \pi_{\ngalS{}{} \intell{}{},\tang} \bigl(\vx';\vx' - \vx \bigr)
,\\
  \PEstimatorSum{\ngalD{}{}}{\ngalS{}{} \intellt{}{}} (r) &= \!
 \int_{\FOV}\idiff{\vx} \! \int_{\FOV}\idiff{\vx'} 
   \Delta\bigl(r, |\vx' - \vx| \bigr)
  \pi_{\ngalD{}{}} \bigl(\vx \bigr) \pi_{\ngalS{}{} \intell{}{},\tang} \bigl(\vx';\vx' - \vx \bigr)
,\\
 \PEstimatorSumPM{\deltashearMatter}{\deltashearMatter} (r) &= \!
 \int_{\FOV}\idiff{\vx} \! \int_{\FOV}\idiff{\vx'} 
   \Delta\bigl(r, |\vx' - \vx| \bigr)
  \left[  
  \pi_{\deltashearMatter,\tang} \bigl(\vx; \vx' - \vx \bigr) \pi_{\deltashearMatter,\tang} \bigl(\vx';\vx' - \vx \bigr)
  \pm
  \pi_{\deltashearMatter,\cross} \bigl(\vx; \vx' - \vx \bigr) \pi_{\deltashearMatter,\cross} \bigl(\vx';\vx' - \vx \bigr)
  \right]
  ,\\
 \PEstimatorSumPM{\deltashearMatter}{{\ngalS{}{} \intell{}{}}} (r) &= \!
 \int_{\FOV}\idiff{\vx} \! \int_{\FOV}\idiff{\vx'} 
   \Delta\bigl(r, |\vx' - \vx| \bigr)
  \left[  
  \pi_{\deltashearMatter,\tang} \bigl(\vx; \vx' - \vx \bigr) \pi_{\ngalS{}{} \intell{}{},\tang} \bigl(\vx';\vx' - \vx \bigr)
  \pm
  \pi_{\deltashearMatter,\cross} \bigl(\vx; \vx' - \vx \bigr) \pi_{\ngalS{}{} \intell{}{},\cross} \bigl(\vx';\vx' - \vx \bigr)
  \right]
  ,\\
 \PEstimatorSumPM{{\ngalS{}{} \intell{}{}}\!}{{\ngalS{}{} \intell{}{}}} (r) &= \!
 \int_{\FOV}\idiff{\vx} \! \int_{\FOV}\idiff{\vx'} 
   \Delta\bigl(r, |\vx' - \vx| \bigr)
  \left[  
  \pi_{\ngalS{}{} \intell{}{}\!,\tang} \bigl(\vx; \vx' - \vx \bigr) \pi_{\ngalS{}{} \intell{}{}\!,\tang} \bigl(\vx';\vx' - \vx \bigr)
  \pm
  \pi_{\ngalS{}{} \intell{}{}\!,\cross} \bigl(\vx; \vx' - \vx \bigr) \pi_{\ngalS{}{} \intell{}{}\!,\cross} \bigl(\vx';\vx' - \vx \bigr)
  \right]
  \!,\!\!  
\end{align} 
where $\FOV=[0,L]^2$, and $\Delta\bigl(r, |\vx' - \vx| \bigr)$ denotes a suitable radial binning function as in the previous Sections. These raw correlations can be efficiently computed by sampling the fields on a regular mesh covering the field of view, and then employing Fast Fourier Transforms and the convolution theorem \citep[see, e.g.,][]{HilbertHartlapSchneider2011}.

Assuming that the simulation produces a fair sample, and that spatial averaging approximates ensemble averaging, the measured raw correlations should be close to their expectation values. With the help of Limber's equation, one obtains for the raw matter density-ellipticity correlation:
\begin{equation}
\begin{split}
& \bEV{\PEstimatorSum{\deltaMatter}{\ngalS{}{} \intellt{}{}} (r)}
\\&=
 \int_{\FOV}\idiff{\vx} \! \int_{\FOV}\idiff{\vx'} 
   \Delta\bigl(r, |\vx' - \vx| \bigr)
  \bEV{\pi_{\deltaMatter} \bigl(\vx \bigr) \pi_{\ngalS{}{} \intell{}{},\tang} \bigl(\vx';\vx' - \vx \bigr)}
\\&=
 \int_{\FOV}\idiff{\vx} \! \int_{\FOV}\idiff{\vx'} 
   \Delta\bigl(r, |\vx' - \vx| \bigr)
\int_0^L\idiff{\chi} \int_0^L\idiff{\chi'}
  \bEV{\deltaMatter \bigl(\vx, \chi \bigr) [\ngalS{}{} \intellt{}{}] \bigl(\vx', \chi';\vx' - \vx \bigr)}
\\&=
 \int_{\FOV}\idiff{\vx} \! \int_{\FOV}\idiff{\vx'} 
   \Delta\bigl(r, |\vx' - \vx| \bigr)
\int_0^L\idiff{\chi} \int_0^L\idiff{\chi'}
\\&\quad\times
  \EV{
    \biggl\{
\prod_{i=1}^{\NgalS{}{}} 
\frac{1}{L^3} 
\!\int_{\FOV} \idiff[2]{\xgalS{(i)}{}} \!\int_0^L\idiff{\chiS^{(i)}}
\left[ 1 + \deltagalS{}{} \bigl( \xgalS{(i)}{},  \chiS^{(i)} \bigr) \right]
\biggr\}
  \deltaMatter \bigl(\vx, \chi \bigr) 
  \sum_{i=1}^{\NgalS{}{}}
  \intellt{}{}\bigl(\xgal{(i)}{},\chiS^{(i)};\xgal{(i)}{}  - \vx \bigr) 
  \DiracDelta \bigl(\vx' - \xgalS{(i)}{} \bigr) \DiracDelta \bigl(\chi' - \chiS^{(i)} \bigr)
  }_{\!\delta, \intell{}{}}
  \!\!\!\!\!
\\&=
 \int_{\FOV}\idiff{\vx} \! \int_{\FOV}\idiff{\vx'} 
   \Delta\bigl(r, |\vx' - \vx| \bigr)
\int_0^L\idiff{\chi} \int_0^L\idiff{\chi'}
  \EV{
  \sum_{i=1}^{\NgalS{}{}}
\frac{1}{L^3} 
\deltaMatter \bigl(\vx, \chi \bigr)
  \left[ 1 + \deltagalS{}{} \bigl( \vx',  \chi' \bigr) \right]
   \intellt{}{}\bigl(\vx', \chi'; \vx' - \vx \bigr) 
  }_{\!\delta, \intell{}{}}
\\&=
 \int_{\FOV}\idiff{\vx} \! \int_{\FOV}\idiff{\vx'} 
   \Delta\bigl(r, |\vx' - \vx| \bigr)
\int_0^L\idiff{\chi} \int_0^L\idiff{\chi'}\,
 \meanngalS{}{} \,
\ccorr{\deltaMatter}{(1 + \deltagalS{}{})\intellt{}{}} \bigl( \vx' - \vx, \chi' - \chi \bigr) 
\\&=
 \int_{\FOV}\idiff{\vx} \! \int_{\FOV}\idiff{\vx'}
   \Delta\bigl(r, |\vx' - \vx| \bigr)
 \meanngalS{}{} L 
\pcorr{\deltaMatter}{(1 + \deltagalS{}{})\intellt{}{}} \bigl( \vx' - \vx \bigr) 
\\&\approx
\RandomEstimatorSum{1}{1} (r)
L \meanngalS{}{} \,\pcorr{\deltaMatter}{(1 + \deltagalS{}{})\intellt{}{}} (r) 
.
\end{split}
\end{equation}

The expectations for the other raw correlations read:
\begin{align}
\bEV{ \PEstimatorSum{\ngalD{}{}}{\ngalS{}{} \intellt{}{}} (r) }
  &\approx
  \RandomEstimatorSum{1}{1} (r)
 \,  L \, \meanngalD{}{}  \, \meanngalS{}{} \, \pcorr{(1 + \deltagalD{}{})}{(1 + \deltagalS{}{})\intellt{}{}} (r) 
,\\
\bEV{ \PEstimatorSumPM{\deltashearMatter}{\deltashearMatter} (r) }
  &\approx
  \RandomEstimatorSum{1}{1} (r)
  \, L \, \pcorrpm{\deltashearMatter}{\deltashearMatter} (r)
,\\
\bEV{ \PEstimatorSumPM{\deltashearMatter}{{\ngalS{}{} \intell{}{}}} (r) }
  &\approx
  \RandomEstimatorSum{1}{1} (r)
 \,  L\,  \meanngalS{}{}\,  \pcorrpm{\deltashearMatter}{(1 + \deltagalS{}{})\intell{}{}} (r)
,\\
\bEV{ \PEstimatorSumPM{{\ngalS{}{} \intell{}{}}\!}{{\ngalS{}{} \intell{}{}}} (r) }
  &\approx  
  \RandomEstimatorSum{1}{1} (r)
  \, L \, \meanngalS{2}{}\,  \pcorrpm{(1 + \deltagalS{}{})\intell{}{}}{(1 + \deltagalS{}{})\intell{}{}} (r)
.
\end{align}

\bsp  
\label{lastpage}
\end{document}